\documentclass[11pt,a4paper]{article}
\usepackage{jheppub}

\DeclareMathOperator{\tr}{tr} % trace
\DeclareMathOperator{\Spin}{Spin} % Spin group
\DeclareMathOperator{\diag}{diag} % diagonal matrix
\makeatletter
\def\Ddots{\mathinner{\mkern1mu\raise\p@
\vbox{\kern7\p@\hbox{.}}\mkern2mu
\raise4\p@\hbox{.}\mkern2mu\raise7\p@\hbox{.}\mkern1mu}}
\makeatother

\title{Static Charges in the Low-Energy
Theory of the S-Duality Twist}
\author[a]{Ori J. Ganor,}
\author[b]{Yoon Pyo Hong,}
\author[a]{Ruza Markov,}
\author[a]{and Hai Siong Tan}
\affiliation[a]{
Department of Physics,
  University of California,\\
Berkeley, CA 94720, U.S.A.}
\affiliation[b]{School of Physics,
Korea Institute for Advanced Study,\\
Seoul 130-722, Korea}
\emailAdd{origa@socrates.berkeley.edu}
\emailAdd{yph@kias.re.kr}
\emailAdd{ruza@berkeley.edu}
\emailAdd{haisiong\_tan@berkeley.edu}

\abstract{
We continue the study of the low-energy limit of $N=4$ super Yang--Mills theory compactified on a circle with S-duality and R-symmetry twists that preserve $N=6$ supersymmetry in 2+1D. We introduce external static supersymmetric quark and anti-quark sources into the theory and calculate the Witten Index of the resulting Hilbert space of ground states on $T^2.$  Using these results we compute the action of simple Wilson loops on the Hilbert space of ground states without sources.
In some cases we find disagreement between our results for the Wilson loop eigenvalues and previous conjectures about a connection with Chern-Simons theory.
}
\keywords{S-Duality, Wilson loops, Chern--Simons theory}
\subheader{
Preprint:
UCB-PTH-11/13,
KIAS-P11069
}

\begin{document}
\maketitle
\flushbottom

% ==============================================================
% ==============================================================
\newcommand{\secref}[1]{\S\ref{#1}}
\newcommand{\figref}[1]{Figure~\ref{#1}}
\newcommand{\appref}[1]{Appendix~\ref{#1}}
\newcommand{\tabref}[1]{Table~\ref{#1}}

\newcommand\SUSY[1]{{${\mathcal{N}}={#1}$}}  % supersymmetry
\newcommand\px[1]{{\partial_{#1}}}

\def\be{\begin{equation}}
\def\ee{\end{equation}}
\def\bear{\begin{eqnarray}}
\def\eear{\end{eqnarray}}
\def\nn{\nonumber}

\newcommand\bra[1]{{\left\langle{#1}\right\rvert}} % bra
\newcommand\ket[1]{{\left\lvert{#1}\right\rangle}} % ket

\newcommand{\C}{\mathbb{C}}
\newcommand{\R}{\mathbb{R}}
\newcommand{\Z}{\mathbb{Z}}

\def\SL{{{\mbox{\rm SL}}}} % SL() group
\def\PSL{{{\mbox{\rm PSL}}}} % PSL() group
\def\gYM{g_{\text{YM}}} % YM coupling constant

\def\zs{{\widehat{\mathcal{S}}}} % S-twist operator
\def\zr{{\widehat{\mathcal{R}}}} % R-twist operator

\def\xa{{\mathbf{a}}} % SL(2,Z) variable
\def\xb{{\mathbf{b}}} % SL(2,Z) variable
\def\xc{{\mathbf{c}}} % SL(2,Z) variable
\def\xd{{\mathbf{d}}} % SL(2,Z) variable

\def\zg{{\mathbf{g}}} % SL(2,Z) element
\def\zx{{\mathbf{x}}} % SL(2,Z) element
\def\bQ{{\overline{Q}}} % complex conjugate supercharge
\def\bpsi{{\overline{\psi}}}
\def\Op{{\mathcal{O}}} % operator
\def\a{\alpha}
\def\b{\beta}
\def\dta{{\dot{\a}}}
\def\dtb{{\dot{\b}}}
\def\lvk{{k}} % Chern--Simons level
\def\pht{{\upsilon}} % phase of $\tau$
\def\ord{{\mathbf{r}}} % $2\pi/\pht$
\def\gtw{{\gamma}} % R-symmetry twist along the circle
\def\ftw{{\varphi}} % phases of eigenvalues of $\gtw$
\def\Zphi{{Z}} % complex scalar

\def\wx{$-$} % table:a wrapped direction
\def\tx{$\div$} % table:a twist direction
\def\nx{$\times$}% table: a direction that doesn't exist
\def\ox{$\vdash\!\!\dashv$} % table:direction for an open brane

\def\Mst{M_{\text{st}}} % string scale
\def\Mpl{M_{\text{p}}} % Planck scale
\def\gst{g_{\text{st}}} % string coupling constant
\def\gIIA{g_{\text{IIA}}} % string coupling constant IIA
\def\gIIB{g_{\text{IIB}}} % string coupling constant IIB
\def\lst{\ell_{\text{st}}} % string length.
\def\lp{\ell_{\text{P}}} % Planck length.
\def\apr{{\a'}} % string scale
\def\Area{{\mathcal{A}}} % area

\def\btau{{\overline{\tau}}} % c.c. of coupling constant

\def\Psym{{\mathcal{U}}} % symmetry $Z_k$ by isometry
\def\Qsym{{\mathcal{V}}} % symmetry $Z_k$ by homology

\def\ef{{\mathbf{e}}} % electric flux
\def\mf{{\mathbf{m}}} % magnetic flux

\def\fpZ{{\zeta}} % fixed point (zero mode)
\def\wMa{{M_a}} % worldsheet winding along a cycle
\def\wMb{{M_b}} % worldsheet winding along b cycle

\def\TdS{{\mathcal{S}}} % T-duality element S,rotation by $\pi/2$
\def\TdT{{\mathcal{T}}} % T-duality element T

\def\WilV{{\mathcal{W}}} % Wilson loop operator

\def\perm{{\sigma}} % permutation
\def\brperm{{\lbrack\perm\rbrack}} % conjugacy class

\def\rhocp{{\rho}} % complex structure of type-IIB $T^2$
\def\brhocp{{\overline{\rho}}} % complex conjugate of $\rhocp$

\def\xL{{L}} % type-IIB side radii in directions 1,2
\def\xR{{R}} % type-IIA/B side radius in direction 3

\def\np{{\tilde{n}}} % single-particle string winding number

\def\Hilb{{\mathcal{H}}}

% ==============================================================

\def\xPhi{{\Phi}}
\def\wxPhi{{\widetilde{\Phi}}}
\def\xF{{\mathcal F}}
\def\xA{{\mathcal A}}
\def\xCh{{\chi}}
\def\Disc{{D}}

\def\WLoop{{C}} % Wilson loop

\def\pvar{{\mathfrak{p}}}
\def\qvar{{\mathfrak{q}}}

\def\DXiv{{\Delta}} % offset of manipulator D3 in direction 4

\newcommand\Sector[1]{\lbrack{#1}\rbrack} % Sector notation
\newcommand\ez[1]{{\{{#1}\}}} % for state notation (z position)
\def\Aqbq{{\mathcal A}_{q\overline{q}}} % pair annihilation operator
\def\OpR{{\mathcal R}} % operator that moves charges
\def\SRt{{\mathcal P}} % SR-twist on spinors
\def\Baa{{\mathcal B}} % space of charge positions.

\def\LoopC{{C}} % closed loop in $T^2$
\def\wn{{\ell}} % winding number.
\def\cycA{{\alpha}} % alpha-cycle on $T^2$
\def\cycB{{\beta}} % beta-cycle on $T^2$
\def\ux{{\xi}} % coordinate in the equivalent 2+1D problem
\def\Kpq{{K}} % level of p-q system for appendix
\def\Aby{{\mathcal A}} % Berry connection

\def\wbq{{w_{\overline{q}}}} % anti-quark position.

\def\wT{{w}} % complex coordinate on $T^2$
\def\bwT{{\overline{w}}} % complex conjugate coordinate on $T^2$

\def\DetM{{\Delta}} % determinant of M

\def\Matx{{\mathcal{M}}} % matrix

\newcommand\XS[1]{{\langle{#1}\rangle}} % sector notation
\def\SRg{{\Omega}} % S-R-twist in type-IIA on spinors

\def\RS{{\Sigma}} % Riemann surface

\def\Zlbd{{\mathcal{L}}} % flat line bundle over $\RS$
\def\Klbd{{\mathcal{K}}} % canonical line bundle
\def\Nlbd{{\mathcal{N}}} % normal line bundle
\def\Olbd{{\mathcal{O}}} % trivial bundle

\def\tenx{{\natural}} % notation for 10 as an index
\def\pSUSY{{\epsilon}}
\def\bpSUSY{{\overline{\epsilon}}}
\def\Xsc{{\Phi}}
\def\Xsp{{\lambda}}
\def\bXsp{{\overline{\lambda}}}
\def\OpAC{{\varrho}} % fermionic zero mode operators

\def\scX{{X}} % scalar field along F1
\def\fpsi{{\psi}} % fermion field along F1

\def\MatXp{{\mathcal{K}}} % matrix coupling
\def\MatXq{{\mathcal{N}}} % matrix coupling

\def\xvar{{\mathfrak{u}}} % discrete $x_1$ coordinate
\def\yvar{{\mathfrak{q}}}

\def\PMat{{\mathfrak{B}}} % Punch matrix (binding matrix)

%%%
\def\Lbd{{\mathcal{L}}} % line bundle
\def\Vbd{{\mathcal{V}}} % vector bundle

\def\spm{{\mathfrak{m}}} % twice spin for CS representations

\def\iSt{{\mathfrak{i}}} % string counter 1,...,n
\def\jSt{{\mathfrak{j}}} % string counter 1,...,n
\def\iStz{{\mathfrak{i}'}} % string counter 0,...,n
\def\jStz{{\mathfrak{j}'}} % string counter 0,...,n
\def\iDb{{\mathfrak{a}}} % D2-brane counter 1,...,m
\def\jDb{{\mathfrak{b}}} % D2-brane counter 1,...,m
\def\iDbz{{\mathfrak{a}'}} % D2-brane counter 0,1,...,m
\def\jDbz{{\mathfrak{b}'}} % D2-brane counter 0,1,...,m
\def\iStIIB{{\mathfrak{c}}} % type-IIB string counter 1,...,2m

\def\xJ{{\mathcal{J}}} % U(1) generator

% ==============================================================
% ==============================================================

% - - - - - - - - - - - - - - - - - - - - - - - - - - - - - - -

% - - - - - - - - - - - - - - - - - - - - - - - - - - - - - - -

% - - - - - - - - - - - - - - - - - - - - - - - - - - - - - - -

% - - - - - - - - - - - - - - - - - - - - - - - - - - - - - - -

% - - - - - - - - - - - - - - - - - - - - - - - - - - - - - - -
%\def\bpXd{{
%\begin{picture}(18,18)
%\color{black}
%\thinlines
%\put(9,1){\vector(0,1){16}}
%\put(1,9){\vector(1,0){16}}
%\color{blue}
%\thicklines
%\put(4,4){\line(1,1){10}}
%\end{picture}
%}}
% - - - - - - - - - - - - - - - - - - - - - - - - - - - - - - -

% ==============================================================

\section{Introduction}
\label{sec:Intro}

For $U(n)$ gauge group, S-duality of \SUSY{4} super Yang--Mills theory \cite{Montonen:1977sn,Witten:1978mh,Osborn:1979tq} asserts that the theory at the complex combination of coupling constant and $\theta$ angle
$$
\tau\equiv \frac{4\pi i}{\gYM^2} + \frac{\theta}{2\pi}
$$
is equivalent to the same theory at complex coupling $-1/\tau.$ The conjecture has passed many tests (see for instance
\cite{Sen:1994yi,Vafa:1994tf,Bershadsky:1995vm,Harvey:1995tg}) and is generally accepted as true, even though no proof exists. Over the years, much insight has been accumulating on the way S-duality works. Some notable breakthroughs include the geometric realization of S-duality in terms of the $(2,0)$-theory \cite{Witten:1995zh,Vafa:1997mh}, the connection with the geometric Langlands program \cite{Kapustin:2006pk}, and the discovery \cite{Gaiotto:2008ak,Gaiotto:2008sa} of the role of certain strongly-coupled 2+1D \SUSY{4} theories \cite{Intriligator:1996ex} as intertwiners between a supersymmetric boundary condition and its S-dual.

Another way to explore S-duality was recently examined in \cite{Ganor:2008hd,Ganor:2010md}. There, an S-duality twist was introduced into a compactification of \SUSY{4} super Yang--Mills (SYM) on $S^1$ in a way that preserves \SUSY{6} supersymmetry in 2+1D. An S-duality twist is an unusual possible boundary condition that is permissible when the complex coupling constant is set to the self-dual value $\tau=i.$ The S-duality twist is then achieved by inserting the transformation that realizes $\tau\rightarrow -1/\tau$ at some point along $S^1.$ We can then further compactify the remaining two spatial dimensions on, say, $T^2.$ The Hilbert space of ground states of this compactification, which was studied in \cite{Ganor:2010md}, provides insight into the operator that realizes S-duality. We refer to the resulting three-dimensional low-energy theory as {\it Tr-S}, because correlation functions of Wilson loops $\langle W(\LoopC_1)\cdots W(\LoopC_l)\rangle_{\text{Tr-S}}$ in this theory can be interpreted as, roughly speaking, a regularized version of the trace $\tr ((-1)^F S R W(C_1)\cdots W(C_l))$, where the trace is taken over the Hilbert space of \SUSY{4} $U(n)$ SYM at the self-dual coupling $\tau=i$, $S$ is the S-duality operator, $R$ is an appropriate $SU(4)$ R-symmetry operator that is inserted in order to preserve \SUSY{6} supersymmetry in 2+1D, $F$ is the fermion number (which is equivalent to a central element of the R-symmetry group), and $W(\LoopC_1), \dots, W(\LoopC_l)$ are Wilson loop operators associated with the loops $\LoopC_1,\dots,\LoopC_l$ in $\R^3$.
%\footnote{We are grateful to E.~Witten for pointing out a missing $(-1)^F$ in an earlier version.}

S-duality is part of a larger $\SL(2,\Z)$ group of dualities, and some of them can be used as twists as well. These arise for the
special coupling $\tau=e^{\frac{i\pi}{3}}$, which is invariant under
$\tau\rightarrow(\tau-1)/\tau$ and $\tau\rightarrow -1/(\tau-1)$, and the corresponding $\SL(2,\Z)$ elements can be used to twist the
boundary conditions. Together with $\tau\rightarrow-1/\tau$, we thus have three $\SL(2,\Z)$ elements to explore as possible twists.\footnote{The twists are in $\SL(2,\Z)$ and not $\PSL(2,\Z)$ because the central element $-1\in\SL(2,\Z)$ acts nontrivially, being equivalent to charge conjugation. For each $\zg\in\SL(2,\Z)$ in the list \eqref{eqn:deflvk} below, one can also use $-\zg$ as a twist, but it is always identical to the inverse of an element that already appears in the list, and the resulting theory with $-\zg$ is always the parity transform of another theory from the list.}
 We denote a general $\SL(2,\Z)$-element by
\be\label{eqn:xabcd} \zg\equiv
\begin{pmatrix} \xa & \xb \\ \xc & \xd \\ \end{pmatrix}
\,,
\qquad
\tau\rightarrow \frac{\xa\tau+\xb}{\xc\tau+\xd}\,,
\ee
and we denote its order in the group by $\ord$ (thus $\zg^\ord=1$).
In \cite{Ganor:2010md} an integer $\lvk$ and a phase $-\pi<\pht<\pi$ were assigned to the three $\SL(2,\Z)$ elements $\zg$ as follows:
\be\label{eqn:phtdef}
e^{i\pht}\equiv\xc\tau+\xd
\,,
\qquad
\lvk\equiv\frac{2-\xa-\xd}{\xc}\,.
\ee
It can easily be checked that in our three cases $\pht$ is real and equal to $2\pi/\ord$, and $\lvk$ is an integer.
Explicitly,
\be\label{eqn:deflvk}
\left.
\begin{array}{llllll}
\lvk=1, & \ord=6, & \pht=\frac{\pi}{3} &\text{for} & \tau = e^{\pi i/3},
&\zg=\left(\begin{array}{rr} 1 & -1 \\ 1 & 0 \\ \end{array}\right)\,;
\\ & & & & \\
\lvk=2, & \ord=4, & \pht=\frac{\pi}{2} &\text{for} & \tau = i,
&\zg=\left(\begin{array}{rr}
 0 & -1 \\ 1 & 0 \\ \end{array}\right)\,;
\\ & & & & \\
\lvk=3, & \ord=3, & \pht=\frac{2\pi}{3} &\text{for} & \tau = e^{\pi i/3},
&
\zg=\left(\begin{array}{rr} 0 & -1 \\ 1 & -1 \\ \end{array}\right)\,.
\\
\end{array}\right\}
\ee

In \cite{Ganor:2010md} the study of the Tr-S theory with gauge group $U(n)$ compactified on $T^2$ was started, and the Hilbert space of ground states was determined. We use the notation $\Hilb(\lvk,n)$ to refer to the Hilbert space of ground states of the theory with $\tau$ and $\zg$ that are determined according to the list \eqref{eqn:deflvk}. In \cite{Ganor:2010md} it was also found convenient to restrict attention to the cases with $n<\ord$, since these have no Coulomb branch, as we review in \secref{sec:Review}. This restriction arises because for $n\ge\ord$ there are elements of the Weyl group $S_n\subset U(n)$ that have order $\ord$, and a gauge transformation by such a Weyl group element can cancel the effect of the R-symmetry twist and produce a zero-mode. For $n<\ord$ there are no such zero modes. We believe that in this case the compactification has a mass-gap, and the 2+1D low-energy theory Tr-S is topological.

The purpose of this paper is to continue to explore the Hilbert space of Tr-S on $T^2$ by introducing supersymmetric static charges corresponding to $m$ pairs of heavy quarks and anti-quarks at fixed locations on $T^2$ and $S^1.$ We will study the resulting Hilbert space of ground states, and show that the Witten Index of this problem can be calculated by counting the states of a simple quantum mechanical system with action
\be\label{eqn:pqIntro}
I=\frac{1}{2\pi}\int\sum_{\iDb,\alpha=1}^{m}\Matx_{a\alpha}\pvar^{\iDb} \dot{\qvar}^{\alpha}\, dt \,,
\ee
%\be\label{eqn:pqIntro}
%I=2\pi\int\left\lbrack \pvar^{(1)} (\dot{\qvar}^{(1)}
%+\dot{\qvar}^{(m')}) + \sum_{j=2}^{m'} \pvar^{(j)} (\dot{\qvar}^{(j)}
%-\dot{\qvar}^{(j-1)}) \right\rbrack dt \,,
%\ee
where $\Matx_{\iDb\alpha}$ is an $m\times m$ matrix with integer entries, and $\pvar^{1},\qvar^{1},\dots,\pvar^{m},\qvar^{m}$ are periodic coordinates with period $2\pi.$ The action \eqref{eqn:pqIntro} describes geometric quantization of $T^{2m}.$

One motivation for introducing static charges into the Tr-S theory---apart from a better understanding of the theory itself---is to clarify the relationship between Tr-S and another (known) topological theory in 2+1D, namely the Chern--Simons theory. In fact, when the gauge group is abelian, in which case the \SUSY{4} SYM is a free theory and we have a complete understanding of its S-duality \cite{Witten:1995gf}, we have an explicit description for Tr-S: it is simply the abelian Chern--Simons theory at level $\lvk$ given in \eqref{eqn:phtdef} (see \S5 of \cite{Ganor:2010md}). This simple picture does not hold true for nonabelian gauge groups, but the result of \cite{Ganor:2010md} suggested that there might still exist a close relationship between the two theories. There, it was observed that the Hilbert space of the Tr-S theory compactified on $T^2$ decomposes into different sectors, and for almost all sectors it was shown that their symmetry operators and behavior under modular transformations of $T^2$ agree with those of the Hilbert spaces of the Chern--Simons theory with appropriate gauge groups and levels. Introduction of static charges then provides a further test on the identification of the Hilbert spaces of the two theories.

Our strategy for extracting the Witten Index of the system with static charges inserted follows closely that of \cite{Ganor:2010md}. Since little is known about the S-duality operator itself for nonabelian gauge groups, we will embed our setting into full type IIB string theory and apply a series of string theory dualities, after which the low energy description of the system is given by the simple quantum mechanical one  in \eqref{eqn:pqIntro}. For abelian gauge group, the result we obtain in this way precisely agrees with what we expect from introducing Wilson line operator in abelian Chern--Simons theory. This is as it must, because we already know that the Tr-S theory \textit{is} Chern--Simons theory in this case. For nonabelian gauge groups, we show that our result passes a nontrivial consistency check in itself, but does not agree with Chern--Simons theory predictions in general. We will provide more discussion on this discrepancy in the concluding section.

The paper is organized as follows.
In \secref{sec:Review} we explain the problem in detail and describe the S-duality and R-symmetry twists, the amount of supersymmetry that is left unbroken, and the absence of the Coulomb branch for $n<\ord.$ We then construct the type-IIA dual of the theory without the charges.
In \secref{sec:StaticCharges} we introduce the supersymmetric static charges and derive their type-IIA dual description. We then explain how the geometric quantization systems of the type \eqref{eqn:pqIntro} arise.
In \secref{sec:U(1)} we use the known solution of the problem with $U(1)$ gauge group, which reduces to $U(1)$ Chern--Simons theory at level $\lvk$ [defined in \eqref{eqn:deflvk}], to demonstrate how the type-IIA dual reproduces the known results about the Hilbert space of ground states of $U(1)$ Chern--Simons theory with charges.
In \secref{sec:U(n)} we move to the case of $U(n)$ gauge group. The goal of this section is to calculate the Witten Index of this system as a function of $n,\lvk,$ and $m.$ We describe the technical aspects of the calculation in detail and summarize the final results in \tabref{tab:WI}. Next, using a Wick-rotation we express the Witten index as a supertrace of a combination of spatial Wilson loops over $\Hilb(\lvk,n)$ (the Hilbert space without external charges). This provides us with a consistency check on the final result, and moreover, allows us to calculate the eigenvalues of the spatial Wilson loop operators on $\Hilb(\lvk,n).$ We then compare the results to Chern--Simons theory as conjectured in \cite{Ganor:2010md}, and show that they do not match.
We conclude in \secref{sec:disc} with a discussion and future directions.

% ==============================================================
\section{Review of the S-duality twist and its type-IIA dual}
\label{sec:Review}

In this section, we carry out a brief review of Tr-S theory, how to realize it in string theory, and how to construct the type-IIA dual of the theory on $T^2.$ We refer our readers to \cite{Ganor:2010md} for more comprehensive details.
At the end of this section we include a small discussion on why we believe Tr-S is topological for $n<\ord.$

% --------------------------------------------------------------
\subsection{Definition of Tr-S}
\label{subsec:defTrS}

By definition, Tr-S is the 2+1D low-energy limit of a compactification of \SUSY{4} super Yang--Mills (SYM) theory on $S^1$ with boundary conditions that include an S-duality twist and an appropriate R-symmetry twist to be discussed below. This compactification was introduced in \cite{Ganor:2008hd,Ganor:2010md}, and similar compactifications have also recently been studied in \cite{Terashima:2011qi,Dimofte:2011jd,Dimofte:2011ju}, where the S-duality twist was referred to as a ``duality wall.''

By itself, S-duality does not commute with supersymmetry \cite{Kapustin:2006pk}, and since we want to preserve some supersymmetry we have to supplement the S-twist with an R-symmetry twist. Therefore, in \cite{Ganor:2010md} we twisted the boundary conditions on $S^1$ further by an element $\gtw$ of the R-symmetry group $SU(4)$, which in a particular basis takes the form:
\be\label{eqn:gtwN=6}
\gtw=
\begin{pmatrix}
e^{\frac{i}{2}\pht} & &  & \\
& e^{\frac{i}{2}\pht} &  & \\
& & e^{\frac{i}{2}\pht}  & \\
& & & e^{-\frac{3i}{2}\pht}  \\
\end{pmatrix} \in SU(4)_R\,,
\ee
where $\pht$ is given by \eqref{eqn:phtdef}. This choice, it turns out, preserves the maximal possible amount of supersymmetry in the presence of S-duality and R-symmetry twists, which is \SUSY{6} in 2+1D. It also preserves a $U(3)\subset SU(4)_R$ R-symmetry. This $U(3)$ can be thought of as the unitary group of rotations that act holomorphically on the transverse $\C^3\simeq\R^6.$

3+1D \SUSY{4} SYM has a Coulomb branch on which the gauge group is broken to $U(1)^n\subset U(n).$ After compactification with the S-twist and R-twist together, the Coulomb branch completely disappears for $n<\ord$ \cite{Ganor:2010md}. This is because a point on the Coulomb branch $(\R^6)^n/S_n$ of \SUSY{4} $U(n)$ SYM is described by an unordered set of noncoincident $n$ points in $\R^6.$ S-duality preserves the point in moduli space, but the R-symmetry twist $\gtw$ acts on it by rotation of $\R^6.$ Since $\gtw$ has order $\ord\equiv2\pi/\pht$ when acting on $\R^6$, we see that we need $n\ge\ord$ for a point on the Coulomb branch to survive the twist. For $n\ge\ord$ the situation is more complicated and some portion of the moduli space of the Coulomb branch survives \cite{Ganor:2010md}. In order to avoid these complications we will restrict the discussion that follows to the case $n<\ord.$

We can easily realize Tr-S in string theory using D$3$-branes. Consider the type-IIB background $\R^{9,1}$ with Cartesian coordinates $x_0,\dots,x_9$, and place $n$ D3-branes at $x_4=x_5=\cdots=x_9=0.$ The type-IIB coupling constant is denoted by $\tau=\chi+\frac{i}{\gIIB}$, where $\gIIB$ is the string coupling constant, and $\chi$ is the R-R scalar. The S-duality transformation $\zg$ of \eqref{eqn:xabcd} then lifts to an S-duality transformation of the full type-IIB string theory (that we also denote by $\zg$), and the R-symmetry element $\gtw$ lifts to a geometrical rotation in the six directions transverse
to the D$3$-branes. We compactify the $x_3$-direction on a circle of radius $\xR$ with boundary conditions given by a simultaneous S-duality twist $\zg$ and a $\gtw\in\Spin(6)$ geometrical twist in directions $x_4,\dots ,x_9$, where $\gtw$ is given by \eqref{eqn:gtwN=6}. This means that as we traverse the $x_3$ circle once, we also apply a $\gtw\in\Spin(6)$ rotation in the transverse directions before gluing $x_3=0$ to $x_3=2\pi\xR$.

We now compactify directions $x_1,x_2$, so that $0\le x_1<2\pi \xL_1$ and $0\le x_2<2\pi \xL_2$ are periodic. This puts the 2+1D field theory on $T^2$ with area $4\pi^2\xL_1\xL_2$ and complex structure $i\xL_2/\xL_1.$ In the limit
\be\label{eqn:LimIIB}
\xL_1, \xL_2, \xR\gg \apr^{1/2}\,,
\ee
where $\apr^{1/2}$ is the type-IIB string scale, we can first reduce
the description of the D$3$-branes to \SUSY{4} $U(n)$ SYM at low energy, and then compactify \SUSY{4} SYM with an S-duality and R-symmetry twist.

% --------------------------------------------------------------
\subsection{Type-IIA dual}
\label{subsec:IIAdual}

We will now transform the type-IIB background, using string dualities, to one where S-duality is realized geometrically. For this we need to consider the opposite limit $\xL_1,\xL_2\rightarrow 0$ with $\xR\rightarrow\infty$ (in the order to be specified below). In this limit, the type-IIB description is strongly coupled,
but we will perform a U-duality transformation as specified in
\tabref{tab:Dualities} to transform the setting to a weakly
coupled type-IIA background, enabling us to easily study the
ground states of the field theory.

\begin{table}[t]
\begin{tabular}{|l|l|cccccccccc|l|}
\hline\hline
Type & Brane &
$1$ & $2$ & $3$ & $4$ & $5$ & $6$ & $7$ & $8$ & $9$ & $10$ &
Apply:\\ \hline\hline
IIB  & D3 &
\wx & \wx & \tx & &&&&&&\nx &
T-duality on $1$ to get $\hookrightarrow$ \\ \hline
IIA  & D2 &
    & \wx & \tx & &&&&&&\nx &
Lift to M-theory  to get $\hookrightarrow$ \\ \hline
M    & M2 &
    & \wx & \tx &     &&&&&&&
Reduction to IIA on $2$  to get $\hookrightarrow$ \\ \hline
IIA  & F1 &
     & \nx & \tx &     &&&&&&&
This is the final step!
\\ \hline\hline
\end{tabular}
\caption{
The sequence of dualities from
$n$ D3-branes in type-IIB to $n$ fundamental strings in
type-IIA. A direction that the corresponding brane or string
wraps with periodic boundary conditions
is represented by \wx, a direction that the object wraps
with twisted boundary conditions is represented by \tx,
and a dimension that doesn't exist in the particular string
theory is denoted by \nx.
All the branes in the table are at the origin
of directions $4,\dots ,9.$
}
\label{tab:Dualities}
\end{table}

The U-duality transformation proceeds as follows.
We first replace type-IIB on a circle of radius $\xL_1$
with M-theory on $T^2$ with complex structure $\tau$
and area
$\Area
=(2\pi)^2\apr^2\tau_2^{-1}\xL_1^{-2}
=(2\pi)^2\Mpl^{-3}\xL_1^{-1}$,
where $\Mpl$ is the $11$-dimensional Planck scale.
We now reduce from M-theory to type-IIA on the circle of radius
$\xL_2$ to get a theory with string coupling constant
$$
\gIIA
=(\Mpl\xL_2)^{3/2}
=\tau_2^{1/2}\xL_1^{1/2}\xL_2^{3/2}\apr^{-1}
\,,
$$
and new string scale
$$
\apr_{\text{IIA}}=\Mpl^{-3}\xL_2^{-1}
=\apr^2\tau_2^{-1}\xL_1^{-1}\xL_2^{-1}
\,.
$$
After these dualities, the D$3$-branes become fundamental type-IIA strings with a total winding number $n$ in the $x_3$ direction. The S-duality twist $\zg$ is now a diffeomorphism of the type-IIA torus (in the $x_{10}x_1$ directions), which can be realized as a rotation by an angle $\pht$ listed in \eqref{eqn:deflvk}.
To make this type-IIA background weakly coupled we assume that the limits are taken in such a way that
\be\label{eqn:LimIIA}
\Area\gg\apr_{\text{IIA}}\,,
\qquad
\gIIA\ll 1\,,
\qquad
\xR\gg\apr_{\text{IIA}}^{1/2}.
\ee
This is a different limit than \eqref{eqn:LimIIB}, but we can use the weakly coupled type-IIA background to study the Hilbert space of supersymmetric ground states, or more precisely the Witten Index of the Tr-S theory on $T^2.$

To describe the basic geometry of the dual type-IIA background, it is convenient to divide the $9$ directions into three groups and view the spatial manifold as an orbifold of $T^2\times\R\times\C^3.$ We regard the $T^2$ as the complex plane modded out by a lattice, $\C/(\Z+\tau\Z)$, and take
\be\label{eqn:zperiods}
z\sim z+1\sim z+\tau\,
\ee
as its coordinate. On $\R$, we take the coordinate
$$
-\infty < x_3 < \infty\,,
$$
and on $\C^3\simeq\R^6$, we take the coordinates to be
$$
(\zeta_1,\zeta_2,\zeta_3)\,,
\qquad
\zeta_1,\zeta_2,\zeta_3\in\C\,.
$$
The orbifold is then represented by the identification
\be\label{eqn:zxzzz}
(z,x_3,\zeta_1,\zeta_2,\zeta_3)\sim
(e^{i\pht}z,x_3+2\pi\xR,
e^{i\pht}\zeta_1,e^{i\pht}\zeta_2,e^{i\pht}\zeta_3)\,.
\ee
Also, the shift $x_3\rightarrow x_3+2\pi\xR$ ensures that the orbifold has no fixed points, and thus the geometry is flat and free of singularities. In particular, the $\zeta_1=\zeta_2=\zeta_3=0$ subspace is a $T^2$-fibration over $S^1$ with structure group $\Z_\ord.$

The ground states that are relevant to our problem are those with a total string winding number $n$ along direction $x_3.$ A state with string winding number $n$ is a $p$-particle (that is, $p$-string) state consisting of $1$-particle states of winding numbers $n_1\ge n_2\ge\dots\ge n_p>0$ with
$n_1+n_2+\cdots+n_p=n.$
Thus, the Hilbert space of ground states decomposes as a direct sum:
\be\label{eqn:HilbSumPart}
\Hilb(\lvk,n) = \bigoplus_{p=1}^n
\left(\bigoplus_{
\begin{subarray}{c}
n_1\ge n_2\ge\dots\ge n_p > 0 \\
n_1+n_2+\cdots+n_p=n
\end{subarray}}
\Hilb(\lvk;n_1,\dots,n_p)\right)\,.
\ee
A crucial point is that the problem of finding the ground states can
be solved using essentially classical geometry:
we simply need to find classical string configurations of minimal length.
Consider a superstring with winding number $\np$ in direction $x_3$. It turns out \cite{Russo:1995ik} that (for $\np\neq 0$) the ground states are bosonic and in the R-R sector. (We will independently verify this in \secref{subsec:fzeromodesn=1}.) For $\np$ that is not divisible by $\ord$, there is a basis of ground states that are in one-to-one correspondence with loops of winding number $\np$ and minimal length in the geometry \eqref{eqn:zxzzz}. In the limit $\apr_{\text{IIA}}\rightarrow 0$, these states reduce to the classical string configurations.

To describe the classical configurations, we can fix an $x_3$ coordinate and specify the points where the classical string intersects the transverse coordinates $T^2\times\C^3$ in
the geometry \eqref{eqn:zxzzz}. At winding number $\np$, the string
intersects $T^2\times\C^3$ at $\np$ (not necessarily distinct) points, and in order to be of minimal length the coordinates of
these points should be independent of $x_3.$ The classical configurations are thus characterized by a set of $\np$ points in $T^2\times\C^3$ that is invariant, as a set, under the orbifold operation
$$
(z,\zeta_1,\zeta_2,\zeta_3)\sim
(e^{i\pht}z,e^{i\pht}\zeta_1,e^{i\pht}\zeta_2,e^{i\pht}\zeta_3)\,.
$$
For $\np$ that is not divisible by $\ord$, there is a finite number of such sets, and they are all localized at the origin of $\C^3$, i.e., at $\zeta_1=\zeta_2=\zeta_3=0.$ They are therefore entirely described by the $z$-coordinates of where the string intersects $T^2$:
$z,e^{i\pht}z, e^{2i\pht}z,\dots,e^{i(\np-1)\pht}z$,
since as we go once around the $x_3$ direction the coordinate $z$
switches to $e^{i\pht} z.$ After $\np$ loops $z$ becomes $e^{i\np\pht}z$, which should be identified with $z$ (up to a shift in $\Z+\Z\tau$) in order to close the string. The classical string configurations are then described by solutions $z=\fpZ_{\wMa,\wMb}$ of
\be\label{eqn:fpZsol}
e^{i\np\pht}\fpZ_{\wMa,\wMb}=\fpZ_{\wMa,\wMb}+\wMa+\wMb\tau\,,
\qquad(\wMa,\wMb\in\Z)
\ee
and we consider two solutions $\fpZ_{\wMa,\wMb}$ and $\fpZ_{\wMa',\wMb'}$ as equivalent if they differ by a lattice element, i.e., if
$\fpZ_{\wMa,\wMb}-\fpZ_{\wMa',\wMb'}\in\Z+\Z\tau.$ In addition, $\fpZ_{\wMa,\wMb}$ and $e^{i\pht}\fpZ_{\wMa,\wMb}$ give equivalent solutions, since the intersection points of the string with $T^2$ are unordered.

There is then only a finite number of inequivalent solutions to \eqref{eqn:fpZsol}, and we have described them in detail in \cite{Ganor:2010md}. The full single-particle string spectrum (including excited states) decomposes into a finite sum of distinct sectors, labeled by $\wMa,\wMb$, and the solution $\fpZ_{\wMa,\wMb}$, which is a point on $T^2$, describes the center of mass of the string in the directions of $T^2.$
Thus, a single-particle ground state with winding number $\np$ can be described by the location of the intersection of the classical string configuration with any particular $T^2$ fiber at a constant $x_3$:
\be\label{eqn:solnot}
\ket{[z, e^{i\pht}z,\dots,e^{(\np-1)i\pht}z]}\,,
\ee
where $z$ coordinates are always taken modulo the lattice $\Z+\Z\tau.$ The multi-string states can subsequently be described by
$$
\ket{\{
[z_1, e^{i\pht}z_1,\dots,e^{(n_1-1)i\pht}z_1],
[z_2, e^{i\pht}z_2,\dots,e^{(n_2-1)i\pht}z_2],
\ldots,
[z_p, e^{i\pht}z_p,\dots,e^{(n_p-1)i\pht}z_p]
\}}\,,
$$
where each $z_i$ is a solution $\fpZ_{\wMa_i,\wMb_i}$ of \eqref{eqn:fpZsol} with $\np\rightarrow n_i,$ and $n=\sum_1^p n_i$ is the total winding number. Also, the number of inequivalent solutions of \eqref{eqn:fpZsol} for $\np=1$ is equal to $\lvk$.
It is a function of $\pht$ alone, as indicated in \eqref{eqn:deflvk}.

%In \cite{Ganor:2010md}, we argued that the ground states of the type-IIA dual
%correspond to classical string configurations of minimal length, and identified linear
%combinations of these ground states with states of Chern--Simons theories of different
%levels and gauge groups. We supported this identification by constructing operators that
%correspond to the symmetries of the background \eqref{eqn:zxzzz}, and matching them with physical operators in the field theory.

% --------------------------------------------------------------
\subsection{$\Z_\lvk$ symmetries}
\label{subsec:ZkSym}

For $\lvk>1$ there are two useful $\Z_\lvk$ symmetries that can be described geometrically in the type-IIA background as follows \cite{Ganor:2010md}:
\begin{enumerate}
\item
The metric \eqref{eqn:zxzzz} has a discrete isometry that is generated by the operator $\Psym$ defined to act as a translation in the $T^2$ fiber:
\be\label{eqn:Psymdef}
\Psym:\qquad
(z,x_3,\zeta_1,\zeta_2,\zeta_3)
\mapsto
(z+\frac{1}{\lvk}(1+\tau),x_3,\zeta_1,\zeta_2,\zeta_3)
\,.
\ee
\item
The first homology group $H_1$ of the space \eqref{eqn:zxzzz} is $\Z\oplus\Z_\lvk$ where $\Z_\lvk$ is generated by the homology class of one of the $1$-cycles of the $T^2$ (in the $z$ direction),
and $\Z$ is generated by a cycle that wraps around the $x_3$ direction at $z=0.$ The homology class of a fundamental string is conserved, and the projection onto the $\Z_\lvk$ factor describes
a conserved quantum number $q\in\Z_\lvk$. We define the operator $\Qsym$ to have the eigenvalue $e^{\frac{2\pi i}{\lvk}q}$ on a state with quantum number $q.$
\end{enumerate}
In other words, $\Psym$ can be viewed as $\Z_\lvk$-momentum, and $\Qsym$ can be viewed as $\Z_\lvk$-winding number. They obey the commutation relations \cite{Ganor:2010md}:
\be\label{eqn:QkPk}
\Qsym^\lvk=\Psym^\lvk=1\,,\qquad
\Qsym\Psym\Qsym^{-1}\Psym^{-1}=e^{\frac{2\pi i n}{\lvk}}
\,.
\ee

This $\Z_\lvk\times\Z_\lvk$ symmetry also has a natural interpretation in terms of the original gauge theory. Its conserved charges can be expressed in terms of S-duality invariant combinations of electric and magnetic fluxes in the center $U(1)\subset U(n)$; we refer the reader to \cite{Ganor:2010md} for details. For the present paper, however, we only need to know the commutation relations of $\Psym$ and $\Qsym$ with the Wilson loop operators. Those follow directly from the relation of $\Psym$ and $\Qsym$ to electric and magnetic fluxes, or can be derived directly in the type-IIA dual. The result will be given later on in the paper, in \eqref{eqn:PQsymWilV}.

For completeness, we also note that in addition to this $\Z_\lvk\times\Z_\lvk$ symmetry we have an $\SL(2,\Z)$ symmetry that acts as the mapping class group of $T^2$ on which the Tr-S theory is defined. In the type-IIA dual picture, the complex structure parameter of this $T^2$ becomes the complexified area modulus of the type-IIA $T^2$ (in $x_{10}x_1$ directions):
\be\label{eqn:rhocp}
\rhocp =
\frac{i}{\alpha'_{\text{IIA}}}
\text{Area}(T^2)
+\frac{1}{2\pi}\int_{T^2}B\,.
\ee
Here $B$ is the NS-NS two-form potential. The $\SL(2,\Z)$ group acts by T-duality, and is generated by
$$
\TdS\rightarrow
\begin{pmatrix} 0 & -1 \\ 1 & 0 \\ \end{pmatrix}
\in\SL(2,\Z)\,,
\qquad
\TdS:\qquad\rhocp\rightarrow-\frac{1}{\rhocp}\,,
$$
and
$$
\TdT\rightarrow
\begin{pmatrix} 1 & 1 \\ 0 & 1 \\ \end{pmatrix}
\in\SL(2,\Z)\,,
\qquad
\TdT:\qquad\rhocp\rightarrow\rhocp+1\,.
$$
In \cite{Ganor:2010md}, this correspondence was used to read off the modular transformation properties of the ground states of Tr-S theory.

% --------------------------------------------------------------
\subsection{Is Tr-S a topological theory?}
\label{subsec:IsTrStopo}

Before we proceed to study static charges for Tr-S on $T^2$, let us explain in more detail why we believe Tr-S is a topological theory. The setting we described above has \SUSY{6} supersymmetry in 2+1D and a $U(3)$ R-symmetry, which is the subgroup of $SU(4)$ that commutes with $\gtw$ in \eqref{eqn:gtwN=6}. If Tr-S is not topological and has propagating degrees of freedom, it must be an interacting \SUSY{6} superconformal field theory. Let us then consider the low-energy limit of Tr-S on $S^1.$ This compactification has \SUSY{(6,6)} supersymmetry in 1+1D.
To gain more insight about this 1+1D theory, let us look at the list of dualities in \tabref{tab:Dualities}, but instead of performing all the dualities all the way to type-IIA at the last line, let us stop one line before the last, at the point where we have M-theory and $n$ M$2$-branes. At this point direction $x_2$ is not yet compact, but directions $x_1, x_{10}$ are compact and form a torus with complex structure $\tau.$ The directions transverse to the M$2$-branes are parameterized by the complex coordinates $(z,\zeta_1, \zeta_2, \zeta_3).$ The M$2$-branes wrap direction $x_3$, and the boundary conditions along $x_3$ are twisted by a geometrical twist, which is a rotation in the directions transverse to the M$2$-branes. This twist acts as
\be\label{eqn:zz123}
(z, \zeta_1, \zeta_2, \zeta_3)\rightarrow
(e^{i\pht}z, e^{i\pht}\zeta_1, e^{i\pht}\zeta_2, e^{i\pht}\zeta_3)
\ee
and corresponds to a rotation by an angle $\pht$ in $4$ transverse planes.

For $n=1$, the twisted boundary conditions create a mass gap of $1/\ord\xR$, where $\ord=2\pi/\pht$ [see the discussion below \eqref{eqn:phtdef}], and the 1+1D low-energy theory has no propagating degrees of freedom. This is consistent with the identification of Tr-S at $n=1$ with abelian Chern--Simons theory, as will be discussed in \secref{sec:U(1)} in more detail, and indeed Chern--Simons theory has no propagating degrees of freedom.

What about the nonabelian case, say $n=2$? In this case we need to understand the low-energy limit describing two M$2$-branes compactified on $S^1$ with transverse directions $T^2\times\C^3$ parameterized by $(z, \zeta_1, \zeta_2, \zeta_3)$ and with boundary conditions twisted by \eqref{eqn:zz123} along $S^1.$ The 1+1D low-energy theory corresponds to configurations where $(z, \zeta_1, \zeta_2, \zeta_3)$ are independent of $x_3.$ Because of the twist, this implies that $\zeta_1=\zeta_2=\zeta_3=0$ and $z$ is a fixed point of the twist. For given $\pht$, the twist has $\lvk$ fixed points on $T^2$, as we explained at the end of \secref{subsec:IIAdual}. It is easy to check that these fixed points are at
\be\label{eqn:zj}
z_j = \frac{j}{\lvk}(1+\tau)\,,\qquad j=0,\dots,\lvk-1.
\ee
The 1+1D low-energy theory therefore has $\lvk(\lvk+1)/2$ sectors. In $\lvk(\lvk-1)/2$ of the sectors the two M$2$-branes sit at different fixed points $z_{j}\neq z_{j'}.$ In this case it is clear that no massless excitations survive the low-energy limit and the low-energy 1+1D theory has no propagating degrees of freedom.
The remaining $\lvk$ sectors have two M$2$-branes at the same $z_j.$
Clearly, all these sectors are equivalent and we can concentrate on one of them, say at $z_0=0.$ Since the M$2$-branes are pinned to the origin, we can safely replace $T^2$ with $\C$ and set $z\rightarrow\zeta_0$, with $\zeta_0\in\C.$ We have now reduced the problem to understanding the compactificiation of two M$2$-branes on $S^1$ with transverse directions $\C^4$ and a twist along $S^1$ given by
\be\label{eqn:z0123}
(\zeta_0, \zeta_1, \zeta_2, \zeta_3)\rightarrow
(e^{i\pht}\zeta_0, e^{i\pht}\zeta_1, e^{i\pht}\zeta_2, e^{i\pht}\zeta_3).
\ee
Up until recently we would have had to proceed indirectly from here, but the recent progress in the low-energy description of M$2$-branes \cite{Bagger:2006sk}-\cite{Bagger:2007vi}, culminating in the discovery of the ABJM action \cite{Aharony:2008ug}, allows us, in principle, to explore this problem directly. We need to take the ABJM action and compactify all fields on $S^1$ with boundary conditions twisted by \eqref{eqn:z0123}. This corresponds to an element of the $SO(8)$ R-symmetry group. However, proceeding to compactify the ABJM theory in this manner involves subtleties that require a separate treatment, which we hope to present elsewhere. Instead, for now we will settle for an indirect approach, modifying the problem a little bit.
Instead of taking the discrete value $\pht=2\pi/\ord$ in \eqref{eqn:z0123}, let us consider the case that $|\pht|\ll 1.$ More precisely, consider the double-scaling limit
$$
\pht\rightarrow 0\,,
\qquad
\xR\rightarrow 0\,,
\qquad
\beta\equiv\frac{\pht}{\xR}\rightarrow\text{const.}
$$
Using a standard technique, we change variables to
$$
z_j'\equiv e^{-\frac{i\pht x_3}{2\pi\xR}}\zeta_j
\,,\qquad j=0,1,2,3,
$$
and write the metric as
\bear
ds^2 &=&
-dx_0^2 + dx_2^2 + dx_3^2 + \sum_{j=0}^3|d\zeta_j|^2
\nn\\ &=&
-dx_0^2 + dx_2^2
+\bigl(1+ \frac{\beta^2}{4\pi^2}\sum_0^3|z'_j|^2\bigr)dx_3^2
+\sum_{j=0}^3|dz'_j|^2
-\frac{\beta}{\pi}dx_3\,\text{Im}\sum_{j=0}^3z'_j {d\overline{z}'_j}
\,.
\nn
\eear
We can now reduce to type-IIA along direction $x_3$ to obtain a ``Melvin background'' \cite{Melvin:1963qx}-\cite{Scherk:1979zr}.
This background has a Ramond-Ramond field strength
$$
F^{RR}=dA^{\text{RR}}=-\frac{\beta}{2\pi}\text{Im}
\sum_{j=0}^3dz'_j\wedge {d\overline{z}'_j}+O(\beta^3)
\,,
$$
and a dilaton
$$
e^{\Phi}=(M_p\xR)^{3/2}\bigl(1+ \frac{\beta^2}{4\pi^2}\sum_0^3|z'_j|^2\bigr)^{3/2}
\,,
$$
where $M_p$ is the 10+1D Planck scale.
This background creates mass terms for all low-energy fields that propagate on a long string in direction $x_2.$ Such a string is pinned to the origin $z'_0=z'_1=z'_2=z'_3$, as is obvious from the M-theory description. In the type-IIA worldsheet description, the fermion mass terms are generated from the coupling of the string modes to the RR field strength, while the bosonic mass terms are generated from the string-frame metric. For small $\beta$ we can trust the approximate perturbative string analysis, and we see that all propagating modes along the remaining noncompact direction $x_2$ have mass of order $\beta=\pht/\xR.$ It is not immediately clear that we can extrapolate this analysis to $\pht=2\pi/\ord$, but we know that for a single string in this background no complications should arise, and in the limit $\xR\rightarrow 0$ string interactions are small.

Furthermore, if Tr-S is a nontrivial SCFT, and if it has a moduli space, then this moduli space must be compact because we have eliminated all noncompact modes along the Coulomb branch via the twist, and because the type-IIA picture shows no trace of noncompact flat directions. But \SUSY{6} supersymmetry in 2+1D is very restrictive and requires the moduli space to be locally flat. It must therefore be of the form $T^{8d}/\Gamma$, where $\Gamma$ is a discrete isometry group, and $d$ is an integer. On the other hand, the unbroken R-symmetry group must act nontrivially on the moduli space (by supersymmetry), but the maximal continuous isometry group of $T^{8d}/\Gamma$ is abelian and cannot contain $SU(3)$, which is a contradiction.

We will proceed under the assumption that Tr-S is topological, but we note that even if this assumption is incorrect, the results of this paper are still meaningful, but they then correspond to the Witten Index of an interacting SCFT, rather than a TQFT. We now proceed to the calculation of the Witten Index.

% ==============================================================
\section{Static charges and their type-IIA dual description}
\label{sec:StaticCharges}

In this paper we study what happens when we insert static charges into the Tr-S theory defined in \secref{sec:Review}. We add $2m$ static external sources to the system at $S^1$ coordinate $x_3=0$. Specifically, we insert $m$ heavy (non dynamical) quarks at the fixed $T^2$ coordinates $(a_1^{(\iDb)}, a_2^{(\iDb)})$  (where $\iDb=1,\dots,m$), and to cancel the net $U(1)$ charge\footnote{
Actually, it is not necessary for the net charge to be zero, thanks to the S-duality twisted boundary conditions in the $x_3$ direction. But the system with nonzero net charge is more complicated and will not be studied here.}
we insert an equal number $m$ of anti-quarks, which we take to be at fixed $T^2$ coordinates $(a_1^{(\iDb+m)}, a_2^{(\iDb+m)}).$ (Here $0\le a_i^{(\iDb)}< 2\pi\xL_i$ ($i=1,2$) are periodic coordinates in type-IIB directions $x_1,x_2.$)

The $\iDb^{th}$ static charge can be incorporated by introducing a matrix element of a timelike Wilson line
\be\label{eqn:WLj}
\text{tr}\,P\exp\left(
i\int_{-\infty}^\infty A_0(t,a_1^{(\iDb)}, a_2^{(\iDb)},0)dt
\right)
\ee
into the path integral. This prescription, however, breaks all the supersymmetry. To preserve some supersymmetry we follow \cite{Rey:1998ik,Maldacena:1998im} and add one of the adjoint scalar fields of \SUSY{4} SYM to $A_0$ in \eqref{eqn:WLj}. For concreteness, we take
\be\label{eqn:WLjSUSY}
\text{tr}\,P\exp\left(
i\int_{-\infty}^\infty
[A_0(t,a_1^{(\iDb)}, a_2^{(\iDb)},0)
+\Phi^9(t,a_1^{(\iDb)}, a_2^{(\iDb)},0)] dt
\right)\,,
\ee
where $\Phi^9$ is the scalar field that corresponds to D$3$-brane fluctuations in the $x_9$ direction. In \secref{subsec:ChargesIIB} we will show that inserting charges that interact with Tr-S as the low-energy limit of \eqref{eqn:WLjSUSY} preserves $4$ real supercharges.

% --------------------------------------------------------------
\subsection{Charges as endpoints of type-IIB strings}
\label{subsec:ChargesIIB}

Our main task now is to identify the type-IIA dual realization of the charges. We start in type-IIB and follow a standard technique to introduce static charges with interactions \eqref{eqn:WLjSUSY} into the type-IIB construction described at the beginning of \secref{sec:Review}.

Following \cite{Rey:1998ik,Maldacena:1998im} we introduce fundamental strings with one endpoint on the D$3$-branes and extending indefinitely in direction $x_9.$ We label the strings by $\iStIIB=1,\dots,2m$ and let the strings labeled by $\iStIIB=1,\dots,m$ extend along $0\le x_9<\infty$ and the strings labeled by $\iStIIB=m+1,\dots,2m$ extend along $-\infty<x_9\le0$. The low-energy description of this system holds the information about the ground states of Tr-S with static charges, and the $(x_1,x_2)$ coordinates of the endpoints of the strings can be set to $(a_1^{(\iStIIB)}, a_2^{(\iStIIB)}).$
We are only interested in the low-energy excitations of the system at energies well below the string scale, as well as the compactification scales:
$$
E\ll \Mst,\frac{1}{\xL_1}\,,\frac{1}{\xL_2}\,,\frac{1}{\xR}
\,.
$$
The semi-infinite strings, in this limit, are very heavy and can be treated semi-classically.

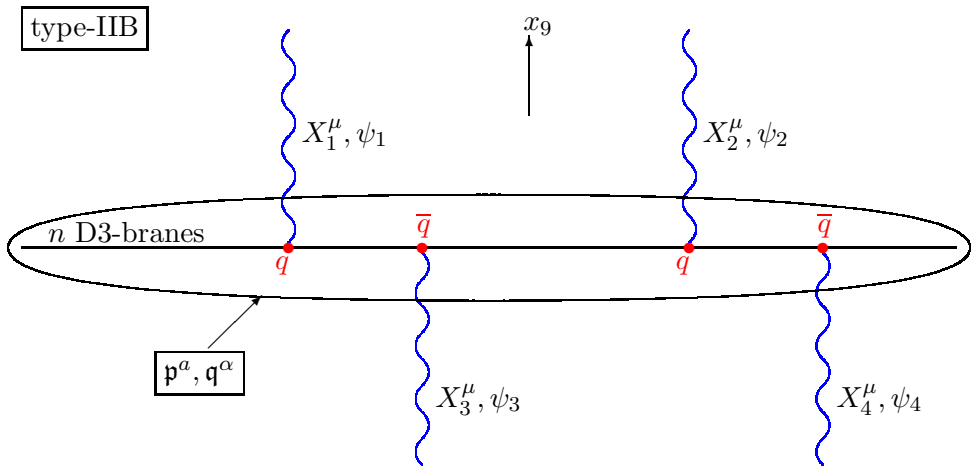
\begin{figure}[t]
\begin{picture}(400,220)

\put(10,10){\begin{picture}(400,200)

\color{black}
\put(0,180){\framebox{type-IIB}}
\thicklines
\color{black}
\put(0,100){\line(1,0){350}}
\put(10,102){$n$ D$3$-branes}

\multiput(100,100)(150,0){2}{\begin{picture}(0,0)
%\thicklines
%\color{black}
%\put(-20,100){\line(1,-1){40}}
%\put(-10,92){D$3$ (manipulator)}

\thinlines
\color{blue}
\multiput(0,2)(0,20){4}{
  \qbezier(0,0)(5,5)(0,10)
  \qbezier(0,10)(-5,15)(0,20)}
%\put(0,80){\circle*{4}}

\color{red}
\put(0,0){\circle*{4}}
\put(-5,-8){$q$}

\end{picture}}

\multiput(150,100)(150,0){2}{\begin{picture}(0,0)
%\thicklines
%\color{black}
%\put(20,-100){\line(-1,1){40}}
%\put(12,-92){D$3$}

\thinlines
\color{blue}
\multiput(0,-2)(0,-20){4}{
  \qbezier(0,0)(5,-5)(0,-10)
  \qbezier(0,-10)(-5,-15)(0,-20)}
%\put(0,-80){\circle*{4}}

\color{red}
\put(0,0){\circle*{4}}
\put(-2,6){$\overline{q}$}

\end{picture}}

\thinlines
\color{black}
\put(190,150){\vector(0,1){30}}
\put(188,182){$x_9$}

\put(155,40){$\scX_3^\mu,\fpsi_3$}
\put(305,40){$\scX_4^\mu,\fpsi_4$}
\put(105,140){$\scX_1^\mu,\fpsi_1$}
\put(255,140){$\scX_2^\mu,\fpsi_2$}

\qbezier(-5,100)(-5,120)(175,120)
\qbezier(-5,100)(-5,80)(175,80)
\qbezier(355,100)(355,120)(175,120)
\qbezier(355,100)(355,80)(175,80)

\put(50,50){\framebox{$\pvar^a,\qvar^\alpha$}}
\put(70,62){\vector(1,1){20}}

\end{picture}}
\end{picture}
\caption{
External quark and anti-quark sources are realized as endpoints of fundamental strings. At low-energy, the strings are described by free 1+1D fields $\scX_\iStIIB^\mu(x_9,t),\fpsi_\iStIIB(x_9,t)$ and the low-energy modes of the compact interacting system of D$3$-branes and charges are described by periodic variables $\pvar^a(t),\qvar^\alpha(t).$
}
\label{fig:qqbar}
\end{figure}

The long-wavelength excitations of each string are described by $8$ free scalars $\scX_\iStIIB^\mu(x_9,t)$
($\mu=1,\dots,8$) and a free Majorana-Weyl fermion $\fpsi_\iStIIB$ satisfying the chirality condition $$\Gamma^{0123456789}\fpsi_\iStIIB=\fpsi_\iStIIB$$
and the free massless Dirac equation along the string:
$$(\Gamma^0\px{t}+\Gamma^9\px{9})\fpsi_\iStIIB=0\,.$$
Their low-energy effective action is of the form
\be\label{eqn:ActionLE}
I = I^{\text{(int)}}+\sum_{\iStIIB=1}^{2m}I_\iStIIB^{(F1)}
\,,
\ee
where $I_\iStIIB$ is the bulk 1+1D action of the free fields $\scX_\iStIIB^\mu,\fpsi_\iStIIB$, and $I^{\text{(int)}}$ is the 0+1D action that couples the fields $\scX_\iStIIB^\mu,\fpsi_\iStIIB$ at $x_9=0$ to the low-energy degrees of freedom of Tr-S theory. In addition to the boundary values of $\scX_\iStIIB^\mu,\fpsi_\iStIIB$, the interaction term $I^{\text{(int)}}$ depends on additional local 0+1D degrees of freedom, whose form we seek to find. (See \figref{fig:qqbar}.)

% --------------------------------------------------------------
% --------------------------------------------------------------
% --------------------------------------------------------------
% --------------------------------------------------------------
% --------------------------------------------------------------

% --------------------------------------------------------------
% --------------------------------------------------------------
% --------------------------------------------------------------
% --------------------------------------------------------------

We will now provide a preview of what $I^{\text{(int)}}$ looks like, and we will explain the derivation at length in the following subsections. The term $I^{\text{(int)}}$ is formulated in terms of additional variables that are localized at the interaction point $x_9=0.$ These variables include a discrete variable that specifies the ``sector,'' with a finite number of sectors altogether. Each sector is then described by $m$ periodic variables $\pvar^a$ ($a=1,\dots,m$) and additional $m'$ periodic variables $\qvar^\alpha$ ($\alpha=1,\dots,m'$), both of which take values in the range $[0,2\pi).$ The number $m'$ of $\qvar^\alpha$'s depends on the sector, but generally $m'\ge m.$ The action $I^{\text{(int)}}$ is then a sum of two terms, which we write schematically as:
$$
I^{\text{(int)}} = I_0 + I_1\,,\qquad
I_0\equiv I_0(\{\pvar^a,\qvar^\alpha\})\,,\qquad
I_1\equiv I_1(\{\pvar^a,\qvar^\alpha\},
\{\scX_\iStIIB^1(0),\scX_\iStIIB^2(0)\}).
$$
The first term $I_0$ describes the local system at $x_9=0$, while $I_1$ is the interaction term that couples the system to the boundary values of the scalar fields on the $2m$ fundamental strings. (The fermions will be discussed later, but are suppressed at the moment.)

For the sectors for which $m'=m$ the configuration space of $\{\pvar^\iDb,\qvar^\alpha\}$ is $T^{2m}$, and the system described by $I_0$ is equivalent to geometric quantization on $T^{2m}$ with the following action:
\be\label{eqn:Mxpq}
I_0^{\text{(g.q.)}} = \frac{1}{2\pi}\int \sum_{a,\alpha=1}^m\Matx_{\iDb\alpha}\pvar^\iDb \dot{\qvar}^\alpha dt
\,,
\ee
where $\Matx_{\iDb\alpha}$ is a nonsingular matrix of integers that we will describe below. As mentioned in the Introduction, the sectors of most interest to us will be of this form. Any additional kinetic terms that are quadratic in $\dot{\qvar}^\alpha,\dot{\pvar}^\iDb$ are irrelevant at low-energy. The other sectors, with $m'>m$, also have a piece of the form $I_0^{\text{(g.q.)}}$ in their action, but it is necessary to include additional kinetic terms.

The coupling term $I_1$ is linear in $\scX_\iStIIB^1,\scX_\iStIIB^2$ and the derivatives $\dot{\qvar}^\alpha,\dot{\pvar}^\iDb$. It is of the form:
\be\label{eqn:I1}
I_1 =
\frac{1}{2\pi}\int\sum_{\iStIIB\alpha}\MatXq_{\iStIIB\alpha}\scX_\iStIIB^1(0)\dot{\qvar}^\alpha dt
+\frac{1}{2\pi}\int\sum_{\iStIIB\iDb}\MatXp_{\iStIIB\iDb}\scX_j^2(0)\dot{\pvar}^\iDb dt
\,,
\ee
where $\MatXq_{\iStIIB\alpha},\MatXp_{\iStIIB\iDb}$ are matrices of integers to be specified later. The remaining fields $\scX^\mu_\iStIIB$ with $\mu=4,\dots,8$ have Dirichlet boundary conditions $\scX^\mu_\iStIIB(0)=0$, while $\scX^3_\iStIIB$ has Neumann boundary conditions. These fields are however irrelevant for our discussion. In \secref{subsec:DualOfCharges} we will explain how the interactions \eqref{eqn:ActionLE}-\eqref{eqn:I1} are derived from the type-IIA dual. But first, let us discuss how much supersymmetry is left.

% --------------------------------------------------------------
\subsection{Supersymmetry}
\label{subsec:SUSYleft}

To see how much supersymmetry is preserved we consider once again the realization of \eqref{eqn:WLjSUSY} in type-IIB. We have fundamental strings that stretch along direction $x_9$ and end on the $n$ D$3$-branes. Let $\Gamma^A$ ($A=0,\dots,9$) be 9+1D Dirac gamma matrices, which we take to be real. Let $\pSUSY=\pSUSY_1 + i\pSUSY_2$ be a complex 9+1D Weyl spinor, where $\pSUSY_1, \pSUSY_2$ are Majorana--Weyl, and $\pSUSY^*=\pSUSY_1 -i\pSUSY_2$ its complex conjugate.

The supersymmetry preserved by the $n$ D$3$-branes is parameterized by those combinations of the supercharges with coefficients $\pSUSY$ that satisfy:
\be\label{eqn:D3-ep}
\Gamma^{0123}\pSUSY =-i\pSUSY
\,.
\ee
The S-R-twist preserves
\be\label{eqn:SR-ep}
\pSUSY = e^{-\frac{i\pht}{2}}
e^{\frac{\pht}{2}
(\Gamma^{45}+\Gamma^{67}+\Gamma^{89})}\pSUSY
\,,
\ee
where the first factor is from the S-twist, and the second from the R-twist,
and the interaction \eqref{eqn:WLjSUSY} preserves the same combinations of supersymmetry generators that a fundamental string in directions $0,9$ preserves, which is given by
\be\label{eqn:F1-ep}
\Gamma^{09}\pSUSY = \pSUSY^*
\,.
\ee

Combining \eqref{eqn:F1-ep} and \eqref{eqn:SR-ep} we find
$$
\pSUSY^* = e^{\frac{i\pht}{2}}
e^{\frac{\pht}{2}
(\Gamma^{45}+\Gamma^{67}-\Gamma^{89})}\pSUSY^*
\,,
$$
while taking complex conjugate of \eqref{eqn:SR-ep} yields (keeping in mind that the gamma matrices are all real)
$$
\pSUSY^* = e^{-\frac{i\pht}{2}}
e^{\frac{\pht}{2}
(\Gamma^{45}+\Gamma^{67}+\Gamma^{89})}\pSUSY^*
\,.
$$
Together, these two equations imply
$$
\Gamma^{89}\pSUSY=i\pSUSY\,.
$$
Then, from \eqref{eqn:SR-ep}, we obtain
\be
\Gamma^{45}\pSUSY=-\Gamma^{67}\pSUSY\,,
\ee
and together with \eqref{eqn:D3-ep} this leaves four linearly independent complex supersymmetry parameters. But \eqref{eqn:F1-ep} then puts a reality condition on these parameters, and leaves four real supercharges unbroken.

Out of the $U(3)$ R-symmetry that is preserved by the R-twist \eqref{eqn:gtwN=6}, the static charges only preserve $SU(2)\times U(1)\subset U(3).$ This is the double-cover of the unitary group $U(2)\simeq [SU(2)\times U(1)]/\Z_2$ that acts as unitary rotations of the variables $x_4+i x_5,$ $x_6 + i x_7,$ and preserves $x_8+ix_9.$ The surviving supercharges transform as a doublet of $SU(2)$ and are neutral under $U(1)$ (which is generated by $\Gamma^{45}+\Gamma^{67}$).

% --------------------------------------------------------------
\subsection{Constructing the type-IIA dual of charges}
\label{subsec:DualOfCharges}

Now we transform the system of D$3$-branes and fundamental strings to type-IIA by applying the U-duality transformation described in \tabref{tab:Dualities}. The $2m$ fundamental strings turn into D$2$-branes, and the $n$ D$3$-branes turn into fundamental strings,
as listed in \tabref{tab:DNS}.
\begin{table}[t]
\begin{tabular}{|l|c|c|c|c|c|c|c|c|c|}
\hline\hline
Brane &
$1$ & $3$ & $4$ & $5$ & $6$ & $7$ & $8$ & $9$ & $10$
\\ \hline\hline
 F1   &
    & \tx &     &     &     &     &     &     &
\\ \hline
 D2   &
    &     &     &     &     &     &     & \ox & \wx
\\ \hline\hline
\end{tabular}
\caption{
Open D$2$-branes are the U-duals of the type-IIB strings.
Here \wx\  denotes a direction that the brane wraps,
\tx\  denotes a direction that the brane/string
wraps with a twist, and \ox\ denote a direction
in which the brane extends but with endpoints (see \secref{subsec:WittenIndex}).
}
\label{tab:DNS}
\end{table}
In the type-IIB picture the strings end on the $n$ D$3$-branes, but in the type-IIA picture the D$2$-branes are too big to end on the $n$ strings. The system must therefore rearrange itself, and we have to pair up each D$2$-brane that corresponds to a quark (extending in the positive $x_9$ direction) with a D$2$-brane that corresponds to an anti-quark (extending in the negative $x_9$ direction), and glue them into a single smooth D$2$-brane. We thus get $m$ D$2$-branes whose worldvolume is equivalent to an infinite cylinder.

\begin{figure}[t]
\begin{picture}(400,200)

\put(10,10){\begin{picture}(390,190)
\color{black}

\color{black}
\put(0,130){\framebox{type-IIA}}
\thicklines
\put(0,0){\vector(0,1){90}}
\put(-2,92){$x_{10}$}
\put(0,0){\vector(1,0){380}}
\put(382,-2){$x_3$}
\put(0,0){\vector(1,1){20}}
\put(22,18){$x_9$}

% NS5-branes
%\thicklines
%\color{green}
%\multiput(50,10)(60,0){5}{\line(0,1){80}}
%\multiput(70,30)(60,0){5}{\line(0,1){80}}

% D2-branes
\color{black}
\thinlines
\multiput(110,10)(60,0){4}{\line(1,1){20}}
\multiput(110,90)(60,0){4}{\line(1,1){20}}
\multiput(110,10)(60,0){4}{\line(0,1){80}}
\multiput(130,30)(60,0){4}{\line(0,1){80}}
\multiput(112,28)(60,0){4}{D$2$}

%\put(265,130){Each D$2$ ends on}
%\put(265,120){NS$5$-branes at these edges.}
%\put(311,91){\vector(-1,-1){1}}
%\qbezier(311,91)(325,105)(325,115)
%\put(289,81){\vector(1,-1){1}}
%\qbezier(289,81)(270,95)(270,115)

\thicklines
\color{blue}
\put(300,50){\vector(1,0){80}}
\color{red}
\put(300,50){\circle*{4}}

\color{red}
\put(300,60){\circle*{4}}
\color{blue}
\multiput(293,60)(3,0){2}{\circle*{1}}
\put(290,60){\line(-1,0){50}}
\multiput(234,60)(3,0){2}{\circle*{1}}
\put(230,60){\line(-1,0){50}}
\color{red}
\put(180,60){\circle*{4}}

\color{red}
\put(180,50){\circle*{4}}
\thicklines
\color{blue}
\multiput(173,50)(3,0){2}{\circle*{1}}
\put(170,50){\line(-1,0){50}}
\color{red}
\put(120,50){\circle*{4}}

\thicklines
\color{blue}
\put(300,40){\vector(1,0){80}}
\multiput(294,40)(3,0){2}{\circle*{1}}
\put(290,40){\line(-1,0){50}}
\color{red}
\put(240,40){\circle*{4}}

\color{red}
\put(240,80){\circle*{4}}
\thicklines
\color{blue}
\multiput(233,80)(3,0){2}{\circle*{1}}
\put(230,80){\line(-1,0){50}}
\multiput(174,80)(3,0){2}{\circle*{1}}
\put(170,80){\line(-1,0){50}}
\multiput(114,80)(3,0){2}{\circle*{1}}
\put(110,80){\line(-1,0){50}}
\put(30,80){\vector(1,0){30}}

%\multiput(54,80)(3,0){2}{\circle*{1}}
%\put(50,80){\vector(-1,0){30}}

\color{red}
\put(120,70){\circle*{4}}
\thicklines
\color{blue}
\multiput(113,70)(3,0){2}{\circle*{1}}
\put(110,70){\line(-1,0){50}}
\put(30,70){\vector(1,0){30}}
%\multiput(54,70)(3,0){2}{\circle*{1}}
%\put(50,70){\vector(-1,0){30}}

\color{blue}
\put(82,84){F$1$}
\put(362,54){F$1$}

%\color{green}
%\put(150,60){\circle*{6}}
\end{picture}}
\end{picture}
\caption{
The type-IIA configuration following the U-duality transformation of \tabref{tab:Dualities}.
The $m$ pairs of type-IIB open strings become $m$ continuous D$2$-branes.  The $n$ D$3$-branes become $n$ fundamental strings, which in the presence of the D$2$-branes can break up into open strings. At least one pair of open strings must be attached to each D$2$-brane. In this example $m=4$ and $n=2.$
}
\label{fig:D2F1}
\end{figure}
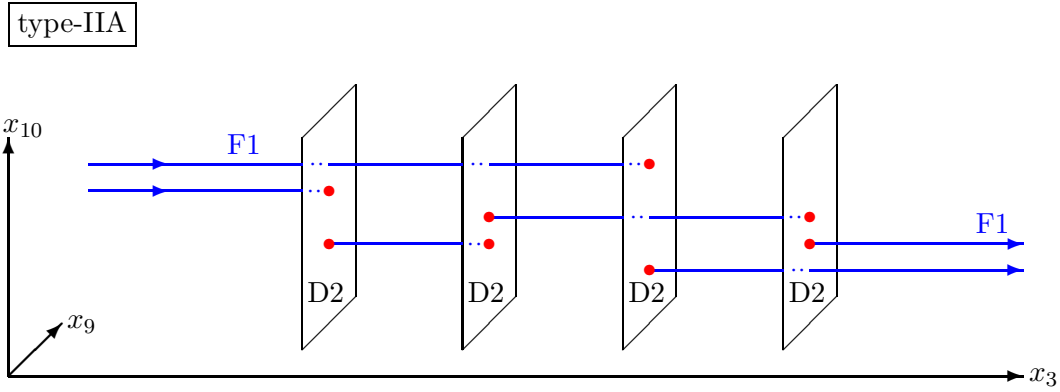

Some of the type-IIA closed strings that we had in \secref{sec:Review} must now be allowed to break into open strings and end on the D$2$-branes. Every D$2$-brane must have at least one such open string attached to it, because otherwise the corresponding type-IIB string would not be bound to any of the $n$ D$3$-branes. For ease of discussion it will be convenient
to slightly separate the D$2$-branes in the $x_3$ direction. The resulting configuration is depicted in \figref{fig:D2F1}.

% --------------------------------------------------------------
% --------------------------------------------------------------
% --------------------------------------------------------------
% --------------------------------------------------------------
% --------------------------------------------------------------

% --------------------------------------------------------------
% --------------------------------------------------------------
% --------------------------------------------------------------
% --------------------------------------------------------------

% --------------------------------------------------------------
\subsection{Local $\pvar^\iDb,\qvar^\alpha$ variables
and their action $I_0$}
\label{subsec:I0}

To understand how the $\pvar^\iDb,\qvar^\alpha$ that appear in \eqref{eqn:Mxpq} arise, it is instructive to start with a simple example: one D$2$-brane ($m=1$) and one string ($n=1$) for the case $\lvk=2$ with $\pht=\frac{\pi}{2}.$ We have one fundamental string wrapping direction $x_3$ and bound to the D$2$-brane, which wraps direction $x_{10}$ and extends in direction $x_9.$ The system is therefore dual to $U(1)$ Tr-S theory with an external quark and anti-quark pair.

\begin{figure}[t]
\begin{picture}(400,150)

\thinlines
\put(30,10){(a)}
\put(10,30){\begin{picture}(100,120)
\multiput(0,0)(80,0){2}{\line(0,1){80}}
\multiput(0,0)(0,80){2}{\line(1,0){80}}

\put(0,-5){\vector(1,0){80}}
\put(82,-7){$x_1$}

\put(-5,0){\vector(0,1){80}}
\put(-9,84){$x_{10}$}

\thicklines
\put(20,0){\line(0,1){80}}
\put(20,20){\circle*{4}}
\put(20,60){\circle*{4}}

\thinlines
\multiput(20,60)(12,12){4}{
\qbezier(0,0)(1,5)(6,6)
\qbezier(6,6)(11,7)(12,12)
}
\multiput(20,20)(-12,-12){4}{
\qbezier(0,0)(-1,-5)(-6,-6)
\qbezier(-6,-6)(-11,-7)(-12,-12)
}
\put(60,90){\vector(1,1){20}}
\put(81,112){$x_3$}

\put(22,40){D$2$}
\put(40,95){F$1$}
\put(-30,-15){F$1$}
\put(10,57){$z'$}
\put(24,17){$z''$}

\end{picture}}

% ----------------------------------
\thinlines
\put(170,10){(b)}
\put(150,30){\begin{picture}(100,80)
\multiput(0,0)(80,0){2}{\line(0,1){80}}
\multiput(0,0)(0,80){2}{\line(1,0){80}}

\put(0,-5){\vector(1,0){80}}
\put(82,-7){$x_1$}

\put(-5,0){\vector(0,1){80}}
\put(-9,84){$x_{10}$}

%\thicklines
%\put(10,0){\line(0,1){5}}
%\put(10,10){\circle{10}}
%\put(7,10){\line(1,0){6}}
%\put(10,7){\line(0,1){6}}
%\put(10,15){\line(0,1){50}}
%\put(10,70){\circle{10}}
%\put(7,70){\line(1,0){6}}
%\put(10,75){\line(0,1){5}}

\thinlines
\put(10,1){\line(0,1){3}}
\put(10,10){\circle{10}}
\put(7,10){\line(1,0){6}}
\put(10,7){\line(0,1){6}}
\multiput(10,16)(0,5){10}{\line(0,1){3}}
\put(10,70){\circle{10}}
\put(7,70){\line(1,0){6}}
\put(10,76){\line(0,1){3}}

\thicklines
\put(20,0){\line(0,1){15}}
\put(20,20){\circle{10}}
\put(17,20){\line(1,0){6}}
\put(20,17){\line(0,1){6}}
\put(20,25){\line(0,1){30}}
\put(20,60){\circle{10}}
\put(17,60){\line(1,0){6}}
\put(20,65){\line(0,1){15}}

%\thinlines
%\multiput(20,1)(0,5){3}{\line(0,1){3}}
%\put(20,20){\circle{10}}
%\put(17,20){\line(1,0){6}}
%\put(20,17){\line(0,1){6}}
%\multiput(20,26)(0,5){6}{\line(0,1){3}}
%\put(20,60){\circle{10}}
%\put(17,60){\line(1,0){6}}
%\multiput(20,66)(0,5){3}{\line(0,1){3}}

%\thicklines
%\put(30,0){\line(0,1){25}}
%\put(30,30){\circle{10}}
%\put(27,30){\line(1,0){6}}
%\put(30,27){\line(0,1){6}}
%\put(30,35){\line(0,1){10}}
%\put(30,50){\circle{10}}
%\put(27,50){\line(1,0){6}}
%\put(30,55){\line(0,1){25}}

\thinlines
\multiput(30,1)(0,5){5}{\line(0,1){3}}
\put(30,30){\circle{10}}
\put(27,30){\line(1,0){6}}
\put(30,27){\line(0,1){6}}
\multiput(30,36)(0,5){2}{\line(0,1){3}}
\put(30,50){\circle{10}}
\put(27,50){\line(1,0){6}}
\multiput(30,56)(0,5){5}{\line(0,1){3}}

\thinlines
\multiput(40,1)(0,5){7}{\line(0,1){3}}
\put(40,40){\circle{10}}
\multiput(40,46)(0,5){7}{\line(0,1){3}}

\thinlines
\multiput(50,1)(0,5){5}{\line(0,1){3}}
\put(50,30){\circle{10}}
\put(47,30){\line(1,0){6}}
\multiput(50,36)(0,5){2}{\line(0,1){3}}
\put(50,50){\circle{10}}
\put(47,50){\line(1,0){6}}
\put(50,47){\line(0,1){6}}
\multiput(50,56)(0,5){5}{\line(0,1){3}}

\thinlines
\multiput(60,1)(0,5){3}{\line(0,1){3}}
\put(60,20){\circle{10}}
\put(57,20){\line(1,0){6}}
\multiput(60,26)(0,5){6}{\line(0,1){3}}
\put(60,60){\circle{10}}
\put(57,60){\line(1,0){6}}
\put(60,57){\line(0,1){6}}
\multiput(60,66)(0,5){3}{\line(0,1){3}}

\thinlines
\put(70,1){\line(0,1){3}}
\put(70,10){\circle{10}}
\put(67,10){\line(1,0){6}}
\multiput(70,16)(0,5){10}{\line(0,1){3}}
\put(70,70){\circle{10}}
\put(67,70){\line(1,0){6}}
\put(70,67){\line(0,1){6}}
\put(70,76){\line(0,1){3}}

\end{picture}}

\end{picture}
\caption{
(a)
A fundamental string (F1) bound to the D$2$-brane.
The D$2$-brane wraps the compact direction $x_{10}$ (the vertical direction) and extends indefinitely in direction $x_9$ (not shown in the picture). The fundamental string is at $x_9=0$ and extends in direction $x_3$ (perpendicular to the plane of the drawing).
Because of the S-duality twist, which in the type-IIA
picture translates to a rotation, the fundamental string's
endpoint $z''$ can be different from its starting point $z'.$
(b) Configurations of the D$2$-brane with the two endpoints
$z',z''$ of the string marked as oppositely charged points. As the
D$2$-brane changes its $x_1$-position, the positions of the charges
change accordingly.
}
\label{fig:StringAndD2}
\end{figure}
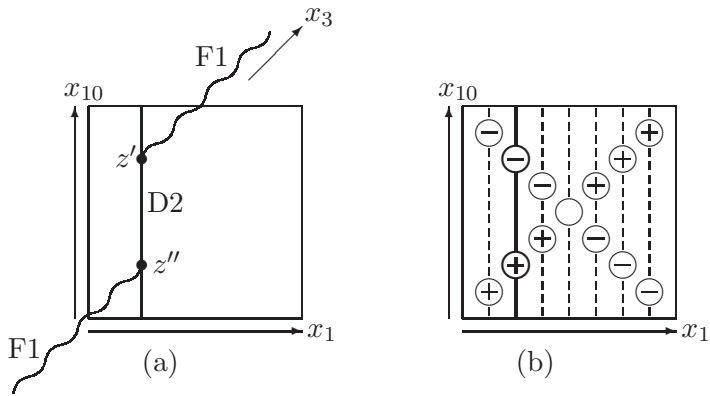

To bind to a D$2$-brane, the closed string must break to become an open string that starts and ends on the D$2$-brane located at $x_3=0$. Let the string start at $T^2$ coordinate $z=x_{10} + i x_{1} \equiv\qvar +i \xvar$
on the D$2$-brane, where $\xvar$ and $\qvar$ are functions of time. To make $\qvar$ and $\xvar$ compact variables with period $2\pi$, we will rescale the $x_{10}$ and $x_1$ coordinates so that from now on they take values in $[0,2\pi)$. To be of minimal length the string must remain at $x_9=0$ and at constant $z$, as $x_3$ varies from $0$ to $2\pi\xR.$ The point with coordinates $x_3=2\pi\xR$ and $z=\qvar+i\xvar$ is equivalent in the geometry of \eqref{eqn:zxzzz} to the point with coordinates $x_3=0$ and $z=e^{i\pht}(\qvar+i\xvar )=-\xvar+i\qvar.$ This point must also be on the D$2$-brane, which wraps direction $x_{10}$ but is at a fixed $x_1$ coordinate. We thus find that $\xvar=\qvar.$ In other words, the bound fundamental string starts
at $z'=(1+i)\qvar$ on the D$2$-brane, and ends at $z''=(-1+i)\qvar$ on the same D$2$-brane (see \figref{fig:StringAndD2}).

% --------------------------------------------------------------
% --------------------------------------------------------------

% --------------------------------------------------------------
% --------------------------------------------------------------

The starting point and endpoint of the string are oppositely charged under the $U(1)$ gauge field that resides on the D$2$-brane.  Let $A_{10}$ be the component of this gauge field in the $x_{10}$ direction.
We can fix the gauge so that $A_{10}$ is independent of
$x_{10}$, and set $\pvar\equiv 2\pi A_{10}$. The starting point and
endpoint of the string are separated along the $x_{10}$ direction by $\Delta x_{10}=2\qvar$, so the action of the system includes the term:
\be\label{eqn:2pdq} I_0\equiv
2\int A_{10} d\qvar = \frac{1}{2\pi}\int 2\pvar d\qvar \,.
\ee
We claim that $I_0$ is the only relevant term at low-energy.  For
example, the kinetic energies of the fundamental string and of the
D$2$-brane are irrelevant at low-energy, because they are proportional to $(\px{0}\qvar)^2$. Indeed, setting the mass dimensions of $\pvar,\qvar$ to zero, we see that the kinetic term has dimension $2$ and is irrelevant in a 0+1D theory.

The action $I_0$ describes the geometric quantization of a torus with the symplectic form $2d\pvar\wedge d\qvar.$ The Hilbert space has two states, and this is indeed the number of states we expect, since the $U(1)$ Tr-S theory is known to be equivalent to Chern--Simons theory at level $\lvk.$ (See \secref{sec:U(1)} for more details.)

We can now turn to the general case.
The 0+1D variables $\pvar^a,\qvar^\alpha$ that appear in \eqref{eqn:Mxpq} arise out of the type-IIA picture as follows. The low-energy description of each of the $m$ D$2$-branes includes a gauge field $A^{(\iDb)}$ ($\iDb=1,\dots,m$). The periodic variable $\pvar^\iDb$ is identified with the holonomy of $A^{(\iDb)}$ around the $x_{10}$ circle at $x_9=0$:
\be\label{eqn:pvaraIIA}
\pvar^\iDb(t) =
\left.\int_0^{2\pi}A_{10}^{(\iDb)}dx_{10}\right\lvert_{x_9=0}
\,.
\ee

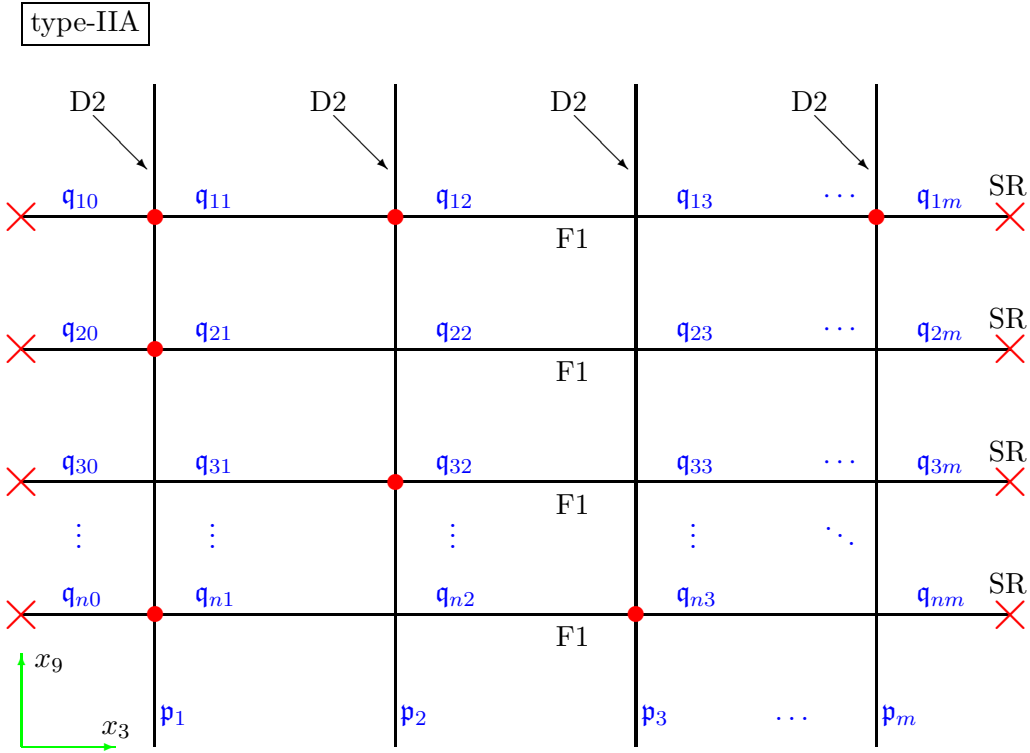
\begin{figure}
\begin{picture}(400,320)
\put(10,10) {\begin{picture}(390,280) \color{black}
    \put(0,270){\framebox{type-IIA}} \thicklines

% F1s
\color{black}
\put(0,50){\line(1,0){370}}
\put(0,100){\line(1,0){370}}
\put(0,150){\line(1,0){370}}
\put(0,200){\line(1,0){370}}

% D2s
\color{black}
\put(50,0){\line(0,1){250}}
\put(140,0){\line(0,1){250}}
\put(230,0){\line(0,1){250}}
\put(320,0){\line(0,1){250}}

% D2-F1 intersections
\color{red}
\put(50,50){\circle*{6}}
\put(50,150){\circle*{6}}
\put(50,200){\circle*{6}}
\put(140,100){\circle*{6}}
\put(140,200){\circle*{6}}
\put(230,50){\circle*{6}}
\put(320,200){\circle*{6}}

% SR twist
\color{red}
\put(-5,45){\line(1, 1){10}}
\put( 5,45){\line(-1, 1){10}}
\put(365,45){\line(1, 1){10}}
\put(375,45){\line(-1, 1){10}}
\put(-5,95){\line(1, 1){10}}
\put( 5,95){\line(-1, 1){10}}
\put(365,95){\line(1, 1){10}}
\put(375,95){\line(-1, 1){10}}
\put(-5,145){\line(1, 1){10}}
\put( 5,145){\line(-1, 1){10}}
\put(365,145){\line(1, 1){10}}
\put(375,145){\line(-1, 1){10}}
\put(-5,195){\line(1, 1){10}}
\put( 5,195){\line(-1, 1){10}}
\put(365,195){\line(1, 1){10}}
\put(375,195){\line(-1, 1){10}}

%labels
\color{black}
\put(362,58){SR}
\put(362,108){SR}
\put(362,158){SR}
\put(362,208){SR}

\color{blue}
\put(15,55){$\yvar_{n0}$}
\put(15,105){$\yvar_{30}$}
\put(15,155){$\yvar_{20}$}
\put(20,75){$\vdots$}
\put(15,205){$\yvar_{10}$}
\put(65,55){$\yvar_{n1}$}
\put(65,105){$\yvar_{31}$}
\put(65,155){$\yvar_{21}$}
\put(70,75){$\vdots$}
\put(65,205){$\yvar_{11}$}
\put(155,55){$\yvar_{n2}$}
\put(155,105){$\yvar_{32}$}
\put(155,155){$\yvar_{22}$}
\put(160,75){$\vdots$}
\put(155,205){$\yvar_{12}$}
\put(245,55){$\yvar_{n3}$}
\put(245,105){$\yvar_{33}$}
\put(245,155){$\yvar_{23}$}
\put(250,75){$\vdots$}
\put(245,205){$\yvar_{13}$}
\put(335,55){$\yvar_{nm}$}
\put(335,105){$\yvar_{3m}$}
\put(335,155){$\yvar_{2m}$}
\put(335,205){$\yvar_{1m}$}
\put(300,205){$\cdots$}
\put(300,105){$\cdots$}
\put(300,155){$\cdots$}
\put(300,75){$\ddots$}
\put(52,10){$\pvar_1$}
\put(142,10){$\pvar_2$}
\put(232,10){$\pvar_3$}
\put(282,10){$\ldots$}
\put(322,10){$\pvar_m$}

\thinlines
\color{black}
\put(200,188){F1}
\put(200,138){F1}
\put(200,88){F1}
\put(200,38){F1}
\put(18,240){D2}
\put(27,238){\vector(1,-1){20}}
\put(108,240){D2}
\put(117,238){\vector(1,-1){20}}
\put(198,240){D2}
\put(207,238){\vector(1,-1){20}}
\put(288,240){D2}
\put(297,238){\vector(1,-1){20}}

%axis
\put(5, 30){$x_9$}
\put(30, 5){$x_3$}
\color{green}
\put(0, 0){\vector(0,1){35}}
\put(0, 0){\vector(1,0){35}}

\end{picture}}
\end{picture}
\caption{A variable $\pvar^\iDb$ ($\iDb=1,\dots,m$) is associated with each D2-brane, ordered in the direction of increasing $x_3$. Let every string break at every brane into a total of $m+1$ open segments. The constant $x_{10}$ coordinate of each of these segments is denoted by $\yvar^{\iSt\iDbz}$ with $\iSt=1,\dots,n$ and $\iDbz=0,\ldots,m$ in the direction of increasing $x_3$. However, the strings actually break only at positions marked with a {\color{red}{\textbullet}} and the main contribution to the action comes from these break points. The strings are coincident, but we separated them in the picture for clarity.} \label{fig:nmconfig}
\end{figure}

In the presence of the D$2$-branes, fundamental strings can break into open strings. A string doesn't have to break at every D$2$-brane, but for ease of notation let us first assume that each of the $n$ fundamental strings does break at every D$2$-brane. We thus have $n \times (m+1)$ open string segments. Each segment must be located at a constant $x_{10}$ in order for its length to be minimal. We denote the $x_{10}$ coordinates of the open string segments by variables $\yvar^{\iSt\iDbz}$ (with $\iSt=1,\dots,n$ and $\iDbz=0,\ldots,m$). The string segments are ordered via the index $\iDbz$ in the direction of increasing $x_3.$ (See \figref{fig:nmconfig} for an illustration.) In addition to the $x_{10}$ coordinates we also need to know the $x_1$ coordinates of the strings. We denote them by $\xvar^{\iSt\iDbz}$:
$$
(x_1,x_{10})\rightarrow (\xvar^{\iSt\iDbz},\yvar^{\iSt\iDbz}).
$$

However, as noted above, an open string doesn't have to break at every D$2$-brane. For a given state, we define the {\it binding matrix} $\PMat$ to be the following $n\times m$ matrix of $0$ and $1$'s that encodes which strings break and which do not:
\be\label{eqn:PMatDef}
\PMat_{\iSt\iDb}=
\begin{cases}
1&  \text{if the $\iSt^{th}$ string is
bound to the $\iDb^{th}$ D$2$-brane},\\
0&  \text{otherwise},
\end{cases}
\ee
for $\iSt=1,\dots,n$ and $\iDb=1,\dots,m.$ We therefore have
$$
\xvar^{\iSt\iDb}=\xvar^{\iSt(\iDb-1)},\quad
\yvar^{\iSt\iDb}=\yvar^{\iSt(\iDb-1)},\qquad
\text{if $\PMat_{\iSt\iDb}=0$},
$$
which indicates that the $\iSt^{th}$ string is continuous at the $\iDb^{th}$ brane. On the other hand, at every break point $(\iSt,\iDb)$ on the $\iDb^{th}$ D$2$-brane the $\iSt^{th}$ string is charged under the gauge field on the $\iDb^{th}$ D$2$-brane $A^{(\iDb)}$, and the $A^{(\iDb)}_{10} dx_{10}$ interaction of the gauge field with the charged particle produces a term proportional to $\pvar^\iDb d\yvar^{\iSt\iDb}$ in the effective action. The sign of this term is positive for one end of the string and negative for the other, as the charges of the two ends are opposite. Thus an open segment of the $\iSt^{th}$ string that starts on the $\iDb^{th}$ D$2$-brane and ends on the $\jDb^{th}$ D$2$-brane contributes $\pvar^\iDb d\yvar^{\iSt\iDb}-\pvar^\jDb d\yvar^{\iSt(\jDb-1)}$ to the action.

On top of this, direction $x_3$ is compact with $x_3\in [0, 2 \pi R]$ and the S-duality and R-symmetry operators act at $x_3=0$. The S-R-twisted boundary conditions induce linear relations between the $x_1$ and $x_{10}$ coordinates of the strings they connect, where we are free to add a permutation $\perm\in S_n$ among the $n$ strings before applying the S-R-twist. A particular sector of the Hilbert space is thus described by the $\PMat_{\iSt\iDb}$ matrix, as well as the permutation $\perm\in S_n.$ We collect this information in a matrix and denote a given sector as
$$
\left\lbrack\begin{array}{cccc|c}
\PMat_{11} & \PMat_{12} & \cdots & \PMat_{1m} & \perm(1) \\
\PMat_{21} & \PMat_{22} & \cdots & \PMat_{2m} & \perm(2) \\
\vdots     & \vdots     & \ddots & \vdots     & \vdots   \\
\PMat_{n1} & \PMat_{n2} & \cdots & \PMat_{nm} & \perm(n) \\
\end{array}\right\rbrack\,.
$$
Obviously, two sectors $[\PMat_{\iSt\iDb}|\perm]$ and $[\PMat_{\perm'(\iSt)\iDb}|{\perm'}\circ\perm\circ\perm'^{-1}]$ are equivalent (with $\perm'\in S_n$).
The boundary conditions can now be written as
\be\label{eqn:SRp}
e^{i\pht}(\yvar^{\iSt m} + \tau\xvar^{\iSt m}) =
\yvar^{\perm(\iSt)0} + \tau\xvar^{\perm(\iSt)0}
\pmod {2\pi(\Z+\tau\Z)}.
\ee
Using \eqref{eqn:phtdef} and $\tau=\frac{\xa\tau+\xb}{\xc\tau+\xd}$ we can rewrite \eqref{eqn:SRp} as an equation with integer coefficients:
\be\label{eqn:SRpZ}
\yvar^{\perm(\iSt)0} = \xd\yvar^{\iSt m}+\xb\xvar^{\iSt m}
\,,\qquad
\xvar^{\perm(\iSt)0} = \xc\yvar^{\iSt m}+\xa\xvar^{\iSt m}
\,.
\ee
For example, for $\pht=\frac{\pi}{2}$ ($\lvk=2$), this becomes
$$
\xvar^{\iSt m}=-\yvar^{\perm(\iSt)0}
\,,\qquad
\yvar^{\iSt m}=\xvar^{\perm(\iSt)0}
\pmod{2\pi\Z}.
$$

Finally, let $\xvar^\iDb$ ($\iDb=1,\dots,m$) be the $x_1$ coordinate of the $\iDb^{th}$ D$2$-brane. Then we have the equations
\be\label{eqn:xconst}
\xvar^\iDb = \xvar^{\iSt\iDb} = \xvar^{\iSt(\iDb-1)}\,,
\qquad
\text{whenever $\PMat_{\iSt\iDb}=1$,}
\ee
since the $\iSt^{th}$ string connects with the $\iDb^{th}$ D2-brane.
The equations \eqref{eqn:SRp}-\eqref{eqn:xconst} reduce the total number of independent $\yvar^{\iSt\iDbz}$ variables. A linearly independent basis can be chosen, and these furnish the $\qvar^{\alpha}$ variables in \eqref{eqn:Mxpq}. In \secref{sec:U(1)}-\secref{sec:U(n)} we will present explicit detailed examples.

% - - - - - - - - - - - - - - - - - - - - - - - - - - - - - - -
\subsubsection*{Congested and decongested matrices}

We say that a binding matrix $\PMat_{\iSt\iDb}$ is {\it congested} if there is at least one $\iDb$ for which there are two distinct $\iSt\neq\jSt$ such that $\PMat_{\iSt\iDb}=\PMat_{\jSt\iDb}=1$. This means that there is at least one D$2$-brane from which at least two strings emanate. If every D$2$-brane has exactly one string emanating from it, we say that the binding matrix is {\it decongested}. In this case, for every $\iDb=1,\dots,m$ there is exactly one $\iSt$ for which $\PMat_{\iSt\iDb}=1.$ The difference between congested and decongested binding matrices will become relevant when we discuss fermionic zero-modes in \secref{subsec:countfm=1}.

% --------------------------------------------------------------
\subsection{The interaction term $I_1$}
\label{subsec:I1}

We will now derive the interaction of the 0+1D variables $\pvar^\iDb,\qvar^\alpha,\xvar^\iDb$ with the 1+1D fields. The low-energy 1+1D fields that are relevant for the present discussion can be described either in type-IIA or in type-IIB. In type-IIB, these fields are the two scalars $\scX^1(x_9,t),\scX^2(x_9,t)$ (with the index $\iStIIB$ of the string suppressed). In type-IIA, the two relevant low-energy fields on the D$2$-brane are the gauge field component $A_{10}$ and the $x_1$ coordinate of the D$2$-brane, which we denote by $\xPhi^1$. Note that direction $1$ in type-IIA is related to direction $1$ in type-IIB via T-duality. Following the U-duality of \tabref{tab:Dualities}, it is then easy to see that $\scX^1,\scX^2$ are the duals (as 1+1D free compact scalar fields) of $\xPhi^1,A_{10}$, respectively:
\be\label{eqn:TDuality}
\partial_9\scX^1=\partial_t\xPhi^1
\,,\quad
\partial_t\scX^1=-\partial_9\xPhi^1
\,,\quad
\partial_9\scX^2=\partial_t A_{10}
\,,\quad
\partial_t\scX^2=\partial_9 A_{10}
\,.
\ee

From this simple observation, it is easy to derive the requisite interaction between $\scX^1,\scX^2$ and $\pvar^\iDb,\qvar^\alpha,\xvar^\iDb.$ In the type-IIA picture $\{\pvar^\iDb,\xvar^\iDb\}$ determine the boundary conditions at $x_9=0$ of $\xPhi^1,A_{10}$. If $\iDb$ is the index of the brane, then by definition, we have the Dirichlet boundary conditions:
\begin{align*}
2\pi A_{10}(x_9=0,t) &= \pvar^\iDb(t)\,,\\
\xPhi^1(x_9=0,t) &= \xvar^\iDb(t)\,,
\end{align*}
where we suppressed brane indices on the left-hand sides.
The duality \eqref{eqn:TDuality} then converts the Dirichlet boundary conditions of type-IIA to Neumann boundary conditions of type-IIB. The latter can be incorporated into the action with the addition of the term
\be\label{eqn:xvarscX}
\int (\xvar^\iDb d\scX^1 + \pvar^\iDb d\scX^2)
\,.
\ee
For most sectors the $\xvar^\iDb$'s can be written as linear combinations of the $\qvar^\alpha$'s by using the various constraints discussed at the end of \secref{subsec:I0}. In these cases the sum of the terms \eqref{eqn:xvarscX} for all the D$2$-branes takes the form of $I_1$ in \eqref{eqn:I1}.

In the present paper we will not have much use for the interaction term $I_1$, since we keep the charge coordinates $\scX^1, \scX^2$ constant. We have nevertheless presented it here for completeness. In a future work we hope to explore the dependence of the Hilbert space on the position of the charges, and the term $I_1$ will then play a central role. We conclude the discussion of the interaction term by presenting a more geometrical interpretation of $I_1$.

% --------------------------------------------------------------
% --------------------------------------------------------------
% --------------------------------------------------------------
% --------------------------------------------------------------

\begin{figure}[t]
\begin{picture}(400,140)

\put(10,10){\begin{picture}(200,120)
\color{red}
\put(60,130){(a)}

% axes
\color{green}
\thinlines
\put(0,90){\vector(0,1){30}}
\put(0,90){\vector(1,0){30}}
\color{black}
\put(-3,122){$x_9$}
\put(32,87){$x_3$}

% junction
\color{black}
\thicklines
\put(90,-10){\line(0,1){125}}
\thinlines
\multiput(90,50)(12,0){7}{\line(1,0){10}}
\multiput(90,50)(-12,0){7}{\line(-1,0){10}}
\put(92,105){D$2$}
\put(25,40){F$1$}
\put(145,40){F$1$}
\color{blue}
\put(130,55){$x_{10}=\qvar_2$}
\put(10,55){$x_{10}=\qvar_1$}
\put(92,90){$x_2 = a_2^{(2)}$}
\put(92,0){$x_2 = a_2^{(1)}$}
\color{red}
\put(90,50){\circle*{5}}

\color{black}
\thinlines
\put(190,-15){\line(0,1){150}}

\put(200,0){\begin{picture}(200,120)
\color{red}
\put(60,130){(b)}

\color{black}
\thicklines
\put(80,-20){\line(0,1){60}}
\put(80,40){\line(2,1){40}}
\put(120,60){\line(0,1){60}}
\thinlines
\multiput(80,40)(-12,0){7}{\line(-1,0){10}}
\multiput(120,60)(12,0){7}{\line(1,0){10}}

\put(63,0){D$2$}
\put(123,110){D$2$}
\put(10,30){F$1$}
\put(170,50){F$1$}
\color{blue}
\put(123,95){$u\rightarrow\infty$}
\put(45,-15){$u\rightarrow 0$}
\put(2,45){$v\rightarrow 0$}
\put(162,65){$v\rightarrow\infty$}
\color{red}
\put(80,40){\circle*{5}}
\put(120,60){\circle*{5}}

\end{picture}}
\end{picture}}

\end{picture}
\caption{ (a) A junction of two open fundamental strings, one starting
  and one ending on a D$2$-brane. (b) The D$2$-brane wraps direction
  $10\equiv\tenx$ and the configuration can be deformed so that
  it lifts to a smooth holomorphic curve in M-theory.}
\label{fig:D2F1junction}
\end{figure}
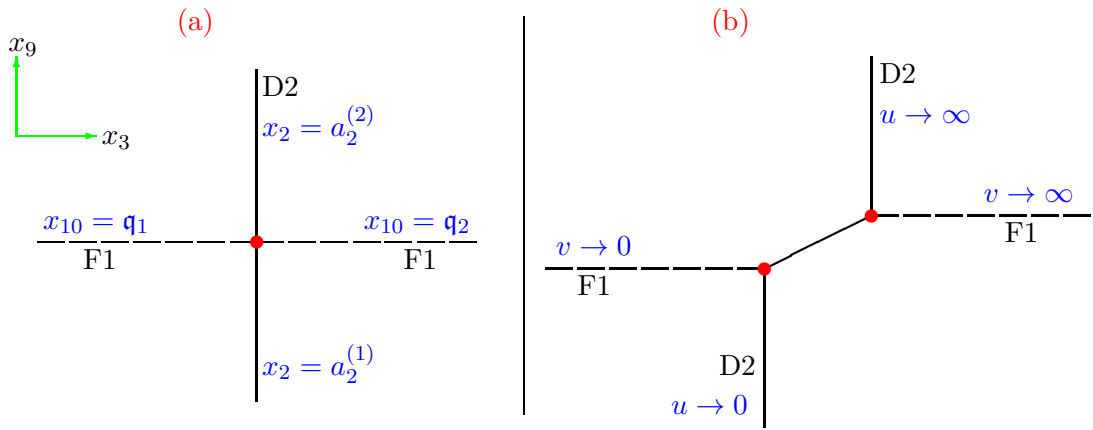
% --------------------------------------------------------------
% --------------------------------------------------------------
% --------------------------------------------------------------
% --------------------------------------------------------------

\subsubsection*{D$2$-F$1$ intersections}
%\label{subsec:D2F1}

The interaction term $\pvar^\iDb d\scX^2$ in \eqref{eqn:xvarscX} has a simple interpretation in terms of the geometry of the lift of the D$2$-branes and F$1$-strings to M-theory. To see this, let us focus on a single F$1$-string and a single D$2$-brane. We start by recalling some facts about this system, following the techniques developed in \cite{Witten:1997sc}. There, a configuration of D$4$-branes and NS$5$-branes was analyzed by lifting it to M-theory. Here, we need to analyze a similar configuration of fundamental strings and D$2$-branes, and we are also going to lift it to M-theory.

The lift essentially brings us back to the second-from-last row of \tabref{tab:Dualities}, and the M-theory direction in the present discussion is therefore denoted by $x_2.$ We assume that $x_2,x_{10}$ are periodic with period $2\pi.$ Let us also assume for simplicity that the length of the $x_3$ direction is very large, so that $-\infty < x_3 < \infty$. (That is, we will suspend the effect of S-R-twist for the moment.) Following \cite{Witten:1997sc}, we define complex variables
$$
v = e^{x_9 + i x_{10}}
\,,\qquad
u = e^{x_3 + i x_2}
\,.
$$
The configuration of an F$1$-string intersecting a D$2$-brane then lifts to a single M$2$-brane which extends along the locus of the complex equation
\be\label{eqn:uCv}
v = \frac{u - e^{i\alpha}}{u-e^{-i\alpha}}
\,,
\ee
where $\alpha$ is a real constant. Equation \eqref{eqn:uCv} is designed so that as $x_9\rightarrow\infty$ (where $v$ has a pole) $x_2=-\alpha$, and as $x_9\rightarrow-\infty$ (where $v$ has a zero) $x_2=\alpha$ (see \figref{fig:D2F1junction}b). Equation \eqref{eqn:uCv} also tells us that as $x_3\rightarrow\infty$ we have $v=1$, while as $x_3\rightarrow-\infty$ we have $v=e^{2i\alpha}.$ If we now let $\qvar^1$, $\qvar^2$ be the $x_{10}$ coordinates of the string between $x_3=\infty$ and $x_3=-\infty$, then we have
\be\label{eqn:Deltaq}
\Delta x_{10}\equiv \qvar^2 - \qvar^1= 2\alpha
=-\scX^2(x_9=+\infty)+\scX^2(x_9=-\infty)
 \mod 2\pi
\,.
\ee

Going to the low-energy limit, we can interpret the boundary condition $\scX^2(x_9=+\infty)$ as the boundary value at $x_9=0$ of the $\scX^2$ field on the $x_9>0$ portion of the string (which we denoted by $\scX^2_\iDb$), and similarly the boundary condition $\scX^2(x_9=-\infty)$ is the low-energy boundary value at $x_9=0$ of the $\scX^2$ field on the $x_9<0$ portion of the string (which we denoted by $\scX^2_{\iDb+m}$). The geometrical equation \eqref{eqn:Deltaq} is therefore consistent with the equations of motion derived from varying $\pvar$ in the action
$$
\int \pvar(\dot{\qvar}^2-\dot{\qvar}^1
+\dot{\scX}^2_\iDb-\dot{\scX}^2_{\iDb+m})\,.
$$
After taking into account the S-R-twist, $\qvar^1$ and $\qvar^2$ are not independent anymore,
and the above expression then becomes the contribution of $\pvar$ to the action \eqref{eqn:ActionLE}.

% --------------------------------------------------------------
\subsection{Bosonic zero modes}
\label{subsec:BosZmodes}

In \secref{subsec:I0} we explained how to determine the bosonic part of the action in terms of the variables $\pvar^\iDb,\xvar^{\iSt\iDbz},$ and $\yvar^{\iSt\iDbz}.$ Generally, this action can be further simplified because equations \eqref{eqn:SRp}-\eqref{eqn:xconst} reduce the total number of independent $\yvar^{\iSt\iDbz}$ variables. So in general, as we shall see in concrete examples in \secref{sec:U(1)}-\secref{sec:U(n)}, some of the $\xvar^{\iSt\iDbz}$ variables can be expressed as linear combinations of the $\yvar^{\iSt\iDbz}$'s, while the remaining $\xvar^{\iSt\iDbz}$'s are reduced to discrete values. After eliminating the $\xvar^{\iSt\iDbz}$'s we get an action of the form \eqref{eqn:Mxpq}. If the corresponding constants $\Matx_{\iDb\alpha}$ that appear in this action form a nonsingular square matrix, quantization of \eqref{eqn:Mxpq} gives rise to a finite dimensional Hilbert space.

However, for some sectors we end up with more than $m$ independent $\yvar^{\iSt\iDbz}$'s and then the procedure of \secref{subsec:I0} yields an expression similar to \eqref{eqn:Mxpq} but where $\Matx_{\iDb\alpha}$ is not a square matrix. The simplest sector for which this happens is for $n=m=2$ with
\be\label{eqn:BZMsector}
[\PMat|\perm] =
\left\lbrack\begin{array}{cc|c}
1 & 1 & 1 \\
1 & 1 & 2 \\
\end{array}\right\rbrack\,.
\ee
In general for $n=2$ there are more than $m$ independent $\yvar^{\iSt\iDbz}$'s when $\PMat$ has at least two columns of the form $(1 1)^\top$ [where $(\cdots)^\top$ denotes the transposed matrix]. As will be explained in more detail in one of the examples of \secref{subsubsec:n=2;m=2}, for $\lvk=2$ the action for the sector \eqref{eqn:BZMsector} is:
\be\label{eqn:I0(n=2,m=2,k=2)BosZmodes}
I_0 =
\frac{1}{2\pi}
\int\left\{
\pvar^1 \left\lbrack  d\left(\yvar^{11} + \yvar^{21}\right)
+ 2d\yvar^{12} \right\rbrack
+\pvar^2 \left\lbrack
  - d\left(\yvar^{11} + \yvar^{21}\right)
+2d\yvar^{12}\right\rbrack
\right\}\,.
\ee
In this sector there are two strings connecting the two D$2$-branes, and the action only depends on $\yvar^{11}$ and $\yvar^{21}$ through the center of mass $Q_{\tiny{com}}\equiv \yvar^{11}+\yvar^{21}$, and is independent of the relative coordinate $Q_{\tiny{rel}}\equiv\yvar^{11}-\yvar^{21}$. Therefore, in order to proceed we need to add a kinetic term, proportional to $\dot{Q}_{\tiny{rel}}^2.$ However, if we are only interested in ground states we may simply rewrite \eqref{eqn:I0(n=2,m=2,k=2)BosZmodes} in terms of $Q_{\tiny{com}}$,
\be\label{eqn:I0(n=2,m=2,k=2)BosZmodesQcom}
I_0 =
\frac{1}{2\pi}
\int\left\{
\pvar^1(dQ_{\tiny{com}}+ 2d\yvar^{12})
+\pvar^2(- dQ_{\tiny{com}}+2d\yvar^{12})
\right\}\,,
\ee
which is of the form \eqref{eqn:Mxpq} with a nonsingular square matrix
$
\Matx=\left( \begin{array}{rr} 1 & 2 \\ -1 & 2  \end{array}\right)
\,.
$
{}From \eqref{eqn:I0(n=2,m=2,k=2)BosZmodesQcom} we can determine the number of ground states, which happens to be $4.$

We will show later on that sectors with bosonic zero modes invariably also possess fermionic zero modes and therefore do not contribute to the Witten Index. In fact, the sectors with bosonic zero modes form a proper subset of the set of congested sectors, and all congested sectors have fermionic zero modes.
In this paper we are only concerned with the Witten Index, and we therefore do not need to consider sectors with bosonic zero modes anymore.

% --------------------------------------------------------------
\subsection{Fermionic zero modes}
\label{subsec:ZeroModes}

So far in this section, we have mainly focused on the bosonic degrees of freedom of the system. While we will not present the explicit form of the fermionic part of the action, it will turn out to be important to understand the fermionic zero modes of the system in each sector described by the binding matrix $\PMat_{\iSt\iDb}$ and permutation $\perm\in S_n$. Specifically, they will be crucial to our argument that only decongested sectors contribute to the Witten Index.\footnote{This will be our only use for the fermionic degrees of freedom, so a reader who doesn't wish to go into the detailed proof of this statement can skip the present section and \secref{subsec:fzeromodesn=1},\secref{subsec:countfm=1}, as well as some portions of \secref{subsec:WittenIndex}.} Therefore, in this subsection, we will discuss various chirality and boundary conditions that these fermionic zero modes have to satisfy.

\begin{figure}
\begin{picture}(400,320)
\put(10,10) {\begin{picture}(390,280) \color{black}
    %\put(0,270){\framebox{type-IIA}}

    \thicklines

% F1s
\color{black}
\put(0,50){\line(1,0){370}}
\put(0,100){\line(1,0){370}}
\put(0,150){\line(1,0){370}}
\put(0,200){\line(1,0){370}}

% D2s
\color{black}
\put(50,0){\line(0,1){250}}
\put(140,0){\line(0,1){250}}
\put(230,0){\line(0,1){250}}
\put(320,0){\line(0,1){250}}

% D2-F1 intersections
\color{red}
\put(50,50){\circle*{6}}
\put(50,150){\circle*{6}}
\put(50,200){\circle*{6}}
\put(140,100){\circle*{6}}
\put(140,200){\circle*{6}}
\put(230,50){\circle*{6}}
\put(320,200){\circle*{6}}

% SR twist
\color{red}
\put(-5,45){\line(1, 1){10}}
\put( 5,45){\line(-1, 1){10}}
\put(365,45){\line(1, 1){10}}
\put(375,45){\line(-1, 1){10}}
\put(-5,95){\line(1, 1){10}}
\put( 5,95){\line(-1, 1){10}}
\put(365,95){\line(1, 1){10}}
\put(375,95){\line(-1, 1){10}}
\put(-5,145){\line(1, 1){10}}
\put( 5,145){\line(-1, 1){10}}
\put(365,145){\line(1, 1){10}}
\put(375,145){\line(-1, 1){10}}
\put(-5,195){\line(1, 1){10}}
\put( 5,195){\line(-1, 1){10}}
\put(365,195){\line(1, 1){10}}
\put(375,195){\line(-1, 1){10}}

%labels
\color{black}
\put(362,58){SR}
\put(362,108){SR}
\put(362,158){SR}
\put(362,208){SR}

\color{blue}
\put(15,55){$\fpsi_{n0}$}
\put(15,105){$\fpsi_{30}$}
\put(15,155){$\fpsi_{20}$}
\put(20,75){$\vdots$}
\put(15,205){$\fpsi_{10}$}
\put(70,55){$\fpsi_{n1}$}
\put(70,105){$\fpsi_{31}$}
\put(70,155){$\fpsi_{21}$}
\put(75,75){$\vdots$}
\put(70,205){$\fpsi_{11}$}
\put(160,55){$\fpsi_{n2}$}
\put(160,105){$\fpsi_{32}$}
\put(160,155){$\fpsi_{22}$}
\put(165,75){$\vdots$}
\put(160,205){$\fpsi_{12}$}
\put(250,55){$\fpsi_{n3}$}
\put(250,105){$\fpsi_{33}$}
\put(250,155){$\fpsi_{23}$}
\put(255,75){$\vdots$}
\put(250,205){$\fpsi_{13}$}
\put(335,55){$\fpsi_{nm}$}
\put(335,105){$\fpsi_{3m}$}
\put(335,155){$\fpsi_{2m}$}
\put(335,205){$\fpsi_{1m}$}
\put(300,205){$\cdots$}
\put(300,105){$\cdots$}
\put(300,155){$\cdots$}
\put(300,75){$\ddots$}
\put(52,25){$\lambda_{1n}$}
%\put(52,75){$\lambda_{13}$}
\put(52,125){$\lambda_{12}$}
\put(52,175){$\lambda_{11}$}
\put(52,225){$\lambda_{10}$}
\put(142,25){$\lambda_{2n}$}
%\put(142,75){$\lambda_{23}$}
\put(142,125){$\lambda_{22}$}
\put(142,175){$\lambda_{21}$}
\put(142,225){$\lambda_{20}$}
\put(232,25){$\lambda_{3n}$}
%\put(232,75){$\lambda_{33}$}
\put(232,125){$\lambda_{32}$}
\put(232,175){$\lambda_{31}$}
\put(232,225){$\lambda_{30}$}
\put(322,25){$\lambda_{mn}$}
%\put(322,75){$\lambda_{m3}$}
\put(322,125){$\lambda_{m2}$}
\put(322,175){$\lambda_{m1}$}
\put(322,225){$\lambda_{m0}$}
\put(282,10){$\ldots$}

\thinlines
\color{black}
\put(200,188){F1}
\put(200,138){F1}
\put(200,88){F1}
\put(200,38){F1}
\put(18,240){D2}
\put(27,238){\vector(1,-1){20}}
\put(108,240){D2}
\put(117,238){\vector(1,-1){20}}
\put(198,240){D2}
\put(207,238){\vector(1,-1){20}}
\put(288,240){D2}
\put(297,238){\vector(1,-1){20}}

%axis
\put(5, 30){$x_9$}
\put(30, 5){$x_3$}
\color{green}
\put(0, 0){\vector(0,1){35}}
\put(0, 0){\vector(1,0){35}}

\end{picture}}
\end{picture}
\caption{The fermionic zero modes are constructed
from solutions of the linear equations for
the boundary conditions of the gluino fields $\lambda_{\iDb\iStz}$
on the D$2$-brane sections (viewed as $1$-dimensional
segments below the $x_{10}$ compactification scale)
and the fermionic modes $\fpsi_{\iSt\iDb}$
on string sections. The S-R-twist is denoted by a $\times$.}
\label{fig:ConfigZeroModes}
\end{figure}
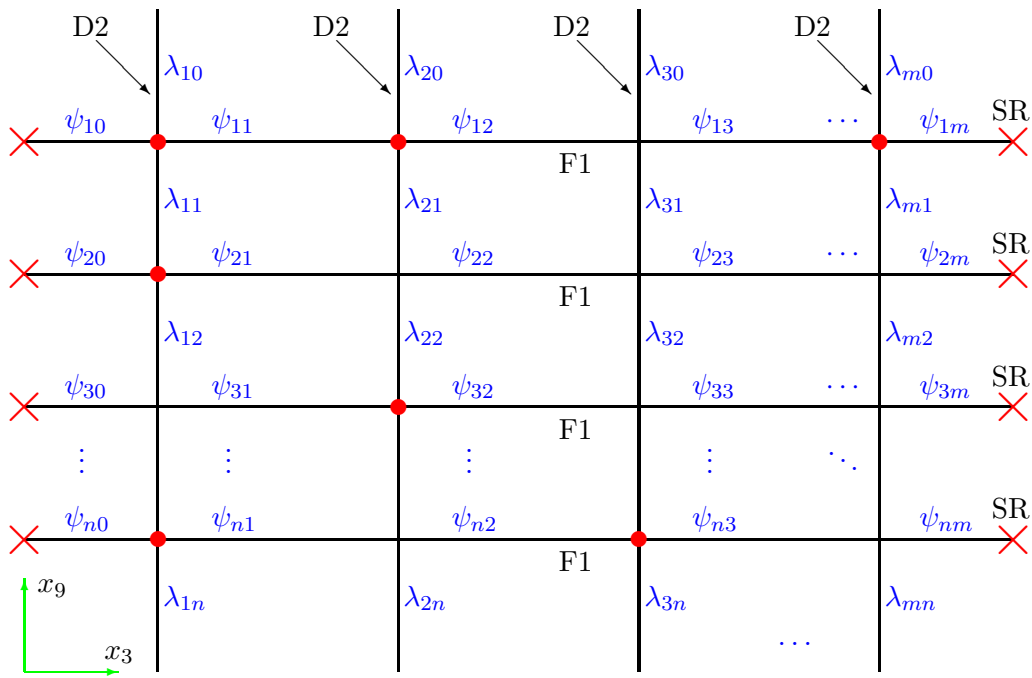

Our conventions for fermions are as follows. In describing the fermionic modes of type-IIA theory we find it more convenient to consider its M-theory lift. Therefore, we denote a fermion by a real $32$-component 10+1D spinor on which the Dirac matrices $\Gamma^0,\dots,\Gamma^9,\Gamma^{\tenx}$ act. (We use the notation $\tenx\equiv 10$ to avoid confusion between $\Gamma^{10}$ and $\Gamma^1\Gamma^0$.) They satisfy the identity
\be\label{eqn:11Dchlty}
\Gamma^{0123456789\tenx}=1.
\ee
The low-energy fermionic modes of the system can then be described in terms of 1+1D fermionic fields that are supported on the type-IIA open strings and on the dimensional reduction (on $x_{10}$ direction) of the D$2$-branes. (See \figref{fig:ConfigZeroModes} for illustration.)
The fermionic field along the $\iSt^{th}$ open F$1$-string between the $\iDbz^{th}$ and $(\iDbz+1)^{st}$ D$2$-branes is denoted by $\fpsi_{\iSt\iDbz}$ (with $\iSt=1,\dots,n$ and $\iDbz=1,\dots,m-1$). For every $\iSt=1,\dots,n$, there is another piece of string starting on the $m^{th}$ D$2$-brane, going through the S-R-twist, and ending on the $1^{st}$ D$2$-brane. To capture the fields on this string using the same notation, we extend the range of $\iDbz$ to $0,\dots,m$ and postulate that the fields $\fpsi_{\iSt 0}$ and $\fpsi_{\iSt m}$ are identified up to the S-R-twist
\be\label{eqn:ZMxSRm}
\fpsi_{\perm(\iSt)0} = e^{\frac{\pht}{2}(
\Gamma^{1\tenx}+\Gamma^{45}+\Gamma^{67}+\Gamma^{89})}
\fpsi_{\iSt m}
\,,
\ee
where we have also allowed the possibility of the action of the permutation $\sigma\in S_n$, as discussed in \secref{subsec:I0}.
We set
\be\label{eqn:OmegaDef}
\mathcal{P}\equiv
e^{-\frac{\pht}{2}(\Gamma^{1\tenx}+\Gamma^{45}+\Gamma^{67}+\Gamma^{89})}\,.
\ee

Generally, the $\fpsi_{\iSt\iDbz}$ fields are functions of $(x_3,t)$ (with the appropriate finite range for $x_3$), but at low-energy only the zero modes are important, so we can assume that $\fpsi_{\iSt\iDbz}$ is independent of $x_3.$
The $\fpsi_{\iSt\iDbz}$ fields also satisfy the obvious boundary conditions that if the $\iSt^{th}$ string is not bound to the $\iDb^{th}$ D$2$-brane then it continuously connects with $\fpsi_{\iSt(\iDb-1)}$:
$$
\fpsi_{\iSt\iDb}=\fpsi_{\iSt(\iDb-1)}\qquad\text{if $\PMat_{\iSt\iDb}=0.$}
$$
All the $\fpsi_{\iSt\iDb}$ fields satisfy the chirality condition
\be\label{eqn:F1ch}
0 = (1+\Gamma^{023})\fpsi_{\iSt\iDb}
\ee
corresponding to the low-energy fields on an M$2$-brane with $x_2$ being the M-theory direction.

% --------------------------------------------------------------
% --------------------------------------------------------------

% --------------------------------------------------------------
% --------------------------------------------------------------
% --------------------------------------------------------------

% --------------------------------------------------------------
% --------------------------------------------------------------
% --------------------------------------------------------------
% --------------------------------------------------------------

Next, we define the fields along the D$2$-branes. Dimensionally reducing along the compact direction of $x_{10}$, the D$2$-branes become 1+1D objects, and we denote by $\lambda_{\iDb\iStz}$ (with $\iDb=1,\dots,m$ and $\iStz=0,\dots,n$) the low-energy fermionic fields supported on the segment of $\iDb^{th}$ D$2$-brane between the $\iStz^{th}$ and $(\iStz+1)^{st}$ F$1$-string, where we use the convention that $\iStz=0$ corresponds to the segment that continues from the first string to $x_9>0$, and $\iStz=n$ continues from the $n^{th}$ string to $x_9<0$ (see \figref{fig:ConfigZeroModes}). These fields are generally functions of $(x_9,t)$,
%where $x_9$ takes values $x_9>0$ for $\iStz=n$ and $x_9<0$ for $\iStz=0$,
but at low energy they can again be assumed to be constant. Formally, the range of $x_9$ for $1\le \iStz<n$ is zero, but because the fields are constant this does not matter. A more elaborate treatment starting with 2D fields that are harmonic functions of $x_9+i x_{10}$, with poles at the intersections with the F$1$-string, will lead to a similar result.

Similarly as for $\fpsi_{\iSt\iDbz}$, we have for $\lambda_{\iDb\iStz}$ the continuity conditions
$$
\lambda_{\iDb\iSt}=\lambda_{\iDb(\iSt-1)}
\qquad\text{if $\PMat_{\iSt\iDb}=0$}
$$
and we also have the chirality conditions:
\be\label{eqn:D2ch}
0 = (1+\Gamma^{09\tenx})\lambda_{\iDb\iStz}.
\ee
At the D$2$-F$1$ junctions where the $\iDb^{th}$ D$2$-brane and $\iSt^{th}$ F1-string intersect (so that $\PMat_{\iSt\iDb}=1$), the zero modes have to satisfy the following boundary conditions:
\be\label{eqn:ZMxJp}
(1-\Gamma^{239\tenx})\fpsi_{\iSt\iDb} =
(1-\Gamma^{239\tenx})\fpsi_{\iSt(\iDb-1)} =
(1-\Gamma^{239\tenx})\lambda_{\iDb\iSt} =
(1-\Gamma^{239\tenx})\lambda_{\iDb(\iSt-1)}
\,,
\ee
and
\be\label{eqn:ZMxJm}
0 =
(1+\Gamma^{239\tenx})(\fpsi_{\iSt(\iDb-1)} -\fpsi_{\iSt\iDb}
+ \Gamma^{39}\lambda_{\iDb\iSt} -\Gamma^{39}\lambda_{\iDb(\iSt-1)})\,.
\ee
These equations are derived by first going to the M-theory picture as in \figref{fig:D2F1junction}, where the D$2$-F$1$ junction is described by a single M$2$-brane, and then deforming the M$2$-brane worldvolume. Details of the derivation of \eqref{eqn:ZMxJp}--\eqref{eqn:ZMxJm} are provided in \appref{app:SpinorsSUSY}.

% --------------------------------------------------------------
\subsection{The eigenvalues of $\mathcal{P}$}
\label{subsec:evP}

The operator $\mathcal{P}$, defined in \eqref{eqn:OmegaDef}, realizes the S-R-twist on the fermionic modes of the fundamental strings in the type-IIA picture. In this subsection we calculate its eigenvalues. This will be important in \secref{subsec:fzeromodesn=1} and \secref{subsec:countfm=1} where we prove the absence of zero modes for certain sectors.

Consider a spinor $\fpsi$ that satisfies $\mathcal{P}\fpsi=\varepsilon\fpsi$ for some eigenvalue $\varepsilon.$ Since $\mathcal{P}$ commutes with $\Gamma^{023}$ we may assume that $\fpsi$ has a specific $\Gamma^{023}$ chirality. We first assume that
$$
\Gamma^{023}\fpsi=\fpsi.
$$
Then, $\Gamma^{0123456789\tenx}=1$ implies
$$
0 = (1-\Gamma^{023})\fpsi =
(1-\Gamma^{1456789\tenx})\fpsi \,,
$$
and hence
$$
\Gamma^{451\tenx}\fpsi
= \Gamma^{451\tenx}\Gamma^{1456789\tenx}\fpsi
= \Gamma^{6789}\fpsi\,.
$$
We can therefore rewrite $\mathcal{P}\fpsi$ as
$$
\mathcal{P}\fpsi =
e^{\frac{\pht}{2}\Gamma^{45}(1-\Gamma^{451\tenx})
+\frac{\pht}{2}\Gamma^{67}(1-\Gamma^{6789})}
\fpsi
=
e^{\frac{\pht}{2}\Gamma^{45}(1-\Gamma^{6789})
+\frac{\pht}{2}\Gamma^{67}(1-\Gamma^{6789})}
\fpsi
\,.
$$
Since $\Gamma^{6789}$ has eigenvalues $\pm 1$, and $\Gamma^{ij}$ has eigenvalues $\pm i$ for all spatial indices $i,j$ ($i\neq j$), we deduce that $\mathcal{P}$ has eigenvalues $1$ and $e^{\pm 2i\pht}$ on the subspace with $\Gamma^{023}$-chirality $1.$

We can similarly analyze the eigenvalues of $\mathcal{P}$ on the subspace of $\fpsi$'s with the opposite $\Gamma^{023}$-chirality, i.e., $\Gamma^{023}\fpsi=-\fpsi.$ On that subspace we find
\be\label{eqn:P6789}
\mathcal{P}\fpsi =
e^{\frac{\pht}{2}\Gamma^{45}(1-\Gamma^{451\tenx})
+\frac{\pht}{2}\Gamma^{67}(1-\Gamma^{6789})}
\fpsi
=
e^{\frac{\pht}{2}\Gamma^{45}(1+\Gamma^{6789})
+\frac{\pht}{2}\Gamma^{67}(1-\Gamma^{6789})}
\fpsi
\,,
\ee
and hence deduce that $\mathcal{P}$ has eigenvalues $e^{\pm i\pht}$. Note that $e^{\pm i\pht}\neq 1$ and so there is no nontrivial solution to $\fpsi=\mathcal{P}\fpsi$ on the subspace with $\Gamma^{023}$-chirality $-1.$ This fact will come in handy later on.

% --------------------------------------------------------------
\subsection{Constructing a Witten Index}
\label{subsec:WittenIndex}

The states of the system discussed in \secref{subsec:ChargesIIB}, while they contain all the information about the Hilbert space of Tr-S theory with external charges, also contain superfluous excitations in the form of long wavelength modes of $\scX_\iStIIB^\mu,\fpsi_\iStIIB$ along the semi-infinite strings. We can eliminate these excitations by imposing appropriate boundary conditions on the modes $\scX_\iStIIB^\mu,\fpsi_\iStIIB$ at some finite distance from the origin, say at $x_9=\pm\DXiv$ (where the $+$ sign is for $\iStIIB\le m$ and the $-$ sign is for $\iStIIB>m$, for some positive constant $\DXiv$). The following boundary conditions preserve the four real supersymmetries left unbroken by \eqref{eqn:D3-ep}-\eqref{eqn:F1-ep}.

In the type-IIB picture, we pick Neumann boundary conditions for
fluctuations in directions $4,\dots,8$:
\be\label{eqn:D5bc45678}
\px{9}\scX_\iStIIB^4(\pm\DXiv) =
\px{9}\scX_\iStIIB^5(\pm\DXiv) =
\px{9}\scX_\iStIIB^6(\pm\DXiv) =
\px{9}\scX_\iStIIB^7(\pm\DXiv) =
\px{9}\scX_\iStIIB^8(\pm\DXiv) = 0
\,,
\ee
Dirichlet boundary conditions in directions $1,2,3$:
\be\label{eqn:D5bc123}
\scX_\iStIIB^1(\pm\DXiv) = a_1^{(\iStIIB)}
\,,\quad
\scX_\iStIIB^2(\pm\DXiv) = a_2^{(\iStIIB)}
\,,\quad
\scX_\iStIIB^3(\pm\DXiv) = 0
\,,
\ee
and supersymmetric boundary conditions for the fermions:
\be\label{eqn:D5bcfpsi}
\Gamma^{045678}\fpsi_\iStIIB(\pm\DXiv) = \fpsi_\iStIIB(\pm\DXiv)
\,.
\ee
These boundary conditions are formally what we would get if we let the string end on a D$5$-brane that extends in directions $4,\dots,8$ and is fixed in directions $1,2,3,9.$ However, we must note that because directions $1,2,3$ are compact, such a D$5$-brane will back-react strongly on the metric and will require additional orientifolds or other objects to make a complete string-theory solution. (This is similar to the situation with D$8$-branes as developed in \cite{Polchinski:1995df}, or with D$7$-branes in \cite{Vafa:1996xn}.) Here, we will simply regard \eqref{eqn:D5bc45678}--\eqref{eqn:D5bcfpsi} as formal boundary conditions that we impose to get rid of unwanted zero modes.

To write down the boundary conditions equivalent to \eqref{eqn:D5bcfpsi} in the type IIA picture, we first decompose the fermionic zero mode $\fpsi_\iStIIB$ on the type-IIB fundamental string into left-moving and right-moving components as $\fpsi_\iStIIB=\fpsi_{\iStIIB+}+\fpsi_{\iStIIB-}$, where
\be
\fpsi_{\iStIIB\pm}=\pm\Gamma^{09}\fpsi_{\iStIIB\pm}=\pm\Gamma^{12345678}\fpsi_{\iStIIB\pm}\,.
\ee
The boundary condition \eqref{eqn:D5bcfpsi} can now be written as a relation between the left-moving and right-moving modes at the end of the string ($x_9=\pm\DXiv$):
\be
\fpsi_{\iStIIB+}=\Gamma^{045678}\fpsi_{\iStIIB-}\,.
\ee
We now follow the dualities of \tabref{tab:Dualities}, each time transforming the fermionic field to a dual field on a dual brane.  %The transformation rules easily follow from the requirements of chirality of the corresponding fields.
%Thus, after T-duality in the $x_1$ direction we change variables to $\fpsi_{\iStIIB-}'\equiv\Gamma^{1\tenx}\fpsi_{\iStIIB-}$ and $\fpsi_{\iStIIB+}'\equiv\fpsi_{\iStIIB+}$, and get the boundary condition for the fermionic fields $\fpsi_{\iStIIB\pm}'$ on the type-IIA string:
%\be\label{eqn:bcf1d2}
%\fpsi_{\iStIIB+}'=-\Gamma^{0145678}\fpsi_{\iStIIB-}'\,,
%\ee
%where we used $\Gamma^{\tenx}\fpsi_{\iStIIB-}'=\fpsi_{\iStIIB-}'$ in type-IIA theory in our convention. (As explained in \appref{appsub:F1D2}, the gluino field on the D2 brane satisfies $\Gamma^{09\tenx} \fpsi_{\iStIIB-}' = - \fpsi_{\iStIIB-}'$.) Lifting to M-theory, we substitute
%$$
%\fpsi_{\iStIIB\pm}'=\tfrac12(1\pm\Gamma^{09})\lambda_{\iDb\iStz}
%=\tfrac12(1\mp\Gamma^{\tenx})\lambda_{\iDb\iStz}\,,
%$$
%where $\lambda_{\iDb\iStz}$ for $\iStz=0,n$ are the fermionic zero modes of the M2-branes. Here, the choice of indices of $\lambda_{\iDb\iStz}$ is that for $j=1,\ldots,m$ we take $\iDb=1,\ldots,m$ with $\iStz=n$, while for $\iStIIB=m+1,\ldots,2m$, we take $\iDb=1,\ldots,m$ with $\iStz=0$. After dimensional reduction on the $x_2$ direction, $\lambda_{a\iStz}$ become the fermionic zero modes of D$2$-branes that we discussed in \secref{subsec:ZeroModes}, and the boundary condition \eqref{eqn:bcf1d2} becomes
The type-IIA boundary conditions then become:
\be\label{eqn:bcotherend}
0=(1-\Gamma^{\tenx})
(1 +\Gamma^{0145678})
\lambda_{\iDb\iStz}
\ee
for $\iStz=0,n$.
We can now define the Witten index $I=\mathrm{Tr}(-1)^F$ of the quantum mechanical system of D$3$-branes and open strings in the type IIB picture, and we can calculate it using the system of D$2$-branes and F$1$-strings in the dual type-IIA picture. With boundary conditions \eqref{eqn:D5bc45678}--\eqref{eqn:D5bcfpsi}, the semi-infinite open strings can be regarded as external charges with no internal dynamics, so the Witten index $I$ will simply count the number of ground states of the theory with external charges inserted.

On the other hand, in the type IIA picture, we have various configurations of D2-branes and F1-strings, which can be classified into sectors described by the binding matrix $\PMat$ and permutation $\perm.$ With the extra conditions \eqref{eqn:bcotherend} we will show (see \secref{subsec:fzeromodesn=1} and \secref{subsec:countfm=1}) that only decongested sectors do not have fermionic zero modes. Such sectors will make a nonzero contribution to the Witten Index.

We also remark that it is possible to create open strings in a supersymmetric configuration by using D$3$-branes instead of D$5$-branes. This will avoid the problem of incompleteness of the background, but will instead create additional fermionic zero-modes to make the Witten Index identically zero. We briefly discuss this construction in \appref{app:tiltedD3}.

% --------------------------------------------------------------
\subsection{Summary of the rules}
\label{subsec:subsummary}

A sector of Tr-S theory with $U(n)$ gauge group in the presence of $m$ external quark and anti-quark pairs is described by a permutation $\perm\in S_n$ and an $n\times m$ binding matrix $\PMat_{\iSt\iDb}$ of $0$'s and $1$'s.
We build the action in terms of periodic variables  $\xvar^{\iSt\iDbz},\yvar^{\iSt\iDbz}$, $\xvar^\iDb$, and $\pvar^\iDb$, with $\iSt=1,\dots,n$, $\iDb=1,\dots,m$, and $\iDbz=0,\dots,m.$ All variables have $2\pi$ periodicity.
The variables are further restricted by linear relations with integer coefficients:
\be\label{eqn:xyc}
\xvar^{\iSt\iDb}=\xvar^{\iSt(\iDb-1)},\quad
\yvar^{\iSt\iDb}=\yvar^{\iSt(\iDb-1)},\qquad
\text{whenever $\PMat_{\iSt\iDb}=0$},
\ee
\be\label{eqn:xconstii}
\xvar^\iDb = \xvar^{\iSt\iDb} = \xvar^{\iSt(\iDb-1)}\,,
\qquad
\text{whenever $\PMat_{\iSt\iDb}=1$,}
\ee
and
\be\label{eqn:SRpZii}
\yvar^{\perm(\iSt)0} = \xd\yvar^{\iSt m}+\xb\xvar^{\iSt m}
\,,\qquad
\xvar^{\perm(\iSt)0} = \xc\yvar^{\iSt m}+\xa\xvar^{\iSt m}
\,.
\ee
The action is
\be\label{eqn:I0}
I_0 =\frac{1}{2\pi}
\int dt \sum_{\iSt=1}^n\sum_{\iDb=1}^m\PMat_{\iSt\iDb}\pvar^\iDb(\dot{\yvar}^{\iSt\iDb}-\dot{\yvar}^{\iSt(\iDb-1)})
\,.
\ee
Furthermore, the coordinates $(\scX_{\iDb}^{1},\scX_{\iDb}^{2})$ of the $m$ quarks and the coordinates $(\scX_{\iDb+m}^{1},\scX_{\iDb+m}^{2})$ of the $m$ anti-quarks are encoded in the action via an extra term
\be\label{eqn:I1ii}
I_1 = \frac{1}{2\pi}\int dt \sum_{\iDb=1}^m \lbrack
(\dot{\scX}_{\iDb}^{1}-\dot{\scX}_{\iDb+m}^{1})\xvar^\iDb
+(\dot{\scX}_{\iDb}^{2}-\dot{\scX}_{\iDb+m}^{2})\pvar^\iDb
\rbrack
\,.
\ee
Finally, we only keep those sectors for which $\sum_\iSt\PMat_{\iSt\iDb}=1$ for all $\iDb=1,\dots,m$ (we called these {\it decongested} sectors), because sectors for which $\sum_\iSt\PMat_{\iSt\iDb}>1$ for some $\iDb$ have zero modes and therefore do not contribute to the Witten Index, while sectors for which $\sum_\iSt\PMat_{\iSt\iDb}=0$ for some $\iDb$ have a D$2$-brane disconnected from the rest of the system.

% ==============================================================
\section{$U(1)$ gauge group}
\label{sec:U(1)}

We will now apply the rules of \secref{subsec:subsummary} to study our setup with $U(1)$ gauge group. We already know that abelian Tr-S is equivalent to $U(1)$ Chern--Simons theory at level $\lvk$ \cite{Ganor:2010md}. This model will therefore provide us with a good example of how the rules of \secref{subsec:subsummary} work. In \secref{subsec:m=1}-\secref{subsec:fzeromodesn=1} we apply the rules of \secref{subsec:subsummary} to $U(1)$ Tr-S with $m$ charge pairs, and in \secref{subsec:CSabelian} we show how they reproduce the predictions from Chern--Simons theory.

%% --------------------------------------------------------------
%\subsection{Geometric quantization of $T^{2m}$}
%\label{subsec:T2m}

In solving the problem, a central role is played by the action \eqref{eqn:pqIntro}. The general actions that we will consider are equivalent to quantum mechanical systems that are obtained by geometric quantization of $T^{2m}.$ They are of the form:
\be\label{eqn:Mxpqgq}
I = \frac{1}{2\pi}\int \Matx_{\iDb\alpha}\pvar^\iDb d\qvar^\alpha
\,,\qquad
\iDb,\alpha=1,\dots,m.
\ee
Here $\pvar^\iDb,\qvar^\alpha$ are periodic coordinates
parameterizing $T^{2m}$, in the range $[0,2\pi)$,
and $\Matx_{\iDb\alpha}$ are the components of a nonsingular
$m\times m$ matrix of integers.
We denote its determinant by
$$
\DetM\equiv\det_{m\times m} (\Matx_{\iDb\alpha})\neq 0.
$$
The dimension of the Hilbert space is then $|\DetM|.$
We denote the inverse matrix by $(\Matx^{-1})^{\alpha\iDb}$,
i.e., $\Matx_{\iDb\alpha}(\Matx^{-1})^{\alpha\jDb}=\delta_\iDb^\jDb.$
Suppose that $(k_1,\dots,k_{m})$ and $(l_1,\dots, l_{m})$ are vectors of integers. Then the operators $\exp(i\sum_\iDb k_\iDb\pvar^{\iDb})$ and $\exp(i\sum_\alpha l_\alpha\qvar^{\alpha})$
are well-defined, and we have the commutation relation
\be\label{eqn:CommRelpq}
e^{i\sum_\iDb k_\iDb\pvar^{\iDb}}
e^{i\sum_\alpha l_\alpha\qvar^{\alpha}}
e^{-i\sum_\iDb k_\iDb\pvar^{\iDb}}
e^{-i\sum_\alpha l_\alpha\qvar^{\alpha}}
=e^{2\pi i \sum_{\alpha,\iDb}(\Matx^{-1})^{\alpha \iDb}k_\iDb l_\alpha}
\,.
\ee
%In particular, if $(k_1,\dots,k_{m})$ is a vector
%of integers such that $\sum_\iDb(\Matx^{-1})^{\alpha\iDb}k_\iDb$
%is an integer for every $\alpha=1,\dots,m$, then
%$\exp(i\sum_\iDb k_\iDb\pvar^\iDb)$ is a central element
%of the algebra of operators, and we can set it equal to $1.$
%In general,
%$\sum_\iDb(\Matx^{-1})^{\alpha\iDb}k_\iDb$ will be a rational number with a minimal denominator that is a divisor of $\DetM.$
%If $\sum_\iDb(\Matx^{-1})^{\alpha\iDb}k_\iDb$  for $\alpha=1,\dots,m$ are not all integers, then $\tr\{\exp(i\sum_\iDb k_\iDb\pvar^\iDb)\}=0$, where the trace is on the entire $|\DetM|$-dimensional
%Hilbert space.
%
%As a special case, we will have the system given by [$\lvk=2$ case of \eqref{eqn:I0(n=1)(m)}]
%\be
%I=2\pi\int\left[ \pvar^{1} (\dot{\qvar}^{1}
%+\dot{\qvar}^{m}) + \sum_{\iDb=2}^{m} \pvar^\iDb (\dot{\qvar}^\iDb
%-\dot{\qvar}^{\iDb-1}) \right] dt \,.
%\ee
%In this case, $|\DetM|=2$ and
%\be\label{eqn:trexpp}
%\tr(e^{i\sum_{\iDb=1}^{m}n_\iDb\pvar^\iDb})=
%\begin{cases}
%2 & \text{if $\sum_\iDb n_\iDb$ is even,} \\
%0 & \text{if $\sum_\iDb n_\iDb$ is odd.} \\
%\end{cases}
%\ee

% --------------------------------------------------------------
\subsection{A quark and anti-quark pair $(m=1)$}
\label{subsec:m=1}
Using the rules summarized in \secref{subsec:subsummary}, we can now write down explicitly the low-energy effective action for the type-IIA configuration that corresponds to inserting a quark and anti-quark pair in Tr-S theory. As we have explained in the previous section, in such a case we have one fundamental string wrapping the $x_3$ circle and attached to one D$2$-brane, forming a bound state with it. The closed string that we would have in the absence of external charges breaks to become an open string starting
and ending on the D$2$-brane.
In the notations explained in \secref{subsec:subsummary}, for fixed charges we have the action
\be
\label{eqn:(m=1)I0}
I_0=\frac{1}{2\pi}\int \pvar^1 \left( d\yvar^{11} - d\yvar^{10} \right),
\ee
and if we wish to allow the charges to move around we need to add the term
\be\label{eqn:(m=1)I1}
I_1 =  - \frac{1}{2\pi}\int
\lbrack \left(\scX_{1}^{1} - \scX_{2}^{1} \right)d\xvar^1 +
\left(\scX_{1}^{2} - \scX_{2}^{2} \right)d\pvar^1 \rbrack
\,.
\ee
To proceed, we recall that the S-R-twisted boundary conditions induce linear relations among the variables, following \eqref{eqn:SRpZii}. For $\lvk=2$ this yields $\xvar^1=\yvar^{11}=-\yvar^{10}\equiv \qvar^1$, and thus \eqref{eqn:(m=1)I0}-\eqref{eqn:(m=1)I1} can be simplified to:
\bear
I &\equiv& I_0 + I_1\,\label{eqn:(m=1)(k=2)}\\
I_0 &=& \frac{1}{2\pi} \int 2\pvar^1 d\qvar^1
\,,\label{eqn:(m=1)(k=2)I0}\\
I_1 &=& - \frac{1}{2\pi} \int\Bigl\{
\bigl\lbrack a^{(1)}_1(t)-a^{(2)}_1(t) \bigr\rbrack d\qvar^1
+\bigl\lbrack  a^{(1)}_2(t)-a^{(2)}_2(t) \bigr\rbrack  d\pvar^1
\Bigr\}
\,,\label{eqn:(m=1)(k=2)I1}
\eear
where, as defined in \secref{sec:Intro}, $\bigl(a^{(j)}_1, a^{(j)}_2\bigr)=(X^1_{j},X^2_{j})$ refer to the $T^2$ coordinates of the quark's and anti-quark's worldlines.

After taking into account similarly the constraints in \eqref{eqn:SRpZ} and \eqref{eqn:xconst} for the cases $\lvk=1,3$, the action for static charges \eqref{eqn:(m=1)I0} can be written collectively for all three values of $\lvk$ as\footnote{For $\lvk=1$, $\yvar^{11}=0$ and $\xvar^1=-\yvar^{10}$, whereas for $\lvk=3$, $-\yvar^{10}=2\yvar^{11}=2\xvar^1$.}
\be
\label{eqn:(m=1)general}
I_0=\frac{\lvk}{2\pi} \int \pvar d\qvar\,.
\ee
This action is of the form \eqref{eqn:Mxpqgq} and describes geometric quantization of $T^2.$ It gives rise to a $\lvk$-dimensional Hilbert space.

We also note that the action \eqref{eqn:(m=1)(k=2)} is of the same form as the action for the ground states of the well-known Landau problem describing the low-energy spectrum of a two-dimensional charged particle moving on a torus with $\lvk$ units of magnetic flux. In this context, the velocity of the quark relative to the anti-quark, which is given by
$$
(\dot{a}^{(1)}_1(t)-\dot{a}^{(2)}_1(t),\,
\dot{a}^{(1)}_2(t)-\dot{a}^{(2)}_2(t))
$$
is interpreted as the electric field in the Landau problem. This is consistent with the interpretation of the quark and anti-quark as charges in $U(1)$ Chern--Simons theory (see \secref{subsec:CSabelian} below).
%The Landau problem is briefly reviewed in \secref{app:pqBerry}, where we show that the Hilbert space of \eqref{eqn:(m=1)(k=2)} is two-dimensional.
To simplify matters, in the following, we consider static charges and let the quarks and anti-quarks be located at the origin.

% --------------------------------------------------------------
\subsection{Multiple quark and anti-quark pairs $(m>1)$}
\label{subsec:m>1}

Generalization of the results of \secref{subsec:m=1} to $m>1$ cases is straightforward. There is only one sector: the permutation $\perm\in S_1$ is the identity, and the $1\times m$ binding matrix $\PMat$ is
$$
\PMat = \begin{pmatrix} 1 & 1 & \cdots & 1 \end{pmatrix}.
$$
The relations \eqref{eqn:xconstii} imply that all $\xvar^{\iSt\iDb}$ variables are equal:
$$
\xvar^{10}=\xvar^{11}=\xvar^{12}=\cdots=\xvar^{1m}=
\xvar^1=\cdots=\xvar^m\equiv\xvar.
$$
Just like what we have done in \secref{subsec:m=1}, we can write down the low energy effective action as
\be\label{eqn:m>1g}
\begin{split}
I = \frac{1}{2\pi} \int & \sum_{\iDb=1}^m
\pvar^\iDb \left( d \yvar^{1\iDb} - \yvar^{1(\iDb-1)} \right) \\
&-\frac{1}{2\pi} \int
\sum_{\iDb=1}^m \left[\left( a^{(\iDb)}_1(t)-a^{(\iDb+m)}_1(t) \right) d\xvar - \left( a^{(\iDb)}_2(t)-a^{(\iDb+m)}_2(t) \right) d\pvar^\iDb\right]\,.
\end{split}
\ee
Furthermore, \eqref{eqn:SRpZii} gives linear constraints among $\xvar, \yvar^{10}, \yvar^{1m}$ which depend on $\lvk$:
\be\label{eqn:SRpZiiU1}
\yvar^{10} = \xd\yvar^{1m}+\xb\xvar
\,,\qquad
\xvar = \xc\yvar^{1m}+\xa\xvar
\,.
\ee
This can be solved using the explicit expressions for $\xa, \xb, \xc, \xd$ in terms of $\lvk$, given in \eqref{eqn:deflvk}, and we get
\be
\yvar^{10}=-\xvar\,,\quad
\yvar^{1m}=0\,,\qquad\text{for $\lvk=1$,}
\ee
and
\be\label{eqn:solxy}
\yvar^{10} = (1-\lvk)\xvar
\,,\quad
\yvar^{1m} = \xvar
\,,\qquad
\text{for $\lvk=2,3$.}
\ee
After taking into account these linear constraints, the action for fixed external charges becomes
\be
I_0=\frac{1}{2\pi}\int\bigl\{
\pvar^{1}(d\qvar^{1} + d\qvar^{m})
+\sum_{\iDb=2}^{m-1} \pvar^\iDb (d \qvar^\iDb - d\qvar^{\iDb-1})
-\pvar^{m}d\qvar^{m-1}\bigr\}\,,
\qquad\text{for $\lvk=1$},
\ee
where we set $\qvar^\iDb=\yvar^{1\iDb}$ for $\iDb=1,\ldots,m-1$ and $\qvar^m=\xvar$, and
\be\label{eqn:I0(n=1)(m)}
I_0 = \frac{1}{2\pi}\int\bigl\{
\pvar^{1}[d\qvar^{1} + (\lvk-1)d\qvar^{m}] +
\sum_{\iDb=2}^{m} \pvar^\iDb (d \qvar^\iDb - d\qvar^{\iDb-1})
\bigr\}
\,,\qquad
\text{for $\lvk=2,3$,}
\ee
where we set $\qvar^\iDb\equiv\yvar^{1\iDb}.$
%\be
%\label{eqn:m>1k=1}
%I_0=\begin{
%\frac{1}{2\pi} \int \sum_{\iDb=1}^{m-1} \pvar^\iDb \left( d \qvar^\iDb - d\qvar^{\iDb-1} \right) - \pvar^{m}d\qvar^{m-1}\,,
%\ee
%\be
%\label{eqn:m>1k=2}
%I_{k=2} = \frac{1}{2\pi} \int \sum_{\iDb=2}^{m} \pvar^\iDb \left( d \qvar^\iDb - d\qvar^{\iDb-1} \right) + \pvar^{1} \left( d\qvar^{1} + d\qvar^{m} \right)\,,
%\ee
%\be
%\label{eqn:m>1k=3}
%I_{k=3} = \frac{1}{2\pi} \int \sum_{\iDb=2}^{m} \pvar^\iDb \left( d \qvar^\iDb - d\qvar^{\iDb-1} \right) + \pvar^{1} \left( d\qvar^{1}+2d\qvar^{m} \right)\,.
%\ee
The action \eqref{eqn:I0(n=1)(m)} is of the form \eqref{eqn:Mxpqgq} with
\be\label{eqn:Matx(m>1)(n=1)}
\Matx=\begin{pmatrix}
1  &  0 &  0 &  0 & \cdots & 0 & \lvk-1 \\
-1 &  1 &  0 &  0 & \cdots & 0 & 0 \\
0  & -1 &  1 &  0 & \cdots & 0 & 0 \\
\vdots  & \ddots  &  \ddots & \vdots &  \vdots \\
0  &  0 &  0 &  0 &  \cdots & -1 & 1 \\
\end{pmatrix}\,,
\ee
and it is also straightforward to write down the corresponding matrix $\Matx$ for $\lvk=1$ case.
Following the discussion below \eqref{eqn:Mxpqgq}, we can immediately compute the dimension of each Hilbert space as $|\DetM| = \lvk$.
%(This is very easy to compute by successively adding the $i^{th}$ row to the $(i+1)^{st}$ row starting from the first row.)

% --------------------------------------------------------------
\subsection{Absence of fermionic zero modes}
\label{subsec:fzeromodesn=1}
In this subsection, we show that there is no fermionic zero mode in the sector discussed in \secref{subsec:m>1}. This is necessary for the consistency of our result, because otherwise the contribution of the sector to the Witten index would be zero, and since this is the only sector for $n=1$ cases, this would mean that the Witten index of $U(1)$ Tr-S theory would be zero too, regardless of the value of $\lvk$. On the other hand, we know that $U(1)$ Tr-S theory is simply $U(1)$ Chern--Simons theory, which contains $\lvk$ bosonic ground states only.

\begin{figure}[t]
\begin{picture}(400,120)

\put(10,10){\begin{picture}(390,100)
\thicklines
\color{black}
\put(0,50){\line(1,0){200}}

% twist
\color{red}
\put(-5,45){\line(1,1){10}}
\put(5,45){\line(-1,1){10}}
\color{red}
\put(195,45){\line(1,1){10}}
\put(205,45){\line(-1,1){10}}

% D2
\color{black}
\put(50,0){\line(0,1){100}}
\put(150,0){\line(0,1){100}}

% D2-F1 intersections
\color{red}
\put(50,50){\circle*{6}}
\put(150,50){\circle*{6}}

% NS5
\thinlines
\color{black}
\put(40,0){\line(1,0){20}}
\put(40,100){\line(1,0){20}}
\put(140,0){\line(1,0){20}}
\put(140,100){\line(1,0){20}}

% NS5 intersections
%\color{red}
%\put(50,0){\circle*{6}}
%\put(150,0){\circle*{6}}
%\put(50,100){\circle*{6}}
%\put(150,100){\circle*{6}}

\color{blue}
\put(25,55){$\psi_{10}$}
\put(95,55){$\psi_{11}$}
\put(170,55){$\psi_{12}$}

\put(52,25){$\lambda_{11}$}
\put(52,75){$\lambda_{10}$}
\put(152,25){$\lambda_{21}$}
\put(152,75){$\lambda_{20}$}

\color{black}
\put(70,40){F$1$-string}
\put(18,90){D$2$}
\put(27,88){\vector(1,-1){20}}
\put(118,90){D$2$}
\put(127,88){\vector(1,-1){20}}
%\put(62,-2){NS5}
%\put(62,98){NS5}
%\put(162,-2){NS5}
%\put(162,98){NS5}
\put(192,58){SR}

\end{picture}}
\end{picture}
\caption{Illustration of fermionic zero modes for $n=1$ and $m=2$ case. The fundamental string breaks at every D$2$-brane intersection.
}
\label{fig:ConfigZeroModes1}
\end{figure}
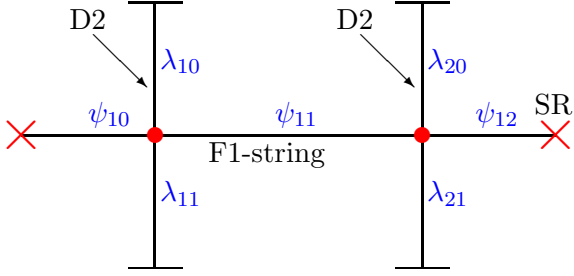

As explained in \secref{subsec:ZeroModes}, the low-energy fermionic modes of our system can be understood via 1+1D fermionic fields $\fpsi_{\iSt\iDbz}$ and $\lambda_{\iDb\iStz}$ supported on the open strings and D$2$-branes. Each $\fpsi_{\iSt\iDbz}$ and $\lambda_{\iDb\iStz}$ satisfies the chirality conditions \eqref{eqn:F1ch} and \eqref{eqn:D2ch}, respectively. In addition, the $\lambda_{\iDb\iStz}$'s at the two far ends of the D$2$-branes (that is, those with $\iStz=0,n$) satisfy the constraint \eqref{eqn:bcotherend}
%\be
%\label{eqn:bcotherend}
%(1-\Gamma^{\tenx})(1+\Gamma^{0145678})\lambda_{\iDb\iStz} = 0\,,
%\ee
which derives from dualizing the boundary conditions for open strings ending on D$5$-branes in the type-IIB setting.
Each intersection of D$2$-F$1$ satisfies the two junction conditions \eqref{eqn:ZMxJp} and \eqref{eqn:ZMxJm}.
Finally, the S-R-twist on the F1-string gives rise to the boundary condition \eqref{eqn:ZMxSRm}.
We now proceed to prove that in the $n=1$ abelian case, these various boundary conditions dictate that we have no fermionic zero modes. (See \figref{fig:ConfigZeroModes1} for notation.)

As a warm-up, we start with the $m=0$ case. In this case there is only one variable $\fpsi_{10}$ which satisfies the boundary conditions \eqref{eqn:ZMxSRm} and the chirality condition \eqref{eqn:F1ch}:
\be\label{eqn:ZM(m=0)}
\fpsi_{10}=\mathcal{P}\fpsi_{10}
\,,\qquad
(\Gamma^{023}+1)\fpsi_{10}=0.
\ee
But in \secref{subsec:evP} we saw that all the eigenvectors of $\mathcal{P}$ with eigenvalue $1$ have the opposite $\Gamma^{023}$ chirality from that of $\fpsi_{10}.$ It follows that\eqref{eqn:ZM(m=0)} does not have any non-trivial solutions and the $m=0$ states have no zero modes.

Now, let us study the $m>0$ case. We first consider the continuity of $\lambda_{\iDb\iStz}$ at each D$2$-F$1$ junction. Define
$$\zeta_\iDb \equiv \lambda_{\iDb 1}-\lambda_{\iDb 0}$$
for $\iDb=1,\ldots,m$.
Since in the abelian case ($n=1$) both $\lambda_{\iDb 0}$ and $\lambda_{\iDb 1}$ satisfy the chirality condition \eqref{eqn:D2ch} and the boundary condition \eqref{eqn:bcotherend}, $\zeta_\iDb$ must also satisfy the equations
\bear
0 &=&
(1+\Gamma^{09\tenx})\zeta_\iDb
\,,
\label{eqn:zetaA}\\
0 &=&
(1-\Gamma^{\tenx})
(1+\Gamma^{0145678})
\zeta_\iDb
\,,
\label{eqn:zetaB}\\
0 &=&
(1-\Gamma^{239\tenx})\zeta_\iDb
\,,
\label{eqn:zetaC}
\eear
where the last equation comes from \eqref{eqn:ZMxJp}.
Equations \eqref{eqn:zetaA} and \eqref{eqn:zetaC} together imply
\be\label{eqn:023z}
\zeta_\iDb = \Gamma^{023}\zeta_\iDb
\,.
\ee
Since in our convention $\Gamma^{0123456789\tenx} = 1$ (see \appref{app:SpinorsSUSY}), this in turn implies
\be
\Gamma^{0145678}\zeta_\iDb=\zeta_\iDb\,,
\ee
and \eqref{eqn:zetaB} becomes
$$
(1-\Gamma^{\tenx})\zeta_\iDb=0\,.
$$
Then \eqref{eqn:zetaC} now reads
\be\label{eqn:239z}
\Gamma^{239}\zeta_\iDb=\zeta_\iDb\,,
\ee
which has no non-trivial solution, because $\Gamma^{239}$ has no real eigenvalues (it squares to $-1$). Therefore, we obtain $\zeta_\iDb=0$, or $\lambda_{\iDb 0}=\lambda_{\iDb 1}$ for all $\iDb=1,\ldots,m$.

Next, we consider the $\fpsi_{1\iDbz}$ fields. Since $\zeta_\iDb=0$, equations \eqref{eqn:ZMxJp}--\eqref{eqn:ZMxJm} now become
$$
(1\pm\Gamma^{239\tenx})(\fpsi_{1,\iDb-1}-\fpsi_{1\iDb})=0\,,
$$
or simply
\be\label{eqn:fpsiallequal}
\fpsi_{10}=\fpsi_{11}=\cdots=\fpsi_{1m}\,.
\ee
On the other hand, we have from the S-R-twist condition $\fpsi_{1m}=\mathcal{P}\fpsi_{10}$, where the operator $\mathcal{P}$ is defined in \eqref{eqn:OmegaDef}, and together with the above equalities,
\be
\label{eqn:twistn=1}
\fpsi_{10}=\mathcal{P}\fpsi_{10}\,.
\ee
At this point, we can again use the fact that all the eigenvectors of $\mathcal{P}$ with eigenvalue $1$ have the opposite $\Gamma^{023}$ chirality from that of $\fpsi_{10}$. Therefore, \eqref{eqn:twistn=1} does not have a non-trivial solution, and hence
\be\label{eqn:fpsiallzero}
\fpsi_{10}=\fpsi_{11}=\cdots=\fpsi_{1m}=0\,.
\ee

Finally, let us define
$$
\xi_\iDb=\lambda_{\iDb 0}+\lambda_{\iDb 1}\,.
$$
With $\zeta_\iDb=\fpsi_{1\iDbz}=0$, we find that $\xi_\iDb$ satisfies the same set of equations \eqref{eqn:zetaA}-\eqref{eqn:zetaC} as $\zeta_\iDb$:
\bear
0 &=&
(1+\Gamma^{09\tenx})\xi_\iDb
\,,
\label{eqn:xiA}\\
0 &=&
(1-\Gamma^{\tenx})
(1 +\Gamma^{0145678})
\xi_\iDb
\,,
\label{eqn:xiB}\\
0 &=&
(1-\Gamma^{239\tenx})\xi_\iDb
\,.
\label{eqn:xiC}
\eear
The first two equations follow from the chirality and boundary conditions for $\lambda_{\iDb\iStz}$, while the third from \eqref{eqn:ZMxJp} and the previous result $\fpsi_{1\iDbz}=0$ for all $\iDbz$. Therefore, we find $\xi_\iDb=0$ as before. Together with $\zeta_\iDb=0$, this implies $\lambda_{\iDb\iStz}=0$ for all $\iDb=1,\ldots,m$ and $\iStz=0,1$.
To summarize, we conclude that there is no fermionic zero mode for the abelian $n=1$ case with an arbitrary number $m$ of D$2$-branes.

% ----------------------------------------------------------------
\subsection{Comparison with Chern--Simons theory results}
\label{subsec:CSabelian}
At this point, it is pertinent to discuss consistency with abelian Chern--Simons theory, which we had explained in \cite{Ganor:2010md} to be the low-energy limit of abelian Tr-S theory. For $U(1)$ Chern--Simons theory the dimension of the Hilbert space with $m$ external charge pairs is always equal to the level $\lvk$, independently of $m.$ This is indeed what we found from the type-IIA dual picture in \secref{subsec:m>1}. There, the dimension $\dim\Hilb(\lvk,n=1,m)$ can be calculated as the determinant of the matrix $\Matx$ that appears in \eqref{eqn:Matx(m>1)(n=1)} and its $\lvk=1$ counterpart, and we indeed find the result $\dim\Hilb(\lvk,n=1,m)=\lvk$ independently of $m.$ The underlying reason for this coincidence is that since there are no fermionic zero modes, as we have proved in \secref{subsec:fzeromodesn=1}, the dimension we calculated is in fact the Witten Index of the system.

To go beyond the mere equality of dimensions of the Hilbert spaces, we can consider for $\lvk>1$ the action of the $\Z_\lvk$ symmetry operators $\Psym,\Qsym$ discussed in \secref{subsec:ZkSym}. At the classical level, the discrete translation $\Psym$ acts as
\be\label{eqn:PsymOnpq}
\Psym:\qquad
\xvar^{\iSt\iDbz}\rightarrow\xvar^{\iSt\iDbz}+\frac{2\pi}{\lvk}
\,,\quad
\yvar^{\iSt\iDbz}\rightarrow\yvar^{\iSt\iDbz}+\frac{2\pi}{\lvk}
\,,\quad
\xvar^\iDb\rightarrow\xvar^\iDb+\frac{2\pi}{\lvk}
\,,\quad
\pvar^\iDb\rightarrow\pvar^\iDb
\,,
\ee
while $\Qsym$, which is related to the homology class of the fundamental string, can be interpreted as electric flux on the D$2$-branes and acts as
\be\label{eqn:QsymOnpq}
\Qsym:\qquad
\xvar^{\iSt\iDbz}\rightarrow\xvar^{\iSt\iDbz}
\,,\quad
\yvar^{\iSt\iDbz}\rightarrow\yvar^{\iSt\iDbz}
\,,\quad
\xvar^\iDb\rightarrow\xvar^\iDb
\,,\quad
\pvar^\iDb\rightarrow\pvar^\iDb+\frac{2\pi}{\lvk}
\,.
\ee
After geometric quantization, the actions \eqref{eqn:PsymOnpq}-\eqref{eqn:QsymOnpq} translate to actions on quantum operators of the system. The action is by conjugation; for example the rightmost expression of \eqref{eqn:QsymOnpq} is to be read as $\Qsym^{-1}\pvar^\iDb\Qsym=\pvar^\iDb+\frac{2\pi}{\lvk}.$

For $n=1$ we find in terms of the variables of \eqref{eqn:I0(n=1)(m)}
\be\label{eqn:PQsymOnpq}
\Qsym^{-1}\pvar^\iDb\Qsym = \pvar^\iDb+\frac{2\pi}{\lvk}
\,,\qquad
\Qsym^{-1}\qvar^\iDb\Qsym = \qvar^\iDb
\,,\qquad
\Psym^{-1}\pvar^\iDb\Psym = \pvar^\iDb
\,,\qquad
\Psym^{-1}\qvar^\iDb\Psym = \qvar^\iDb+\frac{2\pi}{\lvk}
\,,
\ee
for $\iDb=1,\dots,m$.
Using the commutation relation \eqref{eqn:CommRelpq}, together with the inverse of \eqref{eqn:Matx(m>1)(n=1)}:
\be\label{eqn:InvMatx}
\Matx^{-1}=\begin{pmatrix}
\frac{1}{\lvk}  &  \frac{1}{\lvk}-1 &  \frac{1}{\lvk}-1 &  \frac{1}{\lvk}-1 & \cdots & \frac{1}{\lvk}-1 & \frac{1}{\lvk}-1 \\
\frac{1}{\lvk}  &  \frac{1}{\lvk} &  \frac{1}{\lvk}-1 &  \frac{1}{\lvk}-1 & \cdots & \frac{1}{\lvk}-1 & \frac{1}{\lvk}-1 \\
\frac{1}{\lvk}  &  \frac{1}{\lvk} &  \frac{1}{\lvk} &  \frac{1}{\lvk}-1 & \cdots & \frac{1}{\lvk}-1 & \frac{1}{\lvk}-1 \\
\vdots  & \ddots  &  \ddots & \vdots &  \vdots \\
\frac{1}{\lvk}  &  \frac{1}{\lvk} &  \frac{1}{\lvk} &  \frac{1}{\lvk} & \cdots & \frac{1}{\lvk} & \frac{1}{\lvk} \\
\end{pmatrix}\,,
\ee
we find that, up to a possible constant phase, we can identify
\be\label{eqn:PQsymExppq}
\Psym= e^{-i\pvar^1}
\,,\qquad
\Qsym= e^{i\qvar^m}
\,.
\ee

Now, let us consider the effect of the interaction $I_1$ of \eqref{eqn:I1}. For charge positions $a^{(\iStIIB)}_1,a^{(\iStIIB)}_2$ that are independent of time, $I_1$ contributes a total derivative term in \eqref{eqn:m>1g}:
$$
-\frac{1}{2\pi}\sum_{\iDb=1}^m\left(a^{(\iDb)}_1-a^{(\iDb+m)}_1 \right)
\int d\qvar^m
-\frac{1}{2\pi}\int
\sum_{\iDb=1}^m\left(a^{(\iDb)}_2-a^{(\iDb+m)}_2\right) d\pvar^\iDb
\,,
$$
where we used \eqref{eqn:solxy}.
We now claim that the effect of the interaction term $I_1$ is to modify \eqref{eqn:PQsymExppq} to
\be\label{eqn:PQsymExppqII}
\Psym= e^{-i\pvar^1
+\frac{i}{\lvk}\sum_{\iDb=1}^m\left(a^{(\iDb)}_1-a^{(m+\iDb)}_1 \right)}
\,,\qquad
\Qsym= e^{i\qvar^m
+\frac{i}{\lvk}\sum_{\iDb=1}^m\left(a^{(\iDb)}_2-a^{(m+\iDb)}_2 \right)}
\,.
\ee
This is not so obvious for static charge positions, and to see it we actually need to let the position, say $a^{(\iDb)}_1$, vary as a function of time. It can then be checked that an initial state $\ket{i}$ at time $t=-\infty$ evolves into
$$
\ket{f}=
e^{\frac{i}{2\pi}[a^{(\iDb)}_1(\infty)-a^{(\iDb)}_1(-\infty)]\qvar^m}\ket{i}
$$
at $t=\infty$. But in order for $\Psym\ket{i}$ to evolve into $\Psym\ket{f}$, the operator $\Psym$ must add a phase of $\frac{1}{\lvk}a^{(\iDb)}_1(-\infty)$ to $\ket{i}$ and a similar phase of $\frac{1}{\lvk}a^{(\iDb)}_1(\infty)$ to $\ket{f}$, since $\Psym^{-1}\qvar^m\Psym=\qvar^m+\frac{2\pi}{\lvk}$. A similar method can be used to derive the extra phase $\frac{1}{\lvk}\sum_{\iDb=1}^m\left(a^{(\iDb)}_2-a^{(m+\iDb)}_2 \right)$ in the action of $\Qsym.$

Now we can compare the above discussion with Chern--Simons theory.
It was argued in \cite{Ganor:2010md} that in terms of $U(1)$ Tr-S theory (namely Chern--Simons theory) defined on $T^2$ in the $x_1x_2$ directions, $\Psym$ and $\Qsym$ can be understood as gauge transformations with discontinuous gauge parameters:
\be\label{eqn:LambdaLambda}
\Lambda_\Psym = e^{\frac{i}{\lvk}x_1}
\,,\qquad
\Lambda_\Qsym = e^{\frac{i}{\lvk}x_2}
\,,
\ee
where the coordinates $x_1,x_2$ take values in $[0,2\pi)$.
(This was argued by relating $\Psym, \Qsym$ to momentum and winding number in type-IIA and then mapping these quantum numbers to electric fluxes on the D$3$-brane in type-IIB.) Equation \eqref{eqn:LambdaLambda} in conjunction with \eqref{eqn:PQsymExppqII} allows us to directly map states of Chern--Simons theory to states of the system we got from geometric quantization of $T^{2m}$ by, for example, mapping eigenstates of $\Psym$ in \eqref{eqn:LambdaLambda} to eigenstates of $\Psym$ in \eqref{eqn:PQsymExppqII}. Moreover, the extra $a_i^{(\iDb)}$-dependent phase that we got in \eqref{eqn:PQsymExppqII} has a natural interpretation in Chern--Simons theory. This is precisely the phase that we would expect to pick up when acting with a gauge transformation \eqref{eqn:LambdaLambda} on a system that contains $m$ positive charges at positions $(a^{(\iDb)}_1, a^{(\iDb)}_2)$ and $m$ negative charges at $(a^{(\iDb+m)}_1, a^{(\iDb+m)}_2)$, for $\iDb=1,\dots,m.$ This concludes our map from the Hilbert space of the geometric quantization system to the Hilbert space of $U(1)$ Chern--Simons theory.

% ==============================================================
\section{$U(n)$ gauge group}
\label{sec:U(n)}
We now turn to the non-abelian case with a $U(n)$ gauge group. Our goal is to calculate the Witten Index of Tr-S theory on $T^2$ as a function of $\lvk, n$ and the number of charge pairs $m.$ We will begin in \secref{subsec:sectorsU(2)} with a few examples of sectors for $n=2$, including a brief description of all its sectors with low $m$. We then show in \secref{subsec:countfm=1} that only decongested sectors contribute to the Witten Index. This greatly simplifies the computation, since decongested sectors are equivalent to a product of decoupled $U(1)$ Hilbert spaces. We describe the results in \secref{subsec:WittenIndexRes}. A reader who wishes to skip the details is advised to jump directly to \secref{subsec:WittenIndexRes}. In \secref{subsec:WilsonLoops} we test our results by rewriting the Witten Index as a trace of products of Wilson loop operators in Tr-S theory without charges ($m=0$). This provides us with a consistency check, and also allows us to calculate the eigenvalues of Wilson loop operators acting on the $m=0$ Hilbert space. \appref{app:combinatorics} includes some additional details of the combinatorics involved, and for curiosity, we included in \appref{app:GenFun} a combinatorical derivation of the total number of sectors. Interestingly, it is described by a Fibonacci sequence.

% -------------------------------------------------------------
\subsection{Examples of $U(2)$ sectors and states}
\label{subsec:sectorsU(2)}

As explained in \secref{subsec:DualOfCharges},
each sector corresponds to a different choice of the binding matrix $\PMat$ and a permutation $\perm \in S_n$ that accompanies the action of the S-R-twist on the $n$ strings. Since for $n=2$ the permutation group is $S_n\simeq\Z_2$, in the following we shall simply express the permutation as $\perm\in\{1,-1\}$, and write it as a subscript of $\PMat$, so a sector will be denoted as $\PMat_{\pm 1}$. Physically, $\perm=1$ implies that each string's endpoint is connected to its own starting point so that we have two strings, each with winding number $1$ along the $x_3$ circle, whereas $\perm=-1$ means that the string's endpoint is connected to the other string's starting point so that we end up with one string with winding number $2$. If the permutation results in an equivalent configuration, the subscript is omitted. This happens when at least two strings start on the same D$2$-brane (i.e., the binding matrix is congested), and relabeling the string indices results in equivalent sectors with different $\perm$'s (see for example the third configuration in \secref{subsubsec:n=2;m=1}). It is also useful to recall that our conventions are such that all strings begin at $x_3=0$, and the S-R-twist is located at $x_3=2\pi R$.

In this subsection, we will count the states of each sector following the rules of \secref{subsec:subsummary}, and show that in each case the low-energy action can be reduced to the form \eqref{eqn:pqIntro}.

\subsubsection{$m=1$}
\label{subsubsec:n=2;m=1}

%\left(\begin{array}{c} 1 \\  0  \end{array}\right)_{1},\left(\begin{array}{c} 1 \\  0  \end{array}\right)_{-1}.\left(\begin{array}{c} 1 \\  1  \end{array}\right).
We have one D$2$-brane and a string winding number of $n=2$. This yields three sectors described below [where $(\cdots)^\top$ denotes the transposed matrix]. The diagram for each sector is a miniature of \figref{fig:nmconfig}, with the S-R-twists colored to reflect $\perm,$\footnote{For configurations equivalent under $\perm$, we chose $\perm=1$ for our analysis.} while the black circles depict junctions where open strings attach to D2-branes.
\begin{enumerate}
\item
$\PMat = (1\,0)^\top_{1}$
\hfill
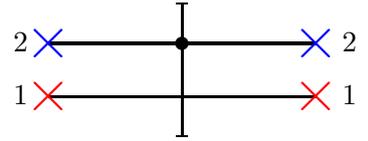
\begin{picture}(125,50)
%\put(0,18){1. Sector $\PMat= \left( \begin{array}{cc} 1 & 1 \\ 0 & 0  \end{array}\right)_{1}$.}
\put(15,0){\begin{picture}(100,50)
\thicklines
%F1
\color{black}
\put(0,15){\line(1,0){100}}
\put(0,35){\line(1,0){100}}
% D2
\color{black}
\put(50,0){\line(0,1){50}}
%SR
\color{red}
\put(-5,20){\line(1,-1){10}}
\put(-5.0,10){\line(1,1){10}}
\color{red}
\put(95,20){\line(1,-1){10}}
\put(95,10){\line(1,1){10}}
\color{blue}
\put(-5,40){\line(1,-1){10}}
\put(-5,30){\line(1,1){10}}
\color{blue}
\put(95,40){\line(1,-1){10}}
\put(95,30){\line(1,1){10}}
%Intersections
\color{black}
\put(50,35){\circle*{5}}
%\put(55,15){\circle*{5}}
%endpoints
\thinlines
\put(48,50){\line(1,0){4}}
\put(48,0){\line(1,0){4}}

\put(-13,12){$1$}
\put(-13,32){$2$}
\put(110,12){$1$}
\put(110,32){$2$}

\end{picture}}
\end{picture}

\noindent
One open string of winding number $1$ bound to the D2-brane and one closed string of winding number $1$.
We saw in \secref{subsec:m=1} that the open string plus the D2-brane system yields $\lvk$ states, while the closed string, being dual to the abelian Chern--Simons theory without charges, also gives rise to $\lvk$ states. In total, we get $\lvk\times \lvk = \lvk^2$ states.

\item
$\PMat = (1\,0)^\top_{-1}$
\hfill
\begin{picture}(125,50)
%\put(0,18){1. Sector $\PMat= \left( \begin{array}{cc} 1 & 1 \\ 0 & 0  \end{array}\right)_{1}$.}
\put(15,0){\begin{picture}(100,50)
\thicklines
%F1
\color{black}
\put(0,15){\line(1,0){100}}
\put(0,35){\line(1,0){100}}
% D2
\color{black}
\put(50,0){\line(0,1){50}}
%SR
\color{red}
\put(-5,20){\line(1,-1){10}}
\put(-5.0,10){\line(1,1){10}}
\color{blue}
\put(95,20){\line(1,-1){10}}
\put(95,10){\line(1,1){10}}
\color{blue}
\put(-5,40){\line(1,-1){10}}
\put(-5,30){\line(1,1){10}}
\color{red}
\put(95,40){\line(1,-1){10}}
\put(95,30){\line(1,1){10}}
%Intersections
\color{black}
\put(50,35){\circle*{5}}
%\put(55,15){\circle*{5}}
%endpoints
\thinlines
\put(48,50){\line(1,0){4}}
\put(48,0){\line(1,0){4}}

\put(-13,12){$1$}
\put(-13,32){$2$}
\put(110,12){$2$}
\put(110,32){$1$}

\end{picture}}
\end{picture}

\noindent
One open string of winding number $2$. Let the string start at $z= \qvar + i\xvar$ and end at $e^{2i\pht} (\qvar + i\xvar)$ (the phase is $2\pht$ because the string passes through the S-R-twist twice before ending on the D2-brane).
For $\lvk=2$, for which $\pht=\tfrac{\pi}{2}$, this means $\xvar=-\xvar$ $\mod 2\pi$, or $\xvar=0$ or $\pi$. For each choice of $\xvar$, the string starts at $x_{10}=\qvar$ on the D2-brane and end at $x_{10}=-\qvar$, so the effective action is
$$
I_{\lvk=2} = \frac{2}{2\pi}
\int \pvar d\qvar,\,\, \text{with $\xvar=0,$ or $\pi$}\,
$$
This gives us $2+2=4$ states. Note that in this case the parameter $\xvar$ is discrete and decoupled from the geometrically quantized $T^2$, which in turn is described only by $(\pvar,\qvar).$

For both $\lvk=1$ and $\lvk=3$, the phase $e^{2i\pht}$ is effectively what we had for $\lvk=3$ case in \secref{subsec:m=1}. Therefore, from similar analysis, the effective action is
$$
I_{\lvk=1,3} = \frac{3}{2\pi}\int \pvar d\qvar\,,
$$
which gives rise to 3 states each.

\item
$\PMat = (1\,1)^\top$
\hfill
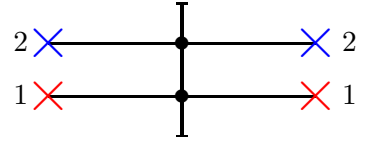
\begin{picture}(125,50)
%\put(0,18){1. Sector $\PMat= \left( \begin{array}{cc} 1 & 1 \\ 0 & 0  \end{array}\right)_{1}$.}
\put(15,0){\begin{picture}(100,50)
\thicklines
%F1
\color{black}
\put(0,15){\line(1,0){100}}
\put(0,35){\line(1,0){100}}
% D2
\color{black}
\put(50,0){\line(0,1){50}}
%SR
\color{red}
\put(-5,20){\line(1,-1){10}}
\put(-5.0,10){\line(1,1){10}}
\color{red}
\put(95,20){\line(1,-1){10}}
\put(95,10){\line(1,1){10}}
\color{blue}
\put(-5,40){\line(1,-1){10}}
\put(-5,30){\line(1,1){10}}
\color{blue}
\put(95,40){\line(1,-1){10}}
\put(95,30){\line(1,1){10}}
%Intersections
\color{black}
\put(50,35){\circle*{5}}
\put(50,15){\circle*{5}}
%endpoints
\thinlines
\put(48,50){\line(1,0){4}}
\put(48,0){\line(1,0){4}}

\put(-13,12){$1$}
\put(-13,32){$2$}
\put(110,12){$1$}
\put(110,32){$2$}

\end{picture}}
\end{picture}

\noindent
Here we have two open strings of winding number $1$ both of which bound to the D$2$-brane. For the D$2$-brane worldvolume gauge field this means that there are twice as many charged particles as in the sector $(1\,0)^\top_{1}$. Therefore, the bosonic part of the action is also twice that of the $(1\,0)^\top_{1}$ sector,  i.e.,
$$\frac{1}{2\pi}\int 2\lvk\pvar d\qvar.$$
This is a {\it congested} sector that also has $4$ real fermionic zero modes, as will be shown in \secref{subsubsec:fzmm=1}.
The bosonic part of the action has $4$ states, but the actual Hilbert space is more complicated because of the fermionic zero modes and because of possible interactions between the bosonic and fermionic modes.
%There are $4$ states for $\lvk=2$, $2$ states for $\lvk=1$, and $6$ states for $\lvk=3$.

\end{enumerate}

% - - - - - - - - - - - - - - - - - - - - - - - - - - - - - - -
\subsubsection{$m=2$}
\label{subsubsec:n=2;m=2}

We have seven sectors in the $m=2$ case, as briefly described below. For each sector, the effective action is given by
$$
I=\frac{1}{2\pi}\int \PMat_{\iSt\iDb}\pvar^\iDb(d\yvar^{\iSt\iDb}-d\yvar^{\iSt(\iDb-1)})\,,
$$
but the $\yvar^{\iSt\iDb}$ variables are constrained by the relations \eqref{eqn:xyc}-\eqref{eqn:SRpZii}.
We derive the dimension of the Hilbert space of each sector in detail for $\lvk=2$, but simply state the results for $\lvk=1,3$.

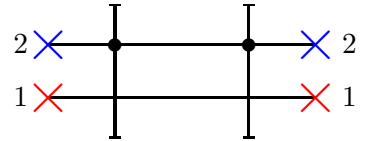
\begin{enumerate}

\item
$\PMat=
\left(\begin{array}{cc} 1 & 1 \\ 0 & 0  \end{array}\right)_{1}$
\hfill
% --------------------------------------------------------------
% --------------------------------------------------------------
\begin{picture}(125,50)
%\put(0,18){1. Sector $\PMat= \left( \begin{array}{cc} 1 & 1 \\ 0 & 0  \end{array}\right)_{1}$.}
\put(15,0){\begin{picture}(100,50)
\thicklines
%F1
\color{black}
\put(0,15){\line(1,0){100}}
\put(0,35){\line(1,0){100}}
% D2
\color{black}
\put(25,0){\line(0,1){50}}
\put(75,0){\line(0,1){50}}
%SR
\color{red}
\put(-5,20){\line(1,-1){10}}
\put(-5.0,10){\line(1,1){10}}
\color{red}
\put(95,20){\line(1,-1){10}}
\put(95,10){\line(1,1){10}}
\color{blue}
\put(-5,40){\line(1,-1){10}}
\put(-5,30){\line(1,1){10}}
\color{blue}
\put(95,40){\line(1,-1){10}}
\put(95,30){\line(1,1){10}}
%Intersections
\color{black}
%\put(25,15){\circle*{5}}
\put(25,35){\circle*{5}}
%\put(75,15){\circle*{5}}
\put(75,35){\circle*{5}}
%endpoints
\thinlines
\put(23,50){\line(1,0){4}}
\put(23,0){\line(1,0){4}}
\put(73,0){\line(1,0){4}}
\put(73,50){\line(1,0){4}}

\put(-13,12){$1$}
\put(-13,32){$2$}
\put(110,12){$1$}
\put(110,32){$2$}

\end{picture}}
\end{picture}

\noindent
% --------------------------------------------------------------
One closed string of winding number $1$ breaks on each of the D$2$-branes to form two open strings, one of which passes through the S-R-twist once. In addition, there is also a closed string.  The open string states were analyzed in \secref{subsec:m>1} to give rise to $\lvk$ states, while the closed string gives rise to $\lvk$ states as well (as mentioned in the first sector of \secref{subsubsec:n=2;m=1}). Thus, there are a total of $\lvk^2$ states in this sector.

\item
$\PMat=
\left(\begin{array}{cc} 1 & 1 \\ 0 & 0  \end{array}\right)_{-1}$
\hfill
% --------------------------------------------------------------
\begin{picture}(125,50)
%\put(0,18){1. Sector $\PMat= \left( \begin{array}{cc} 1 & 1 \\ 0 & 0  \end{array}\right)_{-1}$.}
\put(15,0){\begin{picture}(100,50)
\thicklines
%F1
\color{black}
\put(0,15){\line(1,0){100}}
\put(0,35){\line(1,0){100}}
% D2
\color{black}
\put(25,0){\line(0,1){50}}
\put(75,0){\line(0,1){50}}
%SR
\color{red}
\put(-5,20){\line(1,-1){10}}
\put(-5.0,10){\line(1,1){10}}
\color{blue}
\put(95,20){\line(1,-1){10}}
\put(95,10){\line(1,1){10}}
\color{blue}
\put(-5,40){\line(1,-1){10}}
\put(-5,30){\line(1,1){10}}
\color{red}
\put(95,40){\line(1,-1){10}}
\put(95,30){\line(1,1){10}}
%Intersections
\color{black}
%\put(25,15){\circle*{5}}
\put(25,35){\circle*{5}}
%\put(75,15){\circle*{5}}
\put(75,35){\circle*{5}}
%endpoints
\thinlines
\put(23,50){\line(1,0){4}}
\put(23,0){\line(1,0){4}}
\put(73,0){\line(1,0){4}}
\put(73,50){\line(1,0){4}}

\put(-13,12){$1$}
\put(-13,32){$2$}
\put(110,12){$2$}
\put(110,32){$1$}

\end{picture}}
\end{picture}
% --------------------------------------------------------------

\noindent
One string connects the two D$2$-branes [and hence $\xvar^{11}=\xvar^{12}=\xvar^1=\xvar^2\equiv\xvar$ from \eqref{eqn:xconstii}], while the other starts at the second D2-brane, winds around the $x_3$ circle and passes through the S-R-twist twice before ending on the first D$2$-brane. For $\lvk=2$, the string that starts on the second D$2$-brane at $\yvar^{12}+i\xvar$ ends on the first brane at $-\yvar^{12}-i\xvar$, giving us the constraint $2\xvar=0$ (modulo $2\pi$), which implies
$\xvar=0,$ or $\pi$. The effective action becomes
$$
I_{\lvk=2} = \frac{1}{2\pi}
\int \lbrack \pvar^1 \left( d\yvar^{11} + d\yvar^{12} \right) + \pvar^2 \left( d\yvar^{12} - d\yvar^{11} \right)\rbrack
\,,
$$
which gives us $2$ states for each of the two possible values of $\xvar$. In total, we get $4$ states. For $\lvk=1,3$, the computation is similar to that of the second sector of \secref{subsubsec:n=2;m=1}, and we get 3 states in both cases.

\item
$\PMat=
\left(\begin{array}{cc} 1 & 0 \\ 0 & 1  \end{array}\right)_{1}$
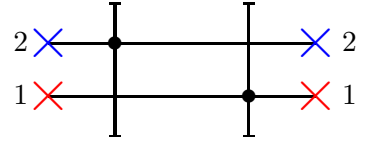
\hfill
% --------------------------------------------------------------
\begin{picture}(125,50)
%\put(0,18){1. Sector $\PMat= \left( \begin{array}{cc} 1 & 0 \\ 0 & 1  \end{array}\right)_{1}$.}
\put(15,0){\begin{picture}(100,50)
\thicklines
%F1
\color{black}
\put(0,15){\line(1,0){100}}
\put(0,35){\line(1,0){100}}
% D2
\color{black}
\put(25,0){\line(0,1){50}}
\put(75,0){\line(0,1){50}}
%SR
\color{red}
\put(-5,20){\line(1,-1){10}}
\put(-5.0,10){\line(1,1){10}}
\color{red}
\put(95,20){\line(1,-1){10}}
\put(95,10){\line(1,1){10}}
\color{blue}
\put(-5,40){\line(1,-1){10}}
\put(-5,30){\line(1,1){10}}
\color{blue}
\put(95,40){\line(1,-1){10}}
\put(95,30){\line(1,1){10}}
%Intersections
\color{black}
%\put(25,15){\circle*{5}}
\put(25,35){\circle*{5}}
\put(75,15){\circle*{5}}
%\put(75,35){\circle*{5}}
%endpoints
\thinlines
\put(23,50){\line(1,0){4}}
\put(23,0){\line(1,0){4}}
\put(73,0){\line(1,0){4}}
\put(73,50){\line(1,0){4}}

\put(-13,12){$1$}
\put(-13,32){$2$}
\put(110,12){$1$}
\put(110,32){$2$}

\end{picture}}
\end{picture}
% --------------------------------------------------------------

\noindent
% The string that starts on the first D$2$-brane at $\xvar'+i\yvar^{11}$ winds around the $x_3$ circle and passes the twist's location once, ending on the first brane at $-\yvar^{11}+i\xvar'$, giving us the constraint $2\xvar'=-\yvar^{11}$. Taking into account the second string that starts on the second D$2$-brane at $\xvar^{2} + i\yvar^{22}$, the action simplifies to
One open string starts and ends on each D$2$-brane, and each string passes through the S-R-twist once. It is easy to see that the two strings are completely decoupled from each other, each being bound to a separate D$2$-brane. We therefore get two decoupled $n=1$ sectors, each with $m=1.$ The action is a sum of two terms:
$$
I_{k=2} = \frac{2}{2\pi}
\int \lbrack \pvar^1 d\yvar^{11} + \pvar^2 d\yvar^{12}  \rbrack
\,,
$$
which gives us $4$ states. A similar computation gives us $1$ and $9$ states for $\lvk=1,3$, respectively.

\item
$\PMat=
\left(\begin{array}{cc} 1 & 0 \\ 0 & 1  \end{array}\right)_{-1}$
\hfill
% --------------------------------------------------------------
\begin{picture}(125,50)
%\put(0,18){1. Sector $\PMat= \left( \begin{array}{cc} 1 & 0 \\ 0 & 1  \end{array}\right)_{-1}$.}
\put(15,0){\begin{picture}(100,50)
\thicklines
%F1
\color{black}
\put(0,15){\line(1,0){100}}
\put(0,35){\line(1,0){100}}
% D2
\color{black}
\put(25,0){\line(0,1){50}}
\put(75,0){\line(0,1){50}}
%SR
\color{red}
\put(-5,20){\line(1,-1){10}}
\put(-5.0,10){\line(1,1){10}}
\color{blue}
\put(95,20){\line(1,-1){10}}
\put(95,10){\line(1,1){10}}
\color{blue}
\put(-5,40){\line(1,-1){10}}
\put(-5,30){\line(1,1){10}}
\color{red}
\put(95,40){\line(1,-1){10}}
\put(95,30){\line(1,1){10}}
%Intersections
\color{black}
%\put(25,15){\circle*{5}}
\put(25,35){\circle*{5}}
\put(75,15){\circle*{5}}
%\put(75,35){\circle*{5}}
%endpoints
\thinlines
\put(23,50){\line(1,0){4}}
\put(23,0){\line(1,0){4}}
\put(73,0){\line(1,0){4}}
\put(73,50){\line(1,0){4}}

\put(-13,12){$1$}
\put(-13,32){$2$}
\put(110,12){$2$}
\put(110,32){$1$}

\end{picture}}
\end{picture}
% --------------------------------------------------------------

\noindent
One open string starts on the first D$2$-brane, winds around the $x_3$ circle once, passing through the S-R-twist, before ending on the second D$2$-brane. For $\lvk=2$, it starts at $\yvar^{11}+i\xvar^{11}$ and ends at $\yvar^{21}+i\xvar^{21}=-\xvar^{11}+i\yvar^{11}$. The other open string starts on the second D$2$-brane at $\yvar^{22}+i\xvar^{22}$, passes the S-R-twist once before ending on the first D$2$-brane at $\yvar^{10}+i\xvar^{10}=-\xvar^{22}+i\yvar^{22}$. Since $-\xvar^{11}+i\yvar^{11}$ and $\yvar^{22}+i\xvar^{22}$ are on the same D$2$-brane, it follows that $\yvar^{11}=\xvar^{22}=-\yvar^{10}$, and since $\yvar^{11}+i\xvar^{11}$ and $-\xvar^{22}+i\yvar^{22}$ are on the same D$2$-brane it follows that $\xvar^{11}=\yvar^{22}=-\yvar^{21}.$ We therefore get the effective action
$$
\frac{2}{2\pi}\int
\lbrack \pvar^1 d\yvar^{11} + \pvar^2 d\yvar^{22}  \rbrack
\,,
$$
which has the same form as that in Sector~3, and thus there are $4$ states. A similar computation gives $3$ states for both $\lvk=1,3$ respectively.

\item
$\PMat=
\left(\begin{array}{cc} 1 & 1 \\ 1 & 0  \end{array}\right)$
\hfill
% --------------------------------------------------------------
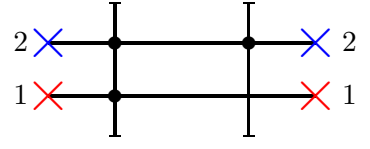
\begin{picture}(125,50)
%\put(0,18){1. Sector $\PMat= \left( \begin{array}{cc} 1 & 1 \\ 1 & 1  \end{array}\right)$.}
\put(15,0){\begin{picture}(100,50)
\thicklines
%F1
\color{black}
\put(0,15){\line(1,0){100}}
\put(0,35){\line(1,0){100}}
% D2
\color{black}
\put(25,0){\line(0,1){50}}
\put(75,0){\line(0,1){50}}
%SR
\color{red}
\put(-5,20){\line(1,-1){10}}
\put(-5.0,10){\line(1,1){10}}
\color{red}
\put(95,20){\line(1,-1){10}}
\put(95,10){\line(1,1){10}}
\color{blue}
\put(-5,40){\line(1,-1){10}}
\put(-5,30){\line(1,1){10}}
\color{blue}
\put(95,40){\line(1,-1){10}}
\put(95,30){\line(1,1){10}}
%Intersections
\color{black}
\put(25,15){\circle*{5}}
\put(25,35){\circle*{5}}
%\put(75,15){\circle*{5}}
\put(75,35){\circle*{5}}
%endpoints
\thinlines
\put(23,50){\line(1,0){4}}
\put(23,0){\line(1,0){4}}
\put(73,0){\line(1,0){4}}
\put(73,50){\line(1,0){4}}

\put(-13,12){$1$}
\put(-13,32){$2$}
\put(110,12){$1$}
\put(110,32){$2$}

\end{picture}}
\end{picture}
% --------------------------------------------------------------

\noindent
One open string stretches between the two D$2$-branes and is located at $\yvar^{11}+i\xvar^1$ for $\lvk=2$. Another string starts on the second D$2$-brane, passes through the S-R-twist and winds around the $x_3$ circle once before ending on the first D$2$-brane. It starts at $\yvar^{12}+i\xvar^2$ and ends at $-\xvar^{2}+i\yvar^{12}$. A third string starts on the first D$2$-brane at $\yvar^{21}+i\xvar^1$, passes the S-R-twist once before ending on the same D$2$-brane at $-\xvar^{1}+i\yvar^{21}$. Taking into account the constraints \eqref{eqn:xyc}-\eqref{eqn:SRpZii}, we get $\yvar^{12}=\yvar^{21}=-\yvar^{20}=-\yvar^{10}=\xvar^1=\xvar^2$, and the bosonic part of the effective action simplifies to
$$
I_{k=2} =  \frac{1}{2\pi}\int
\left\lbrack
\pvar^1 \left( d\yvar^{11} + 3d\yvar^{12} \right) + \pvar^2 \left( d\yvar^{12} - d\yvar^{11} \right)
\right\rbrack
\,.
$$
This is a congested sector which, as will be shown in \secref{subsubsec:fzmm>1}, has $4$ real fermionic zero modes.

%giving us $4$ states.  An analogous computation gives $3$ and $6$ %states for $\lvk=1,3$ respectively.

\item
$\PMat=
\left(\begin{array}{cc} 1 & 1 \\ 0 & 1  \end{array}\right)$
\hfill
% --------------------------------------------------------------
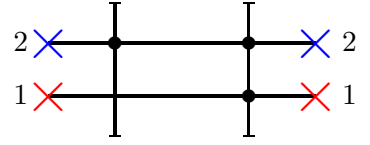
\begin{picture}(125,50)
%\put(0,18){1. Sector $\PMat= \left( \begin{array}{cc} 1 & 1 \\ 0 & 1  \end{array}\right)$.}
\put(15,0){\begin{picture}(100,50)
\thicklines
%F1
\color{black}
\put(0,15){\line(1,0){100}}
\put(0,35){\line(1,0){100}}
% D2
\color{black}
\put(25,0){\line(0,1){50}}
\put(75,0){\line(0,1){50}}
%SR
\color{red}
\put(-5,20){\line(1,-1){10}}
\put(-5.0,10){\line(1,1){10}}
\color{red}
\put(95,20){\line(1,-1){10}}
\put(95,10){\line(1,1){10}}
\color{blue}
\put(-5,40){\line(1,-1){10}}
\put(-5,30){\line(1,1){10}}
\color{blue}
\put(95,40){\line(1,-1){10}}
\put(95,30){\line(1,1){10}}
%Intersections
\color{black}
%\put(25,15){\circle*{5}}
\put(25,35){\circle*{5}}
\put(75,15){\circle*{5}}
\put(75,35){\circle*{5}}
%endpoints
\thinlines
\put(23,50){\line(1,0){4}}
\put(23,0){\line(1,0){4}}
\put(73,0){\line(1,0){4}}
\put(73,50){\line(1,0){4}}

\put(-13,12){$1$}
\put(-13,32){$2$}
\put(110,12){$1$}
\put(110,32){$2$}

\end{picture}}
\end{picture}
% --------------------------------------------------------------

\noindent
Similar to Sector~5, but with the two D$2$-branes exchanged. All results mentioned in Sector~5 are identical to those in this sector.

\item
$\PMat=
\left(\begin{array}{cc} 1 & 1 \\ 1 & 1  \end{array}\right)$
\hfill
% --------------------------------------------------------------
\begin{picture}(125,50)
%\put(0,18){1. Sector $\PMat= \left( \begin{array}{cc} 1 & 1 \\ 1 & 1  \end{array}\right)$.}
\put(15,0){\begin{picture}(100,50)
\thicklines
%F1
\color{black}
\put(0,15){\line(1,0){100}}
\put(0,35){\line(1,0){100}}
% D2
\color{black}
\put(25,0){\line(0,1){50}}
\put(75,0){\line(0,1){50}}
%SR
\color{red}
\put(-5,20){\line(1,-1){10}}
\put(-5.0,10){\line(1,1){10}}
\color{red}
\put(95,20){\line(1,-1){10}}
\put(95,10){\line(1,1){10}}
\color{blue}
\put(-5,40){\line(1,-1){10}}
\put(-5,30){\line(1,1){10}}
\color{blue}
\put(95,40){\line(1,-1){10}}
\put(95,30){\line(1,1){10}}
%Intersections
\color{black}
\put(25,15){\circle*{5}}
\put(25,35){\circle*{5}}
\put(75,15){\circle*{5}}
\put(75,35){\circle*{5}}
%endpoints
\thinlines
\put(23,50){\line(1,0){4}}
\put(23,0){\line(1,0){4}}
\put(73,0){\line(1,0){4}}
\put(73,50){\line(1,0){4}}

\put(-13,12){$1$}
\put(-13,32){$2$}
\put(110,12){$1$}
\put(110,32){$2$}

\end{picture}}
\end{picture}
% --------------------------------------------------------------

\noindent
This is the sector given as an example for bosonic zero modes in \secref{subsec:BosZmodes}.
Each closed string of winding number $1$ breaks on each of the D$2$-branes to form two open strings, one of which passes through the S-R-twist once. One string starts on the second D$2$-brane at $\yvar^{\iSt 2}+i\xvar^1$, passes the twist once before ending on the first D$2$-brane at $-\xvar^{1}+i\yvar^{\iSt 2}$. The remaining string stretches between the two D$2$-branes and is located at $\yvar^{\iSt 1}+i\xvar$, where $\xvar=\xvar^1=\xvar^2$. We have two sets of such strings, for $\iSt=1,2$. Taking into account the constraints, we get $\yvar^{\iSt 0}=-\yvar^{\iSt 2}=-\xvar$ for $\iSt=1,2$, and the bosonic part of the effective action simplifies to the expression given in \eqref{eqn:I0(n=2,m=2,k=2)BosZmodes}.
%which gives us $4$ states. A similar computation gives $2$ and $6$ states for $\lvk=1,3$ respectively.
This is again a congested sector which, as will be shown in \secref{subsubsec:fzmm>1}, has $12$ real fermionic zero modes.

\end{enumerate}
We thus see that the binding matrix $\PMat$ and the permutation $\perm$ can be used to help us visualize the string and brane configurations and determine the effective action rather easily.

% - - - - - - - - - - - - - - - - - - - - - - - - - - - - - - -

\subsubsection{$m=3$}
\label{subsubsec:n=2;m=3}
As a last explicit example, let us enumerate the sectors for $m=3$. It turns out that there are $18$ sectors described by the following set of binding matrices:
\be
\label{eqn:m=3sectors}
\PMat \in \left\{  \begin{array}{lllll}
\left( \begin{array}{ccc} 1&1&1\\0&0&0\end{array}\right)_{\{1,-1\}},&
\left( \begin{array}{ccc} 1&0&0\\0&1&1\end{array}\right)_{\{1,-1\}},&
\left( \begin{array}{ccc} 0&1&0\\1&0&1\end{array}\right)_{\{1,-1\}},&
\left( \begin{array}{ccc} 1&1&0\\0&0&1\end{array}\right)_{\{1,-1\}},&
\left( \begin{array}{ccc} 1&1&1\\1&0&1\end{array}\right),\\
\left( \begin{array}{ccc} 1&1&1\\1&0&0\end{array}\right),&
\left( \begin{array}{ccc} 1&1&1\\0&1&0\end{array}\right)_{1},&
\left( \begin{array}{ccc} 1&1&1\\1&1&0\end{array}\right),&
\left( \begin{array}{ccc} 1&1&0\\1&0&1\end{array}\right),&
\left( \begin{array}{ccc} 1&1&1\\0&1&1\end{array}\right),\\
\left( \begin{array}{ccc} 1&1&0\\0&1&1\end{array}\right)_{1},&
\left( \begin{array}{ccc} 1&1&1\\1&1&1\end{array}\right).&
\left( \begin{array}{ccc} 1&1&1\\0&0&1\end{array}\right),&
\left( \begin{array}{ccc} 1&0&1\\0&1&1\end{array}\right).&{}
\end{array} \right\}
\ee
% ==============================================================
In \eqref{eqn:m=3sectors}, there are $4$ decongested $\PMat$'s of which each permutation $\perm$ gives rise to a distinct sector, and thus these binding matrices generate $8$ sectors in total. Apart from the following pair
$$
\left( \begin{array}{ccc} 1&1&1\\0&1&0\end{array}\right)_{\{1,-1\}}=\left( \begin{array}{ccc} 1&1&0\\0&1&1\end{array}\right)_{\{-1,1\}},
$$
which are equivalent after relabeling of the strings (note that $\perm$ has to be changed as well), the rest of the $\PMat$'s remain invariant under relabeling. There are thus a total of $18$ different sectors for $m=3$.

% -------------------------------------------------------------

\subsection{Counting fermionic zero modes}
\label{subsec:countfm=1}
In order to properly compute the Witten index of our system in the type-IIA picture, we will now count the number of fermionic zero modes in each sector characterized by the binding matrix $\PMat$ and permutation $\perm$. If a sector does not support a fermionic zero mode, then its contribution to the Witten index is just the number of ground states of its Hilbert space; if on the other hand a sector does support fermionic zero modes, then after quantization, its Hilbert space will contain an equal number of bosonic and fermionic ground states, thereby making the net contribution to the Witten index zero. It is therefore crucial in the computation of the Witten index to determine which sectors support fermionic zero modes and which sectors do not.

In \secref{subsubsec:fzmm=1}, we will address this question for the three sectors that arise in the case $n=2$ with $m=1$, as discussed in \secref{subsubsec:n=2;m=1}. This simplest example will serve to illustrate the salient points of the discussion. We will then tackle the cases with general $m$ in \secref{subsubsec:fzmm>1}. For ease of discussion, we will explicitly treat $\lvk=2$ and $n=2$ cases only; generalization to other values of $\lvk$ and $n$ however is straightforward, and leads to the same conclusion.

\subsubsection{$m=1$}
\label{subsubsec:fzmm=1}

\begin{figure}[t]
\begin{picture}(400,170)

\put(50,10){\begin{picture}(150,150)
\put(-10,140){(a)}
\thicklines
%\color{black}
\put(0,50){\line(1,0){100}}
\put(0,100){\line(1,0){100}}

% twist
\color{red}
\put(-5,45){\line(1,1){10}}
\put(5,45){\line(-1,1){10}}
%\color{red}
\put(95,45){\line(1,1){10}}
\put(105,45){\line(-1,1){10}}

\put(-5,95){\line(1,1){10}}
\put(5,95){\line(-1,1){10}}
%\color{red}
\put(95,95){\line(1,1){10}}
\put(105,95){\line(-1,1){10}}
% D2
\color{black}
\put(50,0){\line(0,1){150}}
%\put(150,0){\line(0,1){100}}

% D2-F1 intersections
\color{red}
%\put(50,50){\circle*{6}}
\put(50,100){\circle*{6}}

% NS5
\thinlines
\color{black}
\put(40,0){\line(1,0){20}}
\put(40,150){\line(1,0){20}}
%\put(140,0){\line(1,0){20}}
%\put(140,100){\line(1,0){20}}

% NS5 intersections
%\color{red}
%\put(50,0){\circle*{6}}
%\put(150,0){\circle*{6}}
%\put(50,150){\circle*{6}}
%\put(150,100){\circle*{6}}

\color{blue}
\put(25,55){$\psi_{20}$}
\put(65,55){$\psi_{21}$}
\put(25,105){$\psi_{10}$}
\put(65,105){$\psi_{11}$}

\put(52,25){$\lambda_{12}$}
\put(52,75){$\lambda_{11}$}
\put(52,125){$\lambda_{10}$}
%\put(152,75){$\lambda_2^{(<)}$}

\color{black}
\put(70,40){F1}
\put(70,90){F1}
\put(18,140){D2}
\put(27,138){\vector(1,-1){20}}
%\put(118,90){D2}
%\put(127,88){\vector(1,-1){20}}
%\put(62,-2){NS5}
%\put(62,98){NS5}
%\put(162,-2){NS5}
%\put(162,98){NS5}
\put(92,58){SR}
\put(92,108){SR}
\end{picture}}
\put(250,10){\begin{picture}(150,150)
\put(-10,140){(b)}
\thicklines
%\color{black}
\put(0,50){\line(1,0){100}}
\put(0,100){\line(1,0){100}}

% twist
\color{red}
\put(-5,45){\line(1,1){10}}
\put(5,45){\line(-1,1){10}}
%\color{red}
\put(95,45){\line(1,1){10}}
\put(105,45){\line(-1,1){10}}

\put(-5,95){\line(1,1){10}}
\put(5,95){\line(-1,1){10}}
%\color{red}
\put(95,95){\line(1,1){10}}
\put(105,95){\line(-1,1){10}}
% D2
\color{black}
\put(50,0){\line(0,1){150}}
%\put(150,0){\line(0,1){100}}

% D2-F1 intersections
\color{red}
\put(50,50){\circle*{6}}
\put(50,100){\circle*{6}}

% NS5
\thinlines
\color{black}
\put(40,0){\line(1,0){20}}
\put(40,150){\line(1,0){20}}
%\put(140,0){\line(1,0){20}}
%\put(140,100){\line(1,0){20}}

% NS5 intersections
%\color{red}
%\put(50,0){\circle*{6}}
%\put(150,0){\circle*{6}}
%\put(50,150){\circle*{6}}
%\put(150,100){\circle*{6}}

\color{blue}
\put(25,55){$\psi_{20}$}
\put(65,55){$\psi_{21}$}
\put(25,105){$\psi_{10}$}
\put(65,105){$\psi_{11}$}

\put(52,25){$\lambda_{12}$}
\put(52,75){$\lambda_{11}$}
\put(52,125){$\lambda_{10}$}
%\put(152,75){$\lambda_2^{(<)}$}

\color{black}
\put(70,40){F1}
\put(70,90){F1}
\put(18,140){D2}
\put(27,138){\vector(1,-1){20}}
%\put(118,90){D2}
%\put(127,88){\vector(1,-1){20}}
%\put(62,-2){NS5}
%\put(62,98){NS5}
%\put(162,-2){NS5}
%\put(162,98){NS5}
\put(92,58){SR}
\put(92,108){SR}
\end{picture}}

\end{picture}
\caption{Fermionic zero modes on the D$2$-brane and F$1$-strings for (a) Sectors 1 and 2, and (b) Sector~3 of the $m=1$ cases listed in \secref{subsubsec:n=2;m=1}. In (a), we have trivial identifications $\fpsi_{20}=\fpsi_{21}$ and $\lambda_{10}=\lambda_{11}$.
}
\label{fig:ZeroModesn=2m=1}
\end{figure}
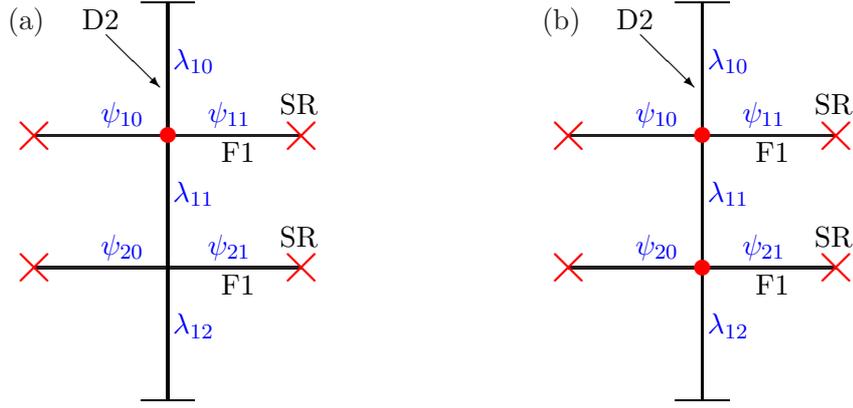

Of the three sectors described in \secref{subsubsec:n=2;m=1}, the first two sectors do not support fermionic zero modes. This essentially follows from our abelian result in \secref{subsec:fzeromodesn=1}. Sector~1 consists of the abelian $n=1$, $m=1$ sector plus a closed string, neither of which supports a fermionic zero mode. The fermionic zero modes of Sector~2 must satisfy the same set of equations as those of the abelian $n=1$, $m=1$ sector, except for those coming from the S-R-twist. In other words, in the notation of \figref{fig:ZeroModesn=2m=1}, the $\fpsi_{1\iDbz}$ for $\iDbz=0,1$ and $\lambda_{1\iStz}$ for $\iStz=0,1$ (note that $\lambda_{11}=\lambda_{12}$ and $\fpsi_{20}=\fpsi_{21}$ in this sector) will satisfy all the equations of \secref{subsec:fzeromodesn=1} with $\mathcal{P}$ replaced by $\mathcal{P}^2$, because the open string starting on the D2-brane passes through the S-R-twist twice before ending on the same D2-brane. Therefore, the boundary condition from the S-R-twist now reads
$$
\fpsi_{11}=\mathcal{P}\fpsi_{20}\equiv\mathcal{P}\fpsi_{21}=\mathcal{P}^2\fpsi_{10}\,,
$$
due to the permutation $\perm=-1$. But the argument otherwise does not change, because the only property of $\mathcal{P}$ that we used there was the fact that it does not have an eigenvalue $+1$, and neither does $\mathcal{P}^2$. We conclude that Sector~2 does not support fermionic zero modes.

% --------------------------------------------------------------
% --------------------------------------------------------------
% --------------------------------------------------------------
It remains to consider Sector~3.
The full set of equations that we need to solve is as follows (see \figref{fig:ZeroModesn=2m=1}). First, we have the chirality conditions:
\bear
0 &=& (1+\Gamma^{023})\psi_{10}
=(1+\Gamma^{023})\psi_{11}
=(1+\Gamma^{023})\psi_{20}
=(1+\Gamma^{023})\psi_{21}
\,,\label{eqn:(n=2)(m=1)psi}\\
0 &=&
(1+\Gamma^{09\tenx})\lambda_{10}
=(1+\Gamma^{09\tenx})\lambda_{11}
=(1+\Gamma^{09\tenx})\lambda_{12}
\,.\label{eqn:(n=2)(m=1)lam}
\eear
Then, we have the boundary conditions at the end of the D$2$-branes \eqref{eqn:bcotherend}:
\be\label{eqn:bcotherendII}
0=(1-\Gamma^{\tenx})
(1 +\Gamma^{0145678})
\lambda_{10}
=(1-\Gamma^{\tenx})
(1 +\Gamma^{0145678})
\lambda_{12}\,.
\ee
Next, we have two junctions with boundary conditions \eqref{eqn:ZMxJp}-\eqref{eqn:ZMxJm}, which read:
\bear
(1-\Gamma^{239\tenx})\psi_{10} &=&
(1-\Gamma^{239\tenx})\psi_{11} =
(1-\Gamma^{239\tenx})\lambda_{10} =
(1-\Gamma^{239\tenx})\lambda_{11}
\label{eqn:ZMxJpI}
\,,\\
0 &=&
(1+\Gamma^{239\tenx})(\psi_{10} -\psi_{11}
+ \Gamma^{39}\lambda_{11} -\Gamma^{39}\lambda_{10})\,.
\label{eqn:ZMxJmI}
\eear
for the first junction, and
\bear
(1-\Gamma^{239\tenx})\psi_{20} &=&
(1-\Gamma^{239\tenx})\psi_{21} =
(1-\Gamma^{239\tenx})\lambda_{11} =
(1-\Gamma^{239\tenx})\lambda_{12}
\label{eqn:ZMxJpII}
\,,\\
0 &=&
(1+\Gamma^{239\tenx})(\psi_{20} -\psi_{21}
+ \Gamma^{39}\lambda_{12} -\Gamma^{39}\lambda_{11})\,.
\label{eqn:ZMxJmII}
\eear
for the second junction. And finally we have the S-R-twist condition \eqref{eqn:ZMxSRm}-\eqref{eqn:OmegaDef}:
\be\label{eqn:ZMxSRmI}
\psi_{11}=\mathcal{P}\psi_{10}
\,,\qquad
\psi_{21}=\mathcal{P}\psi_{20}
\,.
\ee

To solve these equations, we first eliminate the fermionic mode $\lambda_{11}$ which lives on the middle section of the D$2$-brane. To do this, we note that the junction conditions \eqref{eqn:ZMxJpI}-\eqref{eqn:ZMxJmI} together imply
\be\label{eqn:lam11elimI}
\lambda_{11} =
\tfrac{1}{2}(1+\Gamma^{239\tenx})\lambda_{11}
+\tfrac{1}{2}(1-\Gamma^{239\tenx}\lambda_{11}
=
%\tfrac{1}{2}(1+\Gamma^{239\tenx})\Gamma^{39}\psi_{01}
%-\tfrac{1}{2}(1+\Gamma^{239\tenx})\Gamma^{39}\psi_{11}
%+\tfrac{1}{2}(1+\Gamma^{239\tenx})\lambda_{10}
%+\tfrac{1}{2}(1-\Gamma^{239\tenx})\lambda_{10}
%=
\lambda_{10}
+\tfrac{1}{2}(1+\Gamma^{239\tenx})\Gamma^{39}(\psi_{10}-\psi_{11})\,.
\ee
It is not hard to check that if we set $\lambda_{11}$ to the RHS of \eqref{eqn:lam11elimI}, and if we assume that $\lambda_{10},\psi_{10},\psi_{11}$ satisfy the chirality conditions that are required of them in \eqref{eqn:(n=2)(m=1)psi}-\eqref{eqn:(n=2)(m=1)lam}, then $\lambda_{11}$ will automatically satisfy the chirality condition that is required of it in \eqref{eqn:(n=2)(m=1)lam}. It follows that we can safely eliminate $\lambda_{11}$ from the equations using \eqref{eqn:lam11elimI}.
But if we choose to eliminate $\lambda_{11}$ from the second junction \eqref{eqn:ZMxJpII}-\eqref{eqn:ZMxJmII}, we get,
instead of \eqref{eqn:lam11elimI},
\be\label{eqn:lam11elimII}
\lambda_{11} =
%\tfrac{1}{2}(1+\Gamma^{239\tenx})\lambda_{11}
%+\tfrac{1}{2}(1-\Gamma^{239\tenx}\lambda_{11}
%=
%\tfrac{1}{2}(1+\Gamma^{239\tenx})\Gamma^{39}\psi_{10}
%-\tfrac{1}{2}(1+\Gamma^{239\tenx})\Gamma^{39}\psi_{11}
%+\tfrac{1}{2}(1+\Gamma^{239\tenx})\lambda_{12}
%+\tfrac{1}{2}(1-\Gamma^{239\tenx})\lambda_{12}
%=
\lambda_{12}
+\tfrac{1}{2}(1+\Gamma^{239\tenx})\Gamma^{39}(\psi_{21}-\psi_{20})\,.
\ee
Comparing \eqref{eqn:lam11elimI} and \eqref{eqn:lam11elimII}, we obtain
\be\label{eqn:byelimlam11}
\lambda_{10}
+\tfrac{1}{2}(1+\Gamma^{239\tenx})\Gamma^{39}(\psi_{10}-\psi_{11})
=
\lambda_{12}
+\tfrac{1}{2}(1+\Gamma^{239\tenx})\Gamma^{39}(\psi_{21}-\psi_{20})\,.
\ee
Now, we need to solve \eqref{eqn:bcotherendII}-\eqref{eqn:ZMxSRmI} together with \eqref{eqn:byelimlam11} for
$$
\lambda_{10}, \lambda_{12}, \psi_{10},\psi_{11},\psi_{20},\psi_{21}
\,,
$$
that are subject to the chirality conditions specified in \eqref{eqn:(n=2)(m=1)psi}-\eqref{eqn:(n=2)(m=1)lam}.

To proceed, we note that we can combine \eqref{eqn:ZMxJpI} with \eqref{eqn:ZMxJpII} to get
\be\label{eqn:ZMxJpp}
\begin{split}
(1-\Gamma^{239\tenx})\psi_{10}&=
(1-\Gamma^{239\tenx})\psi_{11}=
(1-\Gamma^{239\tenx})\psi_{20}
=
(1-\Gamma^{239\tenx})\psi_{21}\\
&=
(1-\Gamma^{239\tenx})\lambda_{10}=
%(1-\Gamma^{239\tenx})\lambda_{11}=
(1-\Gamma^{239\tenx})\lambda_{12}\,.
\end{split}
\ee
Let
$$
\zeta_1\equiv\lambda_{10}-\lambda_{12}\,.
$$
It satisfies the same set of equations \eqref{eqn:zetaA}--\eqref{eqn:zetaC} of the abelian case: \eqref{eqn:zetaA} because of the chirality condition \eqref{eqn:(n=2)(m=1)lam} on $\lambda_{ab}$, \eqref{eqn:zetaB} because both $\lambda_{10}$ and $\lambda_{12}$ have the boundary conditions \eqref{eqn:bcotherendII} that is dual to type-IIB strings ending on (formal) NS$5$-branes, and \eqref{eqn:zetaC} because of \eqref{eqn:ZMxJpp}. Hence, the same argument we used before, next to \eqref{eqn:023z}-\eqref{eqn:239z}, implies $\zeta_1=0$ again, and so
\be\label{eqn:lam1012}
\lambda_{10}=\lambda_{12}\,.
\ee

Next, substitute \eqref{eqn:lam1012} into \eqref{eqn:byelimlam11} to obtain
\be\label{eqn:sum4psia}
0=(1+\Gamma^{239\tenx})(\psi_{10} -\psi_{11}
+ \psi_{20} -\psi_{21})\,.
\ee
On the other hand, from \eqref{eqn:ZMxJpp} we have
\be\label{eqn:sum4psib}
0=(1-\Gamma^{239\tenx})(\psi_{10} -\psi_{11}
+ \psi_{20} -\psi_{21})\,.
\ee
The two equations \eqref{eqn:sum4psia} and \eqref{eqn:sum4psib} together imply
$$
\psi_{10} -\psi_{11}
+ \psi_{20} -\psi_{21}=0\,,
$$
or
\be\label{eqn:sum4psic}
\psi_{10} + \psi_{20} =\psi_{11}+\psi_{21}\,.
\ee
But from the boundary condition \eqref{eqn:ZMxSRmI} that describes the S-R-twist we can write \eqref{eqn:sum4psic} as:
$$
\psi_{11}+\psi_{21}=\mathcal{P}(\psi_{10}+\psi_{20}),
$$
and since $\mathcal{P}$ does not have an eigenvalue $+1$, we get
\be\label{eqn:sum4psid}
\psi_{10} + \psi_{20} =\psi_{11}+\psi_{21}=0\,.
\ee
Now, from \eqref{eqn:ZMxJpp} we have
$$
(1-\Gamma^{239\tenx})(\psi_{10}-\psi_{20})=0\,,
$$
and together with \eqref{eqn:sum4psid}, we deduce that
\be\label{eqn:fpsichiralzero}
(1-\Gamma^{239\tenx})\psi_{10}=(1-\Gamma^{239\tenx})\psi_{20}=0\,,
\ee
and hence all the other expressions appearing in \eqref{eqn:ZMxJpp} also vanish.

If we now define
$$
\xi_1\equiv\lambda_{10}+\lambda_{12}\,,
$$
the result of the last paragraph implies that $\xi_1$ satisfies the same equations that $\zeta_1$ satisfies in \eqref{eqn:zetaA}-\eqref{eqn:zetaC}, and hence $\xi_1=0$. Therefore, we conclude that
\be\label{eqn:lambda02zero}
\lambda_{10}=\lambda_{12}=0\,.
\ee

At this point, there is essentially only one unknown variable, say $\psi_{11}.$ The other variables $\fpsi_{ij}$ and $\lambda_{11}$ are determined in terms of it by \eqref{eqn:lam11elimII}, \eqref{eqn:sum4psid}, and \eqref{eqn:ZMxSRmI}. The equations it should satisfy are
\be\label{eqn:psi11}
0
= (1-\Gamma^{239\tenx})\mathcal{P}^{-1}\psi_{11}
= (1-\Gamma^{239\tenx})\psi_{11}
= (1+\Gamma^{023})\mathcal{P}^{-1}\psi_{11}
=(1+\Gamma^{023})\psi_{11}
\,.
\ee
where we substituted $\psi_{10}=\mathcal{P}^{-1}\psi_{11}$ from \eqref{eqn:ZMxSRmI}. Next, we recall that for $\lvk=2$ the operator $\mathcal{P}$ realizes a rotation by $\frac{\pi}{2}$ in four transverse 2-planes, and so
$$
\mathcal{P}\Gamma^{239\tenx}\mathcal{P}^{-1}
=-\Gamma^{1238}
\,,\qquad
\mathcal{P}\Gamma^{023}\mathcal{P}^{-1}=\Gamma^{023}.
$$
Using these relations, we can write \eqref{eqn:psi11} as
\be\label{eqn:psi11mod}
\psi_{11}=\Gamma^{239\tenx}\psi_{11}=-\Gamma^{023}\psi_{11}
=-\Gamma^{1238}\psi_{11}.
\ee
We can now work in a basis for which $\Gamma^{23}, \Gamma^{18}, \Gamma^{9\tenx}$ and $\Gamma^0$ are simultaneously diagonal. It is then easy to see that \eqref{eqn:psi11mod} has $4$ linearly independent solutions. This corresponds to $4$ zero modes of our system. These four real zero modes transform as singlets under the $SU(2)$ factor of the $SU(2)\times U(1)$ symmetry group that was mentioned at the end of \secref{subsec:SUSYleft} and they have $\pm 1$ charges under the $U(1)$ factor, which is generated by $\tfrac{i}{2}(\Gamma^{45}+\Gamma^{67})$. These statements are easy to derive from \eqref{eqn:psi11mod}, which together with \eqref{eqn:11Dchlty} implies that $\Gamma^{4567}\psi_{11}=-\psi_{11}$. (The other fermionic fields of the problem are determined in terms of $\psi_{11}$ and are easily seen to satisfy the same chirality condition.) In this subsection we have restricted for simplicity to the $\lvk=2$ case, but the same result of $4$ zero-modes is also true for the other cases $\lvk=1,3.$

% -------------------------------------------------------------

\subsubsection{$m>1$}
\label{subsubsec:fzmm>1}
Having considered the fermionic zero modes for the $n=2$, $m=1$ case, we now move on to consider the cases with general $m$. In this subsection, we prove the following criterion for the existence of fermionic zero modes: \textit{the fermionic zero modes exist precisely in those sectors for which $\PMat_{1\iDb}=\PMat_{2\iDb}=1$ for some $\iDb=1,\ldots,m$.} In other words, they exist if and only if there is at least one D$2$-brane to which both open strings attach. These are what we called {\it congested} sectors.
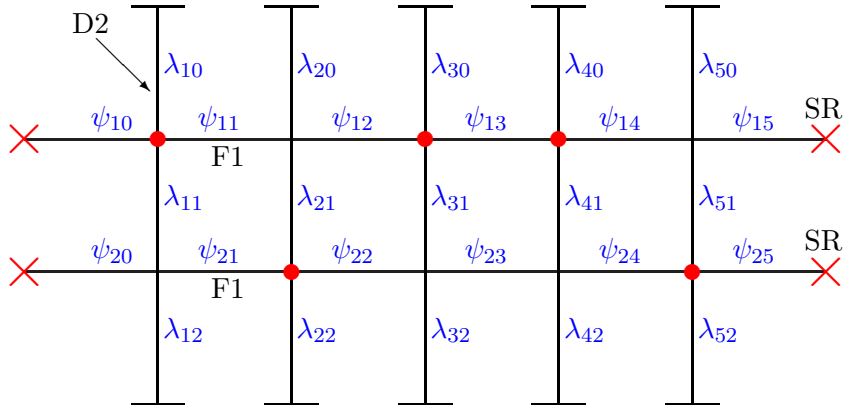
\begin{figure}[t]
\begin{picture}(400,170)

\put(50,10){\begin{picture}(350,150)

\thicklines
%\color{black}
\put(0,50){\line(1,0){300}}
\put(0,100){\line(1,0){300}}

% twist
\color{red}
\put(-5,45){\line(1,1){10}}
\put(5,45){\line(-1,1){10}}
\color{red}
\put(295,45){\line(1,1){10}}
\put(305,45){\line(-1,1){10}}

\put(-5,95){\line(1,1){10}}
\put(5,95){\line(-1,1){10}}
\color{red}
\put(295,95){\line(1,1){10}}
\put(305,95){\line(-1,1){10}}
% D2
\color{black}
\put(50,0){\line(0,1){150}}
\put(100,0){\line(0,1){150}}
\put(150,0){\line(0,1){150}}
\put(200,0){\line(0,1){150}}
\put(250,0){\line(0,1){150}}
%\put(150,0){\line(0,1){100}}

% D2-F1 intersections
\color{red}
%\put(50,50){\circle*{6}}
\put(50,100){\circle*{6}}
\put(100,50){\circle*{6}}
\put(150,100){\circle*{6}}
\put(200,100){\circle*{6}}
\put(250,50){\circle*{6}}

% NS5
\thinlines
\color{black}
\put(40,0){\line(1,0){20}}
\put(40,150){\line(1,0){20}}
\put(90,0){\line(1,0){20}}
\put(90,150){\line(1,0){20}}
\put(140,0){\line(1,0){20}}
\put(140,150){\line(1,0){20}}
\put(190,0){\line(1,0){20}}
\put(190,150){\line(1,0){20}}
\put(240,0){\line(1,0){20}}
\put(240,150){\line(1,0){20}}
%\put(140,0){\line(1,0){20}}
%\put(140,100){\line(1,0){20}}

% NS5 intersections
%\color{red}
%\put(50,0){\circle*{6}}
%\put(50,150){\circle*{6}}
%\put(100,0){\circle*{6}}
%\put(100,150){\circle*{6}}
%\put(150,0){\circle*{6}}
%\put(150,150){\circle*{6}}
%\put(200,0){\circle*{6}}
%\put(200,150){\circle*{6}}
%\put(250,0){\circle*{6}}
%\put(250,150){\circle*{6}}

\color{blue}
\put(25,55){$\psi_{20}$}
\put(65,55){$\psi_{21}$}
\put(25,105){$\psi_{10}$}
\put(65,105){$\psi_{11}$}
\put(115,55){$\psi_{22}$}
\put(165,55){$\psi_{23}$}
\put(115,105){$\psi_{12}$}
\put(165,105){$\psi_{13}$}
\put(215,55){$\psi_{24}$}
\put(265,55){$\psi_{25}$}
\put(215,105){$\psi_{14}$}
\put(265,105){$\psi_{15}$}

\put(52,25){$\lambda_{12}$}
\put(52,75){$\lambda_{11}$}
\put(52,125){$\lambda_{10}$}
\put(102,25){$\lambda_{22}$}
\put(102,75){$\lambda_{21}$}
\put(102,125){$\lambda_{20}$}
\put(152,25){$\lambda_{32}$}
\put(152,75){$\lambda_{31}$}
\put(152,125){$\lambda_{30}$}
\put(202,25){$\lambda_{42}$}
\put(202,75){$\lambda_{41}$}
\put(202,125){$\lambda_{40}$}
\put(252,25){$\lambda_{52}$}
\put(252,75){$\lambda_{51}$}
\put(252,125){$\lambda_{50}$}
%\put(152,75){$\lambda_2^{(<)}$}

\color{black}
\put(70,40){F1}
\put(70,90){F1}
\put(18,140){D2}
\put(27,138){\vector(1,-1){20}}
%\put(118,90){D2}
%\put(127,88){\vector(1,-1){20}}
%\put(62,-2){NS5}
%\put(62,98){NS5}
%\put(162,-2){NS5}
%\put(162,98){NS5}
\put(292,58){SR}
\put(292,108){SR}
\end{picture}}
\end{picture}
\caption{An example of sectors without fermionic zero modes. Only one string is attached to each D$2$-brane.}
\label{fig:n=2nofzm}
\end{figure}

It is easy to see that there is no fermionic zero mode if for each of the $m$ D$2$-branes there is only one string attached to it (see \figref{fig:n=2nofzm} for an example). For those sectors for which the permutation $\perm$ that accompanies the S-R-twist is the identity, we can divide the D$2$-branes into two groups: those connecting to the F$1$-string that is labeled by $\iSt=1$, and those connecting to the F$1$-string labeled by $\iSt=2.$ Each group together with the respective F$1$-string then forms an abelian system discussed in \secref{subsec:fzeromodesn=1}, and hence fermionic zero modes are absent.

For sectors with $\perm=-1$, we can divide the D$2$-brane indices $\iDb=1,\ldots,m$ into two groups so that those with $\iDb=\iDb_1,\ldots,\iDb_k$ attach to the F$1$-string that is labeled by $\iSt=1$, and those with $\iDb=\iDb_{k+1}, \ldots, \iDb_{m}$ to the F$1$-string labeled by $\iSt=2.$ Then the system can again be regarded as an abelian case, the D$2$-branes now being arranged in the new order $\iDb_1,\ldots,\iDb_m$, except for the S-R-twist condition, which should now read
$$
\fpsi_{1\iDb_k}=\mathcal{P}\fpsi_{2\iDb_{k+1}}\,,\quad
\fpsi_{2\iDb_m}=\mathcal{P}\fpsi_{1\iDb_1}\,.
$$
The effect of this new boundary condition is that instead of \eqref{eqn:fpsiallequal}, we get
$$
\fpsi_{1\iDb_1}=\fpsi_{1\iDb_2}=\cdots=\fpsi_{1\iDb_k}=\mathcal{P}\fpsi_{2\iDb_{k+1}}
=\cdots=\mathcal{P}\fpsi_{2\iDb_m}=\mathcal{P}^2\fpsi_{1\iDb_1}\,.
$$
But since $\mathcal{P}^2$ does not have an eigenvalue $+1$, the conclusion \eqref{eqn:fpsiallzero} remains the same, and hence there is no zero mode.

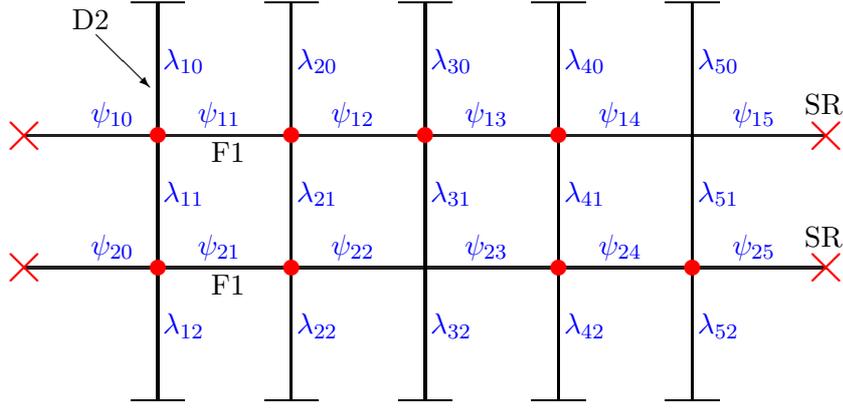
\begin{figure}[t]
\begin{picture}(400,170)

\put(50,10){\begin{picture}(350,150)

\thicklines
%\color{black}
\put(0,50){\line(1,0){300}}
\put(0,100){\line(1,0){300}}

% twist
\color{red}
\put(-5,45){\line(1,1){10}}
\put(5,45){\line(-1,1){10}}
%\color{red}
\put(295,45){\line(1,1){10}}
\put(305,45){\line(-1,1){10}}

\put(-5,95){\line(1,1){10}}
\put(5,95){\line(-1,1){10}}
%\color{red}
\put(295,95){\line(1,1){10}}
\put(305,95){\line(-1,1){10}}
% D2
\color{black}
\put(50,0){\line(0,1){150}}
\put(100,0){\line(0,1){150}}
\put(150,0){\line(0,1){150}}
\put(200,0){\line(0,1){150}}
\put(250,0){\line(0,1){150}}
%\put(150,0){\line(0,1){100}}

% D2-F1 intersections
\color{red}
%\put(50,50){\circle*{6}}
\put(50,100){\circle*{6}}
\put(50,50){\circle*{6}}
\put(100,50){\circle*{6}}
\put(100,100){\circle*{6}}
\put(150,100){\circle*{6}}
\put(200,50){\circle*{6}}
\put(200,100){\circle*{6}}
\put(250,50){\circle*{6}}

% NS5
\thinlines
\color{black}
\put(40,0){\line(1,0){20}}
\put(40,150){\line(1,0){20}}
\put(90,0){\line(1,0){20}}
\put(90,150){\line(1,0){20}}
\put(140,0){\line(1,0){20}}
\put(140,150){\line(1,0){20}}
\put(190,0){\line(1,0){20}}
\put(190,150){\line(1,0){20}}
\put(240,0){\line(1,0){20}}
\put(240,150){\line(1,0){20}}
%\put(140,0){\line(1,0){20}}
%\put(140,100){\line(1,0){20}}

% NS5 intersections
%\color{red}
%\put(50,0){\circle*{6}}
%\put(50,150){\circle*{6}}
%\put(100,0){\circle*{6}}
%\put(100,150){\circle*{6}}
%\put(150,0){\circle*{6}}
%\put(150,150){\circle*{6}}
%\put(200,0){\circle*{6}}
%\put(200,150){\circle*{6}}
%\put(250,0){\circle*{6}}
%\put(250,150){\circle*{6}}

\color{blue}
\put(25,55){$\psi_{20}$}
\put(65,55){$\psi_{21}$}
\put(25,105){$\psi_{10}$}
\put(65,105){$\psi_{11}$}
\put(115,55){$\psi_{22}$}
\put(165,55){$\psi_{23}$}
\put(115,105){$\psi_{12}$}
\put(165,105){$\psi_{13}$}
\put(215,55){$\psi_{24}$}
\put(265,55){$\psi_{25}$}
\put(215,105){$\psi_{14}$}
\put(265,105){$\psi_{15}$}

\put(52,25){$\lambda_{12}$}
\put(52,75){$\lambda_{11}$}
\put(52,125){$\lambda_{10}$}
\put(102,25){$\lambda_{22}$}
\put(102,75){$\lambda_{21}$}
\put(102,125){$\lambda_{20}$}
\put(152,25){$\lambda_{32}$}
\put(152,75){$\lambda_{31}$}
\put(152,125){$\lambda_{30}$}
\put(202,25){$\lambda_{42}$}
\put(202,75){$\lambda_{41}$}
\put(202,125){$\lambda_{40}$}
\put(252,25){$\lambda_{52}$}
\put(252,75){$\lambda_{51}$}
\put(252,125){$\lambda_{50}$}
%\put(152,75){$\lambda_2^{(<)}$}

\color{black}
\put(70,40){F1}
\put(70,90){F1}
\put(18,140){D2}
\put(27,138){\vector(1,-1){20}}
%\put(118,90){D2}
%\put(127,88){\vector(1,-1){20}}
%\put(62,-2){NS5}
%\put(62,98){NS5}
%\put(162,-2){NS5}
%\put(162,98){NS5}
\put(292,58){SR}
\put(292,108){SR}
\end{picture}}
\end{picture}
\caption{An example of sectors {\it with} fermionic zero modes. Both strings attach to the first, second, and fourth D2-branes.}
\label{fig:n=2withfzm}
\end{figure}

So let us now establish the fact that if there is at least one D$2$-brane to which both strings attach, there are fermionic zero modes (see \figref{fig:n=2withfzm} for an example). Let us first note that if only one string is attached to the $\iDb^{th}$ brane, then we have
\be\label{eqn:fpsiequalagain}
\fpsi_{1,\iDb-1}=\fpsi_{1\iDb}\,,\quad \fpsi_{2,\iDb-1}=\fpsi_{2\iDb}\,.
\ee
This follows from the abelian result of \secref{subsec:fzeromodesn=1}: for example, if it is the $\iSt=1$ F$1$-string that attaches to the $\iDb^{th}$ D$2$-brane, then we have a trivial identification $\fpsi_{2,\iDb-1}=\fpsi_{2\iDb}$ (and $\lambda_{\iDb 1}=\lambda_{\iDb 2}$), while the equality $\fpsi_{1,\iDb-1}=\fpsi_{1\iDb}$ follows from the same reasoning that leads to \eqref{eqn:fpsiallequal} in \secref{subsec:fzeromodesn=1}.
If, on the other hand, both strings attach to the $\iDb^{th}$ D2-brane, then we have a weaker identity
\be\label{eqn:fpsisumequalagain}
\fpsi_{1,\iDb-1}+\fpsi_{2,\iDb-1}=\fpsi_{1\iDb}+\fpsi_{2\iDb}\,.
\ee
This follows from the same reasoning that leads to \eqref{eqn:sum4psic} in \secref{subsubsec:fzmm=1}.

From \eqref{eqn:fpsiequalagain} and \eqref{eqn:fpsisumequalagain}, we now get
\be
\fpsi_{10}+\fpsi_{20}=\fpsi_{11}+\fpsi_{21}
=\cdots=\fpsi_{1m}+\fpsi_{2m}\,,
\ee
and from the S-R-twist,
\be
\fpsi_{1m}+\fpsi_{2m}=\mathcal{P}(\fpsi_{10}+\fpsi_{20})\,,
\ee
regardless of the choice of permutation $\perm$.
These two relations together imply that
\be\label{eqn:fpsisum0}
\fpsi_{10}+\fpsi_{20}=\fpsi_{11}+\fpsi_{21}
=\cdots=\fpsi_{1m}
+\fpsi_{2m}=0\,.
\ee

Let us now divide the D$2$-brane indices $\iDb=1,\ldots,m$ so that the subset $\{\iDb_1,\ldots,\iDb_l\}$ refer to those D$2$-branes to which both strings attach. Then we can eliminate the fermionic modes $\lambda_{\iDb_1 1},\dots,\lambda_{\iDb_l 1}$ that reside on the middle D$2$-brane segments, similarly to \eqref{eqn:lam11elimI}.
Then the reasoning leading to \eqref{eqn:lambda02zero} gives us
$$
\lambda_{\iDb_1 0}=\lambda_{\iDb_1 2}
=\lambda_{\iDb_2 0}=\lambda_{\iDb_2 2}
=\cdots
=\lambda_{\iDb_l 0}=\lambda_{\iDb_l 2}=0
\,.
$$
For $\iDb\notin\{\iDb_1,\ldots,\iDb_l\}$, only one string attaches to the $\iDb^{th}$ D$2$-brane, so we have trivial identification $\lambda_{\iDb 1}=\lambda_{\iDb 0}$ or $\lambda_{\iDb 1}=\lambda_{\iDb 2}$. But since $\lambda_{0\iDb}=\lambda_{2\iDb}=0$ from the same argument as in the abelian result, we have $\lambda_{1\iDb}=0$ in either case. Also for $\iDb\notin\{\iDb_1,\ldots,\iDb_l\}$, we have $\fpsi_{\iSt\iDb}=\fpsi_{\iSt,\iDb-1}$ for $\iSt=1,2$ from \eqref{eqn:fpsiequalagain}.
It follows therefore that the only independent variables that we have at this point are
$\fpsi_{1\iDb_1},\dots,\fpsi_{1\iDb_l}$, since $\fpsi_{2\iDb_1},\dots,\fpsi_{2\iDb_l}$ can be recovered from \eqref{eqn:fpsisum0}, and $\fpsi_{\perm(\iSt)0}=\mathcal{P}^{-1}\fpsi_{\iSt m}$. The remaining equations are
\be%\label{eqn:psi11}
0
= (1-\Gamma^{239\tenx})\fpsi_{1\iDb}
=(1+\Gamma^{023})\fpsi_{1\iDb}
\,,\qquad \iDb\in\{\iDb_1,\ldots, \iDb_l\}.
\ee
Similarly to the discussion following \eqref{eqn:psi11}, there are $4$ zero modes for $\psi_{1a_l}$ and $8$ zero modes for each of $\fpsi_{1\iDb_1},\dots,\fpsi_{1\iDb_{l-1}}$. We get $8l-4$ zero modes in total.

% -------------------------------------------------------------

\subsection{The Witten Index -- results}
\label{subsec:WittenIndexRes}

According to the discussion of \secref{subsec:countfm=1} (and its extension to $n>2$), sectors with a congested binding matrix have a nonzero number of zero-modes. They therefore do not contribute to the Witten Index. Only sectors with no zero modes contribute to the Witten Index, and these are precisely the decongested sectors.

By our definition at the end of \secref{subsec:I0}, a decongested sector is a sector whose binding matrix has exactly one `1' in each column. It can be alternatively described as follows. Start with $p$ closed strings of winding numbers $n_1, n_2,\dots, n_p$, such that $n=n_1+\cdots+n_p$, and attach each of the $m$ D$2$-branes to one string. The point of attachment along the string is also important, and since the $x_3$ coordinate of the D$2$-brane is fixed, there are $n_j$ choices for the $j^{th}$ string, since the string passes through the $x_3$ coordinate $n_j$ times. The partition $n=n_1+\cdots+n_p$ determines the conjugacy class $[\perm]$ of the permutation $\perm\in S_n$ (where $\perm$ is represented as a product of cycles and $n_1,\dots,n_p$ are the lengths of the cycles), and the points of attachment determine the binding matrix.
Thus, there are initially $n^m$ choices for the attachment points, but choices that are equivalent up to relabeling of the strings should be counted only once.

Let $m_j$ be the number of D$2$-branes that end up being attached to the $j^{th}$ string. Then $m=m_1+\cdots+m_p$, and it is not difficult to see that the Hilbert space of the corresponding sector is equivalent to a tensor product $\otimes_{j=1}^p\Hilb_j(n_j,m_j)$
of decoupled Hilbert spaces. The dimension of $\Hilb_j(n_j,m_j)$ can be determined explicitly from the {\it effective twist phase} $e^{i n_j\pht}.$ Since we have the constraint $n_j\le n<\ord$, the effective phase is never trivial. In fact, in the cases relevant to us, the effective twist phase takes one of the following seven values:
$$
e^{i n_j\pht}\in\{
\pm i, e^{\pm\frac{\pi i}{3}}, e^{\pm\frac{2\pi i}{3}}, -1
\}.
$$
The first six values are the same twist we got in \secref{sec:U(1)} for $U(1)$ theory at levels $\lvk=\pm 2, \pm 1, \pm 3.$ (It is necessary to keep track of the sign if relative parity is important.)  The dimensions in these cases are given by $\dim\Hilb_j(n_j,m_j)=|\lvk|$, regardless of $m_j$. The last phase $e^{i n_j\pht}=-1$ is a new case that hasn't been discussed in \secref{sec:U(1)}, but appeared for example in the second sectors of \secref{subsubsec:n=2;m=1} and \secref{subsubsec:n=2;m=2} (for $\lvk=2$). As we saw there, it is not hard to check that in this case $\dim\Hilb_j(n_j,m_j)=4$, independently of $m_j, n_j.$

We can now summarize the results for the dimensions in \tabref{tab:DimHilbSectors}. The contribution to the Witten Index is calculated as
$$
I_m(n_1,\dots,n_j)=
\sum_{\{m_j\}:\sum m_j=m}\left(
\prod_{j=1}^p\dim\Hilb_j(n_j,m_j)
\right).
$$
\begin{table}[t]
\begin{center}
\begin{tabular}{l|l|l}
\hline
$\lvk$ & $n=n_1+\cdots+n_p$ & Contribution to Witten Index \\ \hline\hline
$3$ & $1=1$ & $3$ \\ \hline
$3$ & $2=1+1$ & $\frac{9}{2}2^m$\\ \hline
$3$ & $2=2$ & $\frac{3}{2}2^m$\\ \hline
\hline
$2$ & $1=1$ & $2$ \\ \hline
$2$ & $2=1+1$ & $2\cdot 2^m$ \\ \hline
$2$ & $2=2$ & $2\cdot 2^m$ \\ \hline
$2$ & $3=1+1+1$ & $\frac{4}{3}\cdot 3^m+2$ \\ \hline
$2$ & $3=2+1$ & $4\cdot 3^m+2$ \\ \hline
$2$ & $3=3$ & $\frac{2}{3}3^m$ \\ \hline
\hline
$1$ & $1=1$ & 1 \\ \hline
$1$ & $2=1+1$ & $\frac{1}{2}2^m$  \\ \hline
$1$ & $2=2$ & $\frac{3}{2}2^m$ \\ \hline
$1$ & $3=1+1+1$ & $\frac{1}{6}3^m+\frac{1}{2}$ \\ \hline
$1$ & $3=2+1$ & $\frac{3}{2}3^m+\frac{1}{2}$\\ \hline
$1$ & $3=3$ & $\frac{4}{3}3^m$ \\ \hline
$1$ & $4=1+1+1+1$ &
$\frac{1}{24}4^m+\frac{1}{4}2^m+\frac{1}{3}$\\ \hline
$1$ & $4=2+1+1$ & $\frac{3}{4}4^m+2^m$\\ \hline
$1$ & $4=2+2$ & $\frac{9}{8}4^m+\frac{3}{4}2^m$ \\ \hline
$1$ & $4=3+1$ & $\frac{4}{3}4^m+\frac{2}{3}$ \\ \hline
$1$ & $4=4$ & $\frac{3}{4}4^m$ \\ \hline
$1$ & $5 = 1 + 1 + 1 + 1 + 1$ &
$\frac{1}{120} 5^m +\frac{1}{12}3^m +\frac{1}{6}2^m +\frac{3}{8}$
\\ \hline
$1$ & $5 = 2 + 1 +1 + 1$ &
$\frac{1}{4} 5^m + \frac{5}{6}3^m +\frac{1}{2}2^m +\frac{1}{4}$
\\ \hline
$1$ & $5 = 2 + 2 + 1$ &
$\frac{9}{8}5^m +\frac{3}{4}3^m +\frac{9}{8}$
\\ \hline
$1$ & $5 = 3 + 1 + 1$ &
$\frac{2}{3}5^m +\frac{2}{3}3^m + \frac{1}{3}2^m$
\\ \hline
$1$ & $5 = 3 + 2$ & $2\cdot 5^m + \frac{2}{3}3^m +2^m$
\\ \hline
$1$ & $5 = 4 + 1$ & $\frac{3}{4}5^m +\frac{5}{4}$
\\ \hline
$1$ & $5 = 5$ & $\frac{1}{5}5^m$
\\ \hline\hline
\end{tabular}
\end{center}
\caption{
The contribution to the Witten Index of the sector with given $\lvk$ and $n$, and a permutation $\perm$ in the conjugacy class that corresponds to the partition $n=n_1+\cdots+n_p.$
}
\label{tab:DimHilbSectors}
\end{table}

As an example for how the entries in \tabref{tab:DimHilbSectors} were derived, take the case $n=3$ and $\lvk=2$ with partition $3=1+1+1.$ The number of decongested binding matrices is $3^m.$ (This is the total number of ways to attach $m$ D$2$-branes to the strings.) But this over-counts binding matrices that are related by a permutation of the string labels $1,2,3.$ There are three binding matrices in which all $m$ D$2$-branes are attached to the same string (so that we have one row of all 1's and two other rows of all 0's). They are of course equivalent to one another after relabeling the string indices. Excluding these $3$ configurations, to which we shall return later, we are left with $3^m-3$ configurations, and accounting for the relabeling redundancy $3!=6$, we get $\frac{1}{6}(3^m-3)$ inequivalent configurations. For each of this type of configurations, let $m_1, m_2, m_3$ denote the number of D$2$-branes that are attached to the $1^{st}, 2^{nd},$ and $3^{rd}$ string, respectively (so that $m=m_1+m_2+m_3$). For each configuration, the Hilbert space is a product of three Hilbert spaces of the $U(1)$ theory with $2m_1, 2m_2,$ or $2m_3$ charges, respectively. These Hilbert spaces were analyzed in \secref{sec:U(1)} and have $\lvk=2$ states each. So we get a total of $8$ states for each configuration of this form. On the other hand, the remaining $3$ configurations, for which all $m$ D$2$-branes are attached to the same string, are equivalent to each other up to relabeling. The string with its $m$ attached D$2$-branes has a Hilbert space that is equivalent to that of the $U(1)$ theory with $2m$ quarks, and hence possesses $2$ states, while the remaining $2$ unattached strings form a Hilbert space that corresponds to the $\perm=1$ sector of the $U(2)$ Tr-S theory with no charges, and has $3$ states \cite{Ganor:2010md}. The total number of states for this configuration is therefore $2\times 3=6$ and altogether we get the total number of states for $\lvk=2$ and $n=3=1+1+1$:
$$
8\times\tfrac{1}{6}(3^m-3)+6 =
\tfrac{4}{3}3^m+2.
$$

As another example, take $\lvk=1$ and $n=4=2+2.$ The binding matrix has four rows, labeled by string index $i=1,\ldots,4$, and for definiteness we take the permutation $\perm=(12)(34)$. For this discussion it is convenient to pretend that this sector has two closed strings, one formed by connecting $i=1$ and $i=2$ strings, the other by connecting $i=3$ and $i=4$. Each closed string has winding number $2$, to which some D$2$-branes are possibly attached. We start by considering all $4^m$ possible binding matrices, and note that there are $2\times 2^m$ binding matrices for which all D$2$-branes are attached to the same (pretend closed) string. The other $(4^m-2\times 2^m)$ binding matrices have a nonzero number of D$2$-branes attached to each of the two (pretend closed) strings. For this latter type of configurations, the over-counting factor is $8$ because we can exchange the two strings (namely exchanging $\{1,2\}$ and $\{3,4\}$, contributing an over-counting factor of $2$) and within each string we can exchange the two string indices ($i=1$ and $i=2$ for the first string, and similarly $i=3$ and $i=4$ for the second). The total number of inequivalent configurations for which neither of the strings is unattached is therefore $\frac{1}{8}(4^m-2^{m+1})$. Let $m_1>0,$ $m_2>0$ be the number of D$2$-branes attached to each string, with $m=m_1+m_2.$ Each string has an effective twist phase of $e^{2i\pht}=e^{\frac{2\pi i}{3}}$, and therefore its Hilbert space corresponds to the Hilbert space of a $U(1)$ theory with $\lvk'=3$ (which is the value of $\lvk$ for which the phase is $e^{i\pht'}=e^{\frac{2\pi i}{3}}$), and with $2m_1$ or $2m_2$ charges. Each configuration therefore has $3\times 3=9$ states. The remaining configurations have one unattached (pretend closed) string, and all $m$ D$2$-branes are attached to the other (pretend closed) string. There are $2\times 2^m$ such binding matrices, with an over-counting factor of $4=8/2$ (where $2$ is the symmetry factor which corresponds to switching the two string indices of the unattached string), so we get $2^{m-1}$ configurations. Each configuration has a Hilbert space that corresponds to $U(1)$ theory with $\lvk'=3$ and $2m$ charges inserted times the Hilbert space of $U(2)$ theory with $\lvk=1$ in the sector $\perm=(12)$, and no external charges. The latter Hilbert space is 2-dimensional \cite{Ganor:2010md}. The Hilbert space of the configuration thus has a total dimension of $3\times 2=6.$ Altogether we find the total number of states of the $n=4=2+2$ sector of the $\lvk=1$ theory to be:
$$
9\times\tfrac{1}{8}(4^m-2^{m+1})+6\times 2^{m-1}=
\tfrac{9}{8}4^m+\tfrac{3}{4}2^m.
$$

In addition to the sector-by-sector analysis described above, we can also write down closed formulas for certain types of sectors. In general, it is useful to have a formula for the number $f(n,m)$ of non-equivalent decongested binding matrices that have at least one nonzero entry in each of the $n$ rows. This corresponds to the number of configurations for which each of the $n$ strings is attached to at least one D$2$-brane. In \appref{app:fnm} we show that:
\be\label{eqn:fnm}
f(n,m)=\sum_{j=1}^n \frac{(-1)^{n-j}}{j!(n-j)!} j^m\,.
\ee
Using this result we can write a general expression for the Witten Index in the sectors with partition $n=1+1+\cdots+1.$ Such a sector can have $1\le l\le n$ unattached strings. The dimension of the Hilbert space of the unattached strings is the same as the number of states that a system of $l$ identical bosons each occupying one of $\lvk$ states has, which is $\begin{pmatrix}\lvk+l-1 \\ \lvk-1\\ \end{pmatrix}$, while the dimension of the Hilbert space of all $(n-l)$ attached strings is $\lvk^{n-l}.$ So, we get a total of
$$
\sum_{l=0}^{n-1}
\begin{pmatrix}\lvk+l-1 \\ \lvk-1\\ \end{pmatrix}
\lvk^{n-l}f(n-l,m)
%=
%\sum_{j=1}^n
%j^m
%\sum_{l=0}^{n-1}
%\lvk^{n-l}
% \frac{(-1)^{j-1}(\lvk+l-1)!}{(\lvk-1)!(l-1)!j!(n-l-j)!}
=
\sum_{j=1}^n
\left(
\sum_{l=0}^{n-j}
 \frac{(-1)^{n-l-j}\lvk^{n-l}(\lvk+l-1)!}{(\lvk-1)!l!j!(n-l-j)!}
\right)j^m
$$
states. Using this and similar techniques we get the results listed in \tabref{tab:DimHilbSectors}.
We can now add the contribution of the various sectors to the Witten Index for each $\lvk$ and $n$, listed in \tabref{tab:DimHilbSectors}, and obtain the results of \tabref{tab:WI}.
\begin{table}[t]
\begin{center}
\begin{tabular}{l|l|l|l|l}
\hline
$\lvk$ & $n$ & Witten Index &
$\dim\Hilb(\lvk,n,m=0)$ &
$\WilV^\dagger\WilV$ eigenvalues \\ \hline\hline
$3$ & $1$ & $3$ & $3$ & $1_{(3)}$ \\ \hline
$3$ & $2$ & $6\cdot 2^m$ & $9$ & $2_{(6)}, 0_{(3)}$ \\ \hline
\hline
$2$ & $1$ & $2$ & $2$ & $1_{(2)}$ \\ \hline
$2$ & $2$ & $4\cdot 2^m$ & $6$ & $2_{(4)}, 0_{(2)}$ \\ \hline
$2$ & $3$ & $6\cdot 3^m+4$ & $12$ & $3_{(6)}, 1_{(4)}, 0_{(2)}$ \\ \hline
\hline
$1$ & $1$ & $1$ & $1$ & $1_{(1)}$ \\ \hline
$1$ & $2$ & $2\cdot 2^m$ & $3$ & $2_{(2)}, 0_{(1)}$ \\ \hline
$1$ & $3$ & $3\cdot 3^m+1$ & $5$ &
$3_{(3)}, 1_{(1)}, 0_{(1)}$ \\ \hline
$1$ & $4$ & $4\cdot 4^m+2\cdot 2^m+1$ & $10$ &
$4_{(4)}, 2_{(2)}, 1_{(1)}, 0_{(3)}$ \\ \hline
$1$ & $5$ & $5\cdot5^m +3\cdot3^m+2\cdot2^m+3$ & $15$ &
$5_{(5)}, 3_{(3)}, 2_{(2)}, 1_{(3)}, 0_{(2)}$
\\ \hline
\end{tabular}
\end{center}
\caption{
The  Witten Index as a function of $\lvk, n$, and $m$.
The behavior of the Witten Index as a function of the number of quark and anti-quark pairs $m$ allows us to calculate the eigenvalues $\lambda_l$ of the operator $\WilV^\dagger\WilV$ and their multiplicities $N_l$. They are listed in the last column as ${\lambda_1}_{(N_1)}, {\lambda_2}_{(N_2)},\dots$.
}
\label{tab:WI}
\end{table}

% --------------------------------------------------------------

\subsection{Wilson loop operators and their eigenvalues}
\label{subsec:WilsonLoops}

{}From the expressions for the dimensions of Hilbert spaces with static charges we can get information about basic properties of Wilson loop operators along the two (spatial) cycles of $T^2$. Let us define the two basic $1$-cycles of $T^2$, one along the $x_1$ direction, and the other along the $x_2$ direction. We denote the low-energy limits of the two supersymmetric Wilson loop operators that correspond to these cycles, and in the fundamental representation $\Box$ of $U(n)$, by $\WilV_1$ and $\WilV_2$:
\bear
\tr_{\Box}\left(
P e^{i\int_0^{2\pi\xL_1} (A_1(t,x_1,x_2,x_3)
  +\Phi^9(t,x_1,x_2,x_3)) dx_1}\right)
&\xrightarrow{\text{Low energy limit}}&
\WilV_1
\,,\label{eqn:WilV1}\\
\tr_{\Box}\left(
P e^{i\int_0^{2\pi\xL_2} (A_2(t,x_1,x_2,x_3)
  +\Phi^9(t,x_1,x_2,x_3)) dx_2}\right)
&\xrightarrow{\text{Low energy limit}}&
\WilV_2
\,.\label{eqn:WilV2}
\eear
Here $\Phi^9$ is the adjoint scalar from the \SUSY{4} multiplet that corresponds to fluctuations of the D$3$-branes in direction $x_9$ as in \eqref{eqn:WLjSUSY}, and $\WilV_1, \WilV_2$ are operators on the Hilbert space $\Hilb(\lvk,n,m=0)$ (namely, Hilbert space without external charges). Assuming that the low-energy theory is topological we expect $\WilV_1,\WilV_2$ to be independent of $t,x_1,x_2$ altogether.
%Note that $\Phi^9$-dependent term in the operators could have potentially created a divergent path-integral, but it is regulated by quadratic terms in the action that pin the D$3$-branes to the $\Phi^9=0$ origin, and appear due to the R-symmetry twist.

A simple Wick rotation now allows us to derive the eigenvalues of $\WilV_i^\dagger\WilV_i$ from the dimensions of the Hilbert spaces
$\Hilb(\lvk,n,m)$ as a function of $m.$ Obviously, if the theory is topological the eigenvalues are the same for $i=1,2.$ Let us compactify time
on a circle with (supersymmetric) periodic boundary conditions so that $0\le x_0<2\pi T.$ Tr-S theory is now formulated on $T^3$ in directions $x_0,x_1,x_2$. Now insert the $m$ quark and anti-quark pairs.
At this point, if we let $x_2$, for example, play the role of Euclidean time then every quark corresponds to a Wilson loop operator for a loop around direction $x_0.$ In the microscopic 3+1D theory, let $\WilV'$ be such a supersymmetric Wilson loop around direction $x_0$ and at a fixed $x_3.$ In the Hilbert space of Tr-S theory on $T^2$ (in directions $x_0,x_1$), let $\WilV$ be the operator that is the low-energy limit of $\WilV'.$
It is, of course, independent of $x_3.$ Now we can write
\be\label{eqn:dimHWWm}
\dim\Hilb(\lvk,n,m) = \tr\lbrack (\WilV^\dagger\WilV)^m\rbrack
\,.
\ee

Thus, if we calculate $\dim\Hilb(\lvk,n,m)$ for all $m$, we will be able to read off the eigenvalues of $\WilV^\dagger\WilV.$ Since $\WilV^\dagger\WilV$ is a matrix
of dimension $\dim\Hilb(\lvk,n,0)$, it follows that $\dim\Hilb(\lvk,n,m)$ has to be a sum of at most $\dim\Hilb(\lvk,n,0)$ $m$-powers. We therefore expect
\be\label{eqn:dimHWWmExp}
\dim\Hilb(\lvk,n,m) =
\sum_{j=1}^{\dim\Hilb(\lvk,n,0)}\lambda_j(\lvk,n)^m
\,,
\ee
where the $\lambda_j$'s are eigenvalues of $\WilV^\dagger\WilV$ and thus are independent of $m$.
If indeed we can write $\dim\Hilb(\lvk,n,m)$ in the form \eqref{eqn:dimHWWmExp}, that will provide us with a nice test of our construction, and in particular the conjecture that Tr-S is topological. Moreover, we will be able to find the eigenvalues of $\WilV^\dagger\WilV$.

%We can also calculate the eigenvalues of $\WilV_1^\dagger\WilV_2 \WilV_1.$ In fact, we will now show that
%\be\label{eqn:trW1W2m}
%\tr\lbrack
%e^{2\pi i\sum_{j=1}^m\xvar_j}
%\rbrack
%= \tr\lbrack (\WilV_1^\dagger\WilV_2 \WilV_1)^m\rbrack
%\ee
%where the LHS is calculated in terms of the action \eqref{eqn:Mxpq}. This can be seen as follows. Consider a system with $m$ static anti-quarks and $m$ quarks that slowly move from $\scX^1=0$ at some initial time $t_i$ to $\scX^1=2\pi$ at $t_f.$ The interaction \eqref{eqn:xvarscX} can now be integrated to give $e^{2\pi i\sum_{j=1}^m\xvar_j}.$ We now Wick rotate to make time Euclidean and periodic. The partition function of the system is then given by the LHS of \eqref{eqn:trW1W2m}. On the other hand, assuming that the low-energy theory is topological we can switch the role of time and the role of $x_2.$ If $x_2$ plays the role of time, the quarks and anti-quarks worldlines are interpreted as equal-time Wilson loop operators in the Hilbert space of Tr-S formulated on $T^2$ in directions $x_1,t$. There are $m$ Wilson loops in the direction of $x_1$ and $m$ Wilson loops in the diagonal direction $x_1=t$, which can be converted via a diffeomorphism of $T^2$ to a product of the Wilson loops $\WilV_2 \WilV_1.$ In this way we get the RHS of \eqref{eqn:trW1W2m}.
%
Also, note that for $\lvk=1,2$ we have $\WilV=\WilV^\dagger$ for the following reason. For $\lvk=1,2$ we find that the order $\ord=6,4$ of the S-duality twist is even. Thus, the cyclic group $\{1,\zg,\zg^2,\dots,\zg^{\ord-1}\}$ that is generated by the S-duality twist $\zg\in\SL(2,\Z)$ contains $\zg^{\ord/2}$ which is equal to the central element $-I\in\SL(2,\Z)$. This is physically equivalent to the charge-conjugation operator ${\mathcal C}$ [see \eqref{eqn:deflvk}].
Therefore, when we continuously change the $x_3$ position of the Wilson loop $\WilV'$ until it completes $\frac{\ord}{2}$ cycles along the $x_3$ circle, it becomes the charge conjugate $(\WilV')^\dagger.$ Since we assumed that the low energy limit of $\WilV'$ is independent of $x_3$, we find that $\WilV$ is hermitian for those values of $\lvk.$ Thus, for $\lvk=1,2$ the eigenvalues of $\WilV$ are simply the square-roots of the eigenvalues of $\WilV^\dagger\WilV$, and are therefore known up to an overall sign.

% -------------------------------------------------------------

\subsection{Consistency checks}
\label{subsec:Consistency}

Now let us check the consistency of our results.
Comparing the first three columns of \tabref{tab:WI} with the corresponding columns of \tabref{tab:DimHilbSectors} we observe an interesting phenomenon ---
whereas individual sectors in \tabref{tab:DimHilbSectors} do not generally conform to the required form \eqref{eqn:dimHWWmExp}, their total contribution in \tabref{tab:WI} does!
We believe this result is a nontrivial test of our construction and derivation, and we will discuss its meaning further in \secref{subsec:CompareWithCS}. Moreover, from the behavior of the Witten Index as a function of $m$ in \tabref{tab:WI} we can read off the eigenvalues of the Wilson loop operator combination $\WilV^\dagger\WilV$ on the Hilbert space of Tr-S without charges ($m=0$). The results are listed in the last column of \tabref{tab:WI}.
In deriving the eigenvalues of $\WilV^\dagger\WilV$ we matched the expressions for the Witten Index with \eqref{eqn:dimHWWmExp}. In \eqref{eqn:dimHWWmExp} the total number of eigenvalues, taking  multiplicities into account, has to be equal to the dimension of the Hilbert space without charges. These dimensions are listed in the $4^{th}$ column of \tabref{tab:WI} as $\dim\Hilb(\lvk,n,0)$, and in cases where the number of powers appearing in the expression for the Witten Index in the $3^{rd}$ column falls short of $\dim\Hilb(\lvk,n,0)$ we have to add zero eigenvalues. (That the number of powers is always smaller than $\dim\Hilb(\lvk,n,0)$ constitutes another consistency check.) There is, however, an independent check on these results and the number of zero eigenvalues as follows.

For $\lvk>1$, there are symmetry operators $\Psym,\Qsym$ that act on the Hilbert spaces $\Hilb(\lvk,n,0).$ They were introduced in \cite{Ganor:2010md} and reviewed in \secref{subsec:ZkSym}. These operators have a geometrical interpretation in the type-IIA description, but in the original gauge theory description they are understood as large gauge transformations in the $U(1)\subset U(n)$ center. This latter interpretation allows us to immediately write their commutation relations with $\WilV$.
For concreteness, let's assume that $\WilV\equiv\WilV_1$ is a Wilson loop around the $x_1$ direction of the (type-IIB) $T^2$. Then,
\be\label{eqn:PQsymWilV}
\Qsym^{-1}\WilV\Qsym = \WilV\,,
\qquad
\Psym^{-1}\WilV\Psym = e^{\frac{2\pi i}{\lvk}}\WilV\,.
\ee
For $\lvk=2$ we argued above that $\WilV=\WilV^\dagger$, and $\WilV$ is therefore diagonalizable. The second equation of \eqref{eqn:PQsymWilV} then shows that the nonzero eigenvalues of $\WilV$ must come in pairs $(\lambda,-\lambda)$, and so the multiplicities of the nonzero eigenvalues $|\lambda|^2$ of $\WilV^\dagger\WilV$ are all even. For $\lvk=3$ we don't have a similar argument to show that $\WilV$ is diagonalizable, but assuming that it is, the nonzero eigenvalues of $\WilV$ must come in triplets $(\lambda,e^{\frac{2\pi i}{3}}\lambda, e^{-\frac{2\pi i}{3}}\lambda)$ and therefore the multiplicities of the nonzero eigenvalues $|\lambda|^2$ of $\WilV^\dagger\WilV$ must all be divisible by $3.$ This is indeed the case, as we can see from \tabref{tab:WI}.

For $\lvk=n=2$ we can say more. In this case $\Psym$ and $\Qsym$ commute and we can write $\Hilb(\lvk=2,n=2,0)$ as a direct sum $\bigoplus \Hilb_{(u,v)}(2,2,0)$ of simultaneous eigenspaces of $(\Psym,\Qsym)$, with eigenvalues $(u=\pm 1,v=\pm 1)$. It is easy to check that
$$
\dim\Hilb_{(+1,+1)}=\dim\Hilb_{(+1,-1)}=\dim\Hilb_{(-1,+1)}=2
\,,\qquad
\dim\Hilb_{(-1,-1)}=0.
$$
Now take a state $\ket{\psi}\in\Hilb_{(+1,-1)}$ and consider the $(\Psym,\Qsym)$ eigenvalues of $\WilV\ket{\psi}.$ By \eqref{eqn:PQsymWilV} they must be $(-1,-1)$, and since $\Hilb_{(-1,-1)}$ is trivial it follows that $\WilV\ket{\psi}=0.$ Therefore, $\WilV$ is identically zero on the two-dimensional subspace $\Hilb_{(+1,-1)}.$ It follows that $\WilV$ has at least two eigenvalues that are identically zero, and so does $\WilV^\dagger\WilV.$ From \tabref{tab:WI} we see that this is indeed the case, and that the multiplicity of the zero eigenvalue of $\WilV^\dagger\WilV$ is exactly $2.$

% -------------------------------------------------------------

\subsection{Comparison with Chern--Simons theory}
\label{subsec:CompareWithCS}

\begin{table}[t]
\begin{tabular}{l|l|l}
\hline\hline
$\pht=\frac{\pi}{3}$
& $n=1$ &
$U(1)_1$  \\
\cline{2-3}
$(\lvk=1)$
& $n=2$ &
$U(2)_{2,1}\oplus U(2)_{2,-3}$ \\
\cline{2-3}
& $n=3$ &
$U(3)_{3,1}\oplus [U(1)_1\times U(2)_{2,-3}]\oplus U(3)_{3,-2}$  \\
\cline{2-3}
& $n=4$ &
$U(4)_{4,1}\oplus 2[U(2)_{2,1}\times U(2)_{2,-3}]
\oplus [U(1)_1 \times U(3)_{3,-2}]\oplus\Hilb_{(2,2)}$ \\
\cline{2-3}
& $n=5$ &
$U(5)_{5,1}\oplus U(5)_{5,1}\oplus 2[U(3)_{3,1}\times U(2)_{2,-3}]
\oplus [U(1)_1\times\Hilb_{(2,2)}]\oplus $  \\
& & $[U(2)_{2,1}\times U(3)_{3,-2}]
\oplus [U(2)_{2,-3}\times U(3)_{3,-2}]  $ \\
\hline\hline
$\pht=\frac{\pi}{2}$
& $n=1$ &
$U(1)_2$ \\
\cline{2-3}
$(\lvk=2)$
& $n=2$ &
$U(2)_{4,2}\oplus U(2)_{4,-2}$ \\
\cline{2-3}
& $n=3$ &
$U(3)_{6,2}\oplus [U(1)_2\times U(2)_{4,-2}]\oplus U(3)_{6,-1}$ \\
\hline\hline
$\pht=\frac{2\pi}{3}$
& $n=1$ &
$U(1)_3$ \\
\cline{2-3}
$(\lvk=3)$
& $n=2$ &
$U(2)_{6,3}\oplus U(2)_{6,-1}$ \\
\hline\hline
\end{tabular}
\caption{
The results of \cite{Ganor:2010md} regarding the equivalence of the Hilbert spaces of Tr-S and Chern--Simons theory on $T^2$ as representations of the mapping class group $\SL(2,\Z)$ together with $\Psym,\Qsym.$ The notation $U(n)_{\lvk',\lvk''}$ corresponds to a Chern--Simons theory where the $U(1)$ part is at level $\lvk'$and the $SU(n)$ part is at level $\lvk''.$ One of the sectors ($4=2+2$) for $n=4$ and $\lvk=1$ could not be matched with a Chern--Simons theory and is therefore written explicitly as $\Hilb_{(2,2)}.$ It also appears in the $5=2+2+1$ decomposition of the $n=5$ theory.
}
\label{tab:lvkpp}
\end{table}

So far we have found the Witten Indices of Tr-S theory on $T^2$ in individual sectors, listed in \tabref{tab:DimHilbSectors}, and their sum over all sectors, listed in \tabref{tab:WI}. We have also seen that the results pass some nontrivial consistency checks in \secref{subsec:Consistency}. These results are however supposed to provide some clues about what Tr-S theory is. Is it a known theory, or is it an entirely new theory? How should we interpret the results from  \tabref{tab:DimHilbSectors}?

As a first step, we have to know whether different ``sectors'' correspond to different theories, or whether they are part of the same theory. Following the results in \cite{Ganor:2010md} regarding the Hilbert spaces $\Hilb(\lvk,n,m=0)$ and their decomposition as representations of the mapping class group $\SL(2,\Z)$ of (the type-IIB) $T^2$, it was proposed there that a sector $[\perm]$ corresponds to a superselection sector of Tr-S theory on $\R^{2,1}$ --- perhaps a discrete remnant of an expectation value of a Wilson loop along the compact $x_3$ direction. Furthermore, it was observed in \cite{Ganor:2010md} that strictly as representations of $\SL(2,\Z)$ and operators $\Psym,\Qsym$, the Hilbert spaces of most of the sectors are equivalent to the Hilbert spaces of (pure) Chern--Simons theories at various levels and with various gauge groups that are in general subgroups of $U(n).$ We have reproduced the general results of \cite{Ganor:2010md} in \tabref{tab:lvkpp}. For $n=2$, for example, the breakdown into individual sectors is reproduced in \tabref{tab:lvkpp(n=2)}. (Note that, as explained in \cite{Ganor:2010md}, the Chern--Simons theory level of the $U(1)$ part of the gauge group is given by $\lvk'=n\lvk$ in all cases.)  Naturally, it was then conjectured that Tr-S in each of these sectors is equivalent to the Chern--Simons theory at the corresponding level and with the corresponding gauge group. But given the results of \tabref{tab:DimHilbSectors}, we can now take a critical look at some of these conjectures.

\begin{table}
\begin{tabular}{l|l|l}
\hline\hline
$\lvk$ & $n=n_1+\cdots+n_p$ & Hilbert space
\\
\hline\hline
$1$ & $2=1+1$ & $U(2)_{2,1}$ \\
$1$ & $2=2$ & $U(2)_{2,-3}$ \\
\hline
$2$ & $2=1+1$ & $U(2)_{4,2}$ \\
$2$ & $2=2$ & $U(2)_{4,-2}$ \\
\hline
$3$ & $2=1+1$ & $U(2)_{6,3}$ \\
$3$ & $2=2$ & $U(2)_{6,-1}$ \\
\hline\hline
\end{tabular}
\caption{
The $n=2$ results of \cite{Ganor:2010md}, sector by sector. Each Hilbert space of a Tr-S sector is equivalent, as a representation of the mapping class group $\SL(2,\Z)$ and $\Psym,\Qsym$, to a corresponding Hilbert space of Chern--Simons theory. The notation $U(2)_{\lvk',\lvk''}$ corresponds to a Chern--Simons theory where the $U(1)$ part is at level $\lvk'$ and the $SU(2)$ part is at level $\lvk''.$
}
\label{tab:lvkpp(n=2)}
\end{table}

To extract useful information out of \tabref{tab:DimHilbSectors} we need to know how to match a sector of Tr-S with $m>0$ charge pairs to a sector of Tr-S with no charges $(m=0).$ A sector with $m=0$ is described entirely by the conjugacy class $[\perm]$ of the permutation $\perm\in S_n,$ or alternatively, by the partition $n=n_1+n_2+\cdots+n_p.$ A sector with $m>0$, on the other hand, is described by $[\perm]$ together with a binding matrix $\PMat_{ia}$, up to relabeling of string indices $i$, and for general sectors, combinations $(\PMat,[\perm])$ and $(\PMat',[\perm'])$ with different conjugacy classes ($[\perm]\neq[\perm']$) may be equivalent. In general, therefore, we cannot unambiguously assign a sector of $m=0$ theory to a given sector with $m>0$. This is also clear because the $m>0$ sectors have open strings while the $m=0$ sectors only have closed strings.

However, if we restrict to {\it decongested} sectors we can overcome this problem. Since a decongested sector has exactly one pair of open strings ending on each D$2$-brane, we can formally align and recombine without ambiguity these two ends to form a configuration of closed strings, thereby creating a unique $m=0$ sector out of a decongested $m>0$ sector. In fact, the ``pretend closed'' terminology of \secref{subsec:WittenIndexRes} and the partitions $n=n_1+\cdots+n_p$ appearing in \tabref{tab:DimHilbSectors} took advantage of this fact.

But now we face a serious obstacle. It was argued in \secref{subsec:Consistency} that any sector whose entry in \tabref{tab:DimHilbSectors} does not conform to \eqref{eqn:dimHWWmExp} --- one for which the coefficient of any $m^{th}$ power in its contribution to the Witten Index is not an integer --- cannot possibly be a stand-alone theory. For consistency we have to, at the very least, combine sectors so that their total contribution to the Witten Index will be of the form \eqref{eqn:dimHWWmExp}. Thus, for example, both of the $\lvk=n=2$ sectors might be individual theories corresponding to different ``superselection'' sectors. But the $\lvk=2$ and $n=3$ sectors corresponding to the partitions $3=1+1+1$ and $3=3$ cannot be separate theories. Similarly, the $\lvk=1$ and $\lvk=3$ sectors with partitions $2=1+1$ and $2=2$ cannot be separate theories either. This, we have to admit, is evidence against at least some of the conjectures that are implicit in \tabref{tab:lvkpp}.

So, still focusing on the $U(2)$ case, let us assume that we need to combine both $2=1+1$ and $2=2$ sectors for $\lvk=1,3$, and let us remain agnostic about whether we need to combine or not the two sectors for the $\lvk=2$ case. Let us proceed and ask whether in this way Tr-S theory can still be a pure Chern--Simons theory in these cases. What can we learn from \tabref{tab:WI}? We are going to make the assumption that if indeed Tr-S is identified with pure Chern--Simons theory then the Wilson loop operator $\WilV$ is identified with a Wilson loop in Chern--Simons theory (wound around one of the nontrivial $1$-cycles of $T^2$). We will now compare the information from \tabref{tab:WI} about the eigenvalues of Wilson loops with what we know about $U(2)$ Chern--Simons theory.
\subsubsection*{Wilson loop operators in $U(2)$ Chern--Simons theory}
Let us begin by reviewing the known $U(2)$ Chern--Simons results. The Hilbert space of $SU(2)$ Chern--Simons theory on a torus $T^2$ at level $\lvk''$ is $(\lvk''+1)$-dimensional, and as explained in \cite{Witten:1988hf, Verlinde:1988sn}, there exists a canonical basis in which basis states are labeled by $SU(2)$ spin $j=0,\tfrac12,\ldots, \tfrac{\lvk''}{2},$ once we choose a basis of $1$-cycles $a$ and $b$ for the first homology group $H_1(T^2;\Z)$ of the torus. When we think of $T^2$ as the boundary of a solid torus, the $a$-cycle is the one that becomes contractible inside the solid torus, while the $b$-cycle remains non-trivial. The state labeled by spin $j$ is then defined in terms of the wavefunction whose value is given by the path integral of Chern--Simons theory on the solid torus with a Wilson loop in the spin $j$ representation inserted along the $b$-cycle. We will denote such basis states of the Hilbert space by $\ket{\spm}$, with $\spm\equiv 2j=0,\ldots,\lvk''$.

The action of a Wilson loop operator $W^{(n_a,n_b)}$ in any representation of $SU(2)$ that winds around the torus $n_a$ times along the $a$-cycle and $n_b$ times along the $b$-cycle was given in \cite{Labastida:1990bt}. For our present purpose, we need the result for the Wilson loop in the fundamental representation with $n_a=1$, $n_b=0$:
\be
W\equiv W^{(1,0)}=\sum_\spm 2\cos\frac{\pi(\spm+1)}{\lvk''+2}
\ket{\spm}\bra{\spm}\,.
\ee
On the other hand, the Hilbert space of $U(1)$ Chern--Simons theory at level $\lvk'$ is $\lvk'$-dimensional, and the Wilson loops act as
\be\label{eqn:WilVs}
W^{(1,0)} = \sum_{p=0}^{\lvk'-1}
e^{\frac{2\pi i}{\lvk'}p}\ket{p}\bra{p}
\,,\qquad
W^{(0,1)} =
\sum_{p=0}^{\lvk'-1}\ket{p+1}\bra{p}
\,,
\ee
where $\ket{p}$ for $p=0,\ldots,\lvk'-1$ are the basis states.

We can now combine the results for the $U(1)$ and $SU(2)$ theories to construct the Hilbert space for the $U(2)$ theory. The Hilbert space, denoted by $U(2)_{\lvk',\lvk''}$, can be obtained by first taking the tensor product of the Hilbert space of $U(1)$ theory at level $\lvk'=2\lvk$ and that of $SU(2)$ at level $\lvk''$, and then restricting to the subspace where a certain ``large'' gauge transformation acts trivially. This is because the group $U(2)$ is not simply the product of $U(1)$ and $SU(2)$, but rather $U(2)=[U(1)\times SU(2)]/\Z_2$, where $\Z_2$ is the center of $SU(2)$.

Specifically, let us first consider the following ``illegal'' gauge transformations of the $U(1)$ gauge theory on $T^2$:
\be
\Lambda'_1(x_1,x_2)=e^{ix_1/2}\,,\quad \Lambda'_2(x_1,x_2)=e^{ix_2/2}\,.
\ee
Here, $x_1$, $x_2$ are periodic coordinates on the torus with $0\leq x_i\leq 2\pi$, $i=1,2$. Since $\Lambda'_i$ ($i=1,2$) changes its value from $+1$ to $-1\in\Z_2$ as $x_i$ changes from $0$ to $2\pi$, it is not a genuine gauge transformation, and hence acts nontrivially on the physical Hilbert space. If we let $\Omega'_1$, $\Omega'_2$ be the corresponding operators on the Hilbert space of $U(1)$ Chern--Simons theory at level $\lvk'=2\lvk$, then their action on the basis states $\ket{p}$ defined in \eqref{eqn:WilVs} is given by
\be
\Omega'_1\ket{p}=\ket{p+\lvk}\,,\quad \Omega'_2\ket{p}=(-1)^p\ket{p}\,.
\ee
We can similarly define the ``illegal'' gauge transformations for the $SU(2)$ theories:
\be
\Lambda''_i(x_1,x_2)=\diag(e^{ix_i/2},e^{-ix_i/2})\,,\quad i=1,2\,.
\ee
They also change their values from the identity to $-1\in\Z_2$ as $x_i$ change from $0$ to $2\pi$. The corresponding operators $\Omega''_i$ ($i=1,2$) act on the Hilbert space of $SU(2)$ Chern--Simons theory at level $\lvk''>0$ by
\be
\Omega''_1\ket{\spm}=\ket{\lvk''-\spm}\,,\quad
\Omega''_2\ket{\spm}=(-1)^\spm\ket{\spm}\,.
\ee
In both $U(1)$ and $SU(2)$ theories, the action of $\Omega_2'$ and $\Omega_2''$ is easy to understand from the definition of the basis states $\ket{p}$ and $\ket{\spm}$, and then the action of $\Omega_1'$ and $\Omega_1''$ can be inferred from the modular transformation properties of the basis states.

We can now consider the $U(2)$ gauge theory on $T^2$ and perform the transformations $\Lambda'_i$ and $\Lambda''_i$ simultaneously. The point is that while they are ``illegal'' gauge transformations when applied separately, they together become a genuine $U(2)$ gauge transformation, as can be seen explicitly from the above expressions. Therefore, the Hilbert space $U(2)_{\lvk',\lvk''}$ is the subspace of the tensor product of the Hilbert spaces of $U(1)_{\lvk'}$ and $SU(2)_{k''}$ theories on which the operators $\Omega'_i\otimes\Omega''_i$ act trivially. We can then read off the action of the Wilson loop operators on this subspace from those of the $U(1)_{\lvk'}$ and $SU(2)_{\lvk''}$ theories. The results for the cases listed in \tabref{tab:lvkpp(n=2)} are as follows. (In the following, we consider only the action of $W=W^{(1,0)}$, the Wilson loop going around the $a$-cycle once, but $W^{(0,1)}$ is related to $W$ by modular transformation.)
\begin{itemize}
\item $\lvk=1$:
For $U(2)_{2,1}$, the invariant subspace is one-dimensional, spanned by
$$
\ket{0}_{U(1)}\otimes\ket{0}_{SU(2)}+
\ket{1}_{U(1)}\otimes\ket{1}_{SU(2)}\,.
$$
The Wilson loop operator is just the identity: $\WilV=1$.

For $U(2)_{2,3}$, the invariant subspace is two-dimensional, spanned by
\begin{equation*}
\begin{split}
&\ket{0}_{U(1)}\otimes\ket{0}_{SU(2)}+
\ket{1}_{U(1)}\otimes\ket{3}_{SU(2)}\,,\\
&\ket{0}_{U(1)}\otimes\ket{2}_{SU(2)}+
\ket{1}_{U(1)}\otimes\ket{1}_{SU(2)}\,.
\end{split}
\end{equation*}
The Wilson loop operator is given in this basis by
$$
\WilV=\diag(\phi,\phi-1),
$$
where $\phi=\tfrac{1}{2}(1+\sqrt{5})$ is the ``golden ratio.''
On the other hand, according to \tabref{tab:WI} the Tr-S results are:
$$
\WilV=\diag(\pm\sqrt{2}, \pm\sqrt{2},0),
$$
and they clearly don't agree with Chern--Simons results for $U(n)_{2,\lvk''}$ for any $\lvk''$. (We have only explicitly written down the cases $\lvk''=1,3$ above, since they appear in \tabref{tab:lvkpp(n=2)}, but it can be easily checked that other values don't give the right answer either.)

\item $\lvk=2$:
For $U(2)_{4,2}$, the invariant subspace is three-dimensional, spanned by
\begin{equation*}
\begin{split}
&\ket{0}_{U(1)}\otimes\ket{0}_{SU(2)}+
\ket{2}_{U(1)}\otimes\ket{2}_{SU(2)}\,,\\
&\ket{1}_{U(1)}\otimes\ket{1}_{SU(2)}+
\ket{3}_{U(1)}\otimes\ket{1}_{SU(2)}\,,\\
&\ket{0}_{U(1)}\otimes\ket{2}_{SU(2)}+
\ket{2}_{U(1)}\otimes\ket{0}_{SU(2)}\,.
\end{split}
\end{equation*}
The Wilson loop operator in this basis is given by
$$
\WilV=\diag(\sqrt{2},0,-\sqrt{2})\,,
$$
which is also the result for $U(2)_{4,-2}$.
Thus, the conjectures from \tabref{tab:lvkpp(n=2)} of $U(2)_{4,2}$ and $U(2)_{4,-2}$ for the sectors $2=1+1$ and $2=2$, respectively, are in precise agreement with the eigenvalues of $\WilV$ that we calculated and summarized in \tabref{tab:WI}.

\item $k=3$:
For $U(2)_{6,1}$ the invariant subspace is three-dimensional, spanned by
\begin{equation*}
\begin{split}
&\ket{0}_{U(1)}\otimes\ket{0}_{SU(2)}+
\ket{3}_{U(1)}\otimes\ket{1}_{SU(2)}\,,\\
&\ket{1}_{U(1)}\otimes\ket{1}_{SU(2)}+
\ket{4}_{U(1)}\otimes\ket{0}_{SU(2)}\,,\\
&\ket{2}_{U(1)}\otimes\ket{0}_{SU(2)}+
\ket{5}_{U(1)}\otimes\ket{1}_{SU(2)}\,.
\end{split}
\end{equation*}
The Wilson loop operator in this basis is given by
$$
\WilV=\diag(1,-\omega,\omega^2)\,,
$$
where $\omega=e^{\pi i/3}$.

For $U(2)_{6,3}$ the invariant subspace is six-dimensional, spanned by
\begin{equation*}
\begin{split}
&\ket{0}_{U(1)}\otimes\ket{0}_{SU(2)}+
\ket{3}_{U(1)}\otimes\ket{3}_{SU(2)}\,,\\
&\ket{0}_{U(1)}\otimes\ket{2}_{SU(2)}+
\ket{3}_{U(1)}\otimes\ket{1}_{SU(2)}\,,\\
&\ket{1}_{U(1)}\otimes\ket{1}_{SU(2)}+
\ket{4}_{U(1)}\otimes\ket{2}_{SU(2)}\,,\\
&\ket{1}_{U(1)}\otimes\ket{3}_{SU(2)}+
\ket{4}_{U(1)}\otimes\ket{0}_{SU(2)}\,,\\
&\ket{2}_{U(1)}\otimes\ket{0}_{SU(2)}+
\ket{5}_{U(1)}\otimes\ket{3}_{SU(2)}\,,\\
&\ket{2}_{U(1)}\otimes\ket{2}_{SU(2)}+
\ket{5}_{U(1)}\otimes\ket{1}_{SU(2)}\,.
\end{split}
\end{equation*}
The Wilson loop operator in this basis is given by
$$
\WilV=\diag(
\phi, 1-\phi, -\omega(1-\phi), -\omega\phi,
\omega^2\phi, \omega^2 (1-\phi))\,,
$$
where $\phi$ is the golden ratio as before.
The Tr-S eigenvalues that we expect have to have an absolute value of $\sqrt{2}$ or $0$, and so we don't find an agreement in this case either.

\end{itemize}

In the above, we explicitly compared the eigenvalues of Wilson loop operators only for the gauge group $U(2)$, but we can do similar computations for other gauge groups as well using the formula of \cite{Labastida:1990bt}. We find that in general the results of \tabref{tab:WI} do not agree with the eigenvalues of Wilson loop operators in Chern--Simons theories with gauge group listed in \tabref{tab:lvkpp}, except for the $\lvk=2$, $n=2$ case discussed above.

% =============================================================

\section{Discussion}
\label{sec:disc}

We have computed the Witten Index of Tr-S theory on $T^2$ with charges, and we have used the results to calculate the eigenvalues of simple Wilson loop operators in the theory. The Witten Index of the $U(n)$ theory with parameter $\lvk$ and $2m$ charges is listed in \tabref{tab:WI}. We found that for gauge group $U(2)$ and for the $\lvk=2$ case (the basic S-duality twist corresponding to $\tau\rightarrow -1/\tau$) the results are consistent with a conjecture put forward in \cite{Ganor:2010md} relating the theory to two $U(2)$ Chern--Simons theories with $U(1)\subset U(2)$ at level $4$ and the $SU(2)$ at levels $\pm 2.$ This would imply that the low-energy theory has two ``superselection'' sectors.
On the other hand, we saw that in most other cases of $n$ and $\lvk$, the simple decomposition into superselection sectors labeled by a conjugacy class in the permutation group $S_n$ (as conjectured in \cite{Ganor:2010md}) is inconsistent with the form of the Witten Index results, and several sectors have to be combined together to yield a consistent theory. What this theory is we do not know, but we were able to show that it is inconsistent with pure Chern--Simons theory, at least at low levels.

A physical perspective for understanding the discrepancy is plausibly as follows. Well-defined Wilson loops in the four-dimensional \SUSY{4} SYM theory with twisted boundary conditions flow, in the low-energy limit, to a S-duality invariant operator $\WilV_{\text{inv}}$ that is a linear combination of Wilson loops and dual monopole operators. For example, as first briefly discussed in Section 6.6 of \cite{Ganor:2010md}, for $k=2$ the relevant operator supported on any curve $C$ flows as\footnote{For other values of $\lvk$, $\WilV_{\text{inv}}$ involves mixed Wilson-'t Hooft operators. See, for example, \cite{Kapustin:2005py} for an illuminating discussion.}
$$
\WilV_{\text{inv}}(C,x_3) \longrightarrow
\WilV(C,x_3)
+\mathcal{M}(C,x_3)
+\WilV(C,x_3)^\dagger
+\mathcal{M}(C,x_3)^\dagger
$$
where $\mathcal{M}$ is the dual 't Hooft operator. It is thus possible that Wilson loops in Tr-S theory correspond to an appropriate dimensional reduction of $\WilV_{\text{inv}}$. As a simple check, we note that for the abelian case, computing $\tr\lbrack (\WilV_{\text{inv}}^\dagger\WilV_{\text{inv}})^m\rbrack$ yields the correct index $\lvk$. Moreover, for the only non-abelian case which agrees with Chern-Simons theory, namely $\lvk=n=2$, the expectation values of monopole operators and Wilson loops are identical as first explained in \cite{Moore:1989yh}. To phrase it simply, the discrepancy between Tr-S and Chern-Simons theory may be understood physically as coming from non-trivial electromagnetic boundary conditions that descend from the four-dimensional twisted theory.

We conjecture that Tr-S is a topological theory for $n<\ord$, and we presented some arguments in favor of this in \secref{subsec:IsTrStopo}. Another possible test of this could be to look for BPS states that carry nonzero momentum along $T^2.$ If the low-energy theory is topological we would expect to find only states with energies of the order of $1/\xR.$ In the type-IIA dual the momentum quantum numbers become D$0$-brane and D$2$-brane charge (where the D$2$ branes wrap directions $1,10$). It would be interesting to study the bound states of D$0$-branes with the $n$ fundamental strings. In the limit $\xR\rightarrow 0$ this system can be mapped to a sector of a $U(1)$ dipole-theory \cite{Bergman:2000cw,Bergman:2001rw}. We hope to explore this further in a separate work.

To calculate the Witten Index we divided the Hilbert space into ``sectors'' according to the pattern of closed and open strings of the dual type-IIA system. We saw that only a subset of sectors contributes to the Witten Index -- the ``decongested'' sectors. The remaining (congested) sectors have fermionic zero modes, and they do not contribute to the Index.
It would be interesting to explore these sectors further.
For example, we noted that the supersymmetric system of charges has a global $U(1)$ symmetry that is generated by the element $\xJ$ of rotations in transverse directions that acts on spinors as $\tfrac{i}{2}(\Gamma^{45}+\Gamma^{67}).$ Since $\xJ$ commutes with all the surviving supersymmetry generators it is possible to generalize the Index to
$$
I(u)\equiv\tr\left[(-)^F e^{i u\xJ}\right].
$$
This modified index receives contributions only from ground states, but can get contributions from some congested sectors as well.
At the end of \secref{subsubsec:fzmm=1} we gave an example of a congested sector with $2$ complex fermionic zero modes that are all charged under $\xJ$. Quantizing these gives a Hilbert space with $4$ states with $\xJ$ charges $-1,0,0,+1$ and which contributes a term proportional to $(2-2\cos u)$ to the index.
It is possible, however, that the fermionic zero modes interact with the bosonic modes and calculating $I(u)$ therefore requires a separate treatment and will not be pursued here.

Taking a different approach, it would be interesting to construct Tr-S directly in terms of the duality-generating theories $T(U(n))$ defined in \cite{Gaiotto:2008ak}. For example, Gaiotto and Witten argued that S-duality for $SU(2)$ is generated by starting with $T(SU(2))$, which they identified with the strongly-coupled low-energy limit of the 2+1D \SUSY{4} theory of two equally charged hypermultiplets coupled to a $U(1)$ vector-multiplet. This theory has a manifest $SU(2)\times U(1)$ global symmetry, but as conjectured in \cite{Intriligator:1996ex} and further explained in \cite{Hanany:1996ie}, the low-energy limit has an enhanced $SU(2)\times SU(2)$ symmetry. The S-duality twist, according to Gaiotto and Witten, is then realized by gauging one $SU(2)$ with the original gauge field (at $x_3=0$ in our context) and the other $SU(2)$ with the dual gauge field (the one at $x_3=2\pi\xR$). It would be interesting to derive our results for the Witten Index directly from this construction. The computation is not so trivial, of course, because the $T(SU(2))$ theory is strongly coupled.

Recently, Terashima and Yamazaki \cite{Terashima:2011qi} studied a related compactification with an S-duality twist but only \SUSY{2} supersymmetry in 2+1D. They computed the partition function of the theory on $S^3$ and related it to $\SL(2,\R)$ Chern--Simons theory. It would be interesting to understand if this construction can be modified to provide information on the \SUSY{6} setting that we studied in this paper.

In \cite{Ganor:2010md} another way to reproduce Tr-S from the $T(SU(n))$ theories was also offered. This made use of the low-energy limit of a D$3$-brane boundary on a $(p,q)$ $5$-brane, as constructed by Gaiotto and Witten using $T(SU(n))$ \cite{Gaiotto:2008ak}. The starting point for \cite{Ganor:2010md} in this context was the $(2,0)$-theory wrapping the three-dimensional submanifold of the space \eqref{eqn:zxzzz} that is defined by $\zeta_1=\zeta_2=\zeta_3=0.$ Recently, a beautiful picture of the low-energy limit of the $(2,0)$-theory compactified on a general three-dimensional manifold has emerged \cite{Dimofte:2011jd}-\cite{arXiv:1110.2115}. It would be interesting to analyze Tr-S from that perspective as well.

If Tr-S is topological then correlation functions of the low-energy limits of Wilson loops, discussed in \secref{subsec:WilsonLoops}, construct knot and link invariants. The general question, to which this paper provides only partial answers in special cases, is what are these invariants. Recently, there have been exciting new developments in the realization of knot invariants in terms of field theories and string theory (see \cite{Gukov:2004hz} -\cite{Diaconescu:2011xr} for a sample of the recent literature). A better understanding of Tr-S might provide new physical constructions of knot invariants. We hope to explore more general Wilson loops in future papers.

% ==============================================================

\acknowledgments
We are very grateful to Mina Aganagic, Tudor Dimofte, Petr Ho\v{r}ava, Yu Nakayama, Kevin Schaeffer, and Edward Witten for helpful discussions.
This work was supported in part by the
Berkeley Center of Theoretical Physics,
in part by the U.S. National Science Foundation
under grant PHY-04-57315,
and in part by the Director,
Office of Science, Office of High Energy Physics,
of the U.S. Department of Energy under Contract
No. DE-AC03-76SF00098. The work of YPH was supported in part by the National Research Foundation of Korea
(NRF) under grant number 2010-0013526.

% ==============================================================

\begin{appendix}

% ==============================================================

\section{Supersymmetry and fermionic zero modes -- details}
\label{app:SpinorsSUSY}

In this appendix we expand on various statements made
in \secref{subsec:SUSYleft}, \secref{subsec:WittenIndex}, and \secref{subsec:fzeromodesn=1} about the amount
of supersymmetry preserved by intersections of
strings and branes and the fermionic modes
that describe these systems at low-energy.
%% --------------------------------------------------------------
%\subsection{Spinor conventions}
%\label{appsub:SinorConventions}
Our conventions are as follows.
10+1D directions are denoted by
$$
I,J,K,\dots = 0,\dots,10\equiv\tenx
\,,
$$
and we use $\tenx\equiv 10$ in indices of Dirac matrices
to avoid confusion with $1,0.$
We work in Minkowski signature
$$
\eta_{IJ}dx^I dx^J = -dx_0^2+dx_1^2+\cdots + dx_{\tenx}^2
\,.
$$
All our spinors, whether in 10+1D M-theory on 9+1D type-IIA/B
are $32$-component Majorana spinors on which
the $11$-dimensional Dirac matrices $\Gamma^I$ can act.
When we need type-IIA spinors, we will specify which direction
is eliminated (as the ``M-theory direction'').
For example, the $2^{nd}$ row of  \tabref{tab:Dualities}
is obtained from the $3^{rd}$ by eliminating direction $10$,
so the resulting type-IIA spinors $\pSUSY$
are still $32$-component Majorana spinors, but they can be decomposed into left-chirality and right-chirality spinors:
$$
\pSUSY = \pSUSY_{+} + \pSUSY_{-}\,,\qquad
\pSUSY_{\pm}\equiv \frac{1}{2}(1\pm \Gamma^\tenx)\pSUSY.
$$
We will construct type-IIB spinors by performing T-duality
on another direction.
For example, the $1^{st}$ row of \tabref{tab:Dualities}
is obtained from the $2^{rd}$ by T-duality on direction $1$,
so we can define the complex
Weyl type-IIB SUSY parameters as
$$
\pSUSY_{\text{IIB}}\equiv
\pSUSY_{+} + i\Gamma^1\pSUSY_{-}
\,.
$$

%% --------------------------------------------------------------
%\subsection{10+1D Spinors}
%\label{appsub:Spinors}

%We will describe in more detail our convention for 10+1D spinors.
The Dirac matrices $\Gamma^I$ are real and satisfy
$$
\{\Gamma^I,\Gamma^J\} = 2\eta^{IJ}\,,\qquad
\Gamma^{0123456789\tenx} = 1
\,.
$$
%We can pick a representation for which $\Gamma^1,\dots,\Gamma^\tenx$
%are symmetric matrices and $\Gamma^0$ is anti-symmetric
%\cite{Green:1987sp}.
%Our spinors are anti-commuting $32$-component Majorana spinors
%$\pSUSY.$ The conjugate spinors are defined by
%$$
%\bpSUSY \equiv\pSUSY^t\Gamma^0
%$$
%and satisfy
%$$
%\bpSUSY_1\Gamma^{I_1}\Gamma^{I_2}\cdots\Gamma^{I_k}\pSUSY_2
%=
%(-1)^k
%\bpSUSY_2\Gamma^{I_k}\Gamma^{I_{k-1}}\cdots\Gamma^{I_1}\pSUSY_1
%\,,
%$$
%and the Fierz identity
%$$
%\pSUSY_1\bpSUSY_2
%=-\sum_{k=0}^5\frac{1}{32 k!}(-1)^{k(k-1)/2}
%(\bpSUSY_2\Gamma_{I_1\dots I_k}\pSUSY_1)\Gamma^{I_1\dots I_k}
%\,,
%$$

%% --------------------------------------------------------------
%\subsection{M$2$-brane}
%\label{appsub:M2}
Now take an M$2$-brane in directions $0,9,\tenx.$
We denote
$$
\mu,\nu,\dots = 0,9,\tenx;\qquad
a,b,c,\dots = 1,\dots,8
\,.
$$
The M$2$-brane low-energy fields are
the scalars $\Xsc^a$ ($a=1,\dots,8$) and
the spinors $\Xsp$, which satisfy the chirality condition
$$
\Gamma^{09\tenx}\Xsp = -\Xsp
\,.
$$
Let $\pSUSY$ be the 10+1D SUSY parameter and set
$$
\pSUSY_l\equiv \tfrac{1}{2}(1-\Gamma^{09\tenx})\pSUSY
\,,\qquad
\pSUSY_r\equiv \tfrac{1}{2}(1+\Gamma^{09\tenx})\pSUSY
$$
The SUSY transformations are
$$
\delta\Xsp = \pSUSY_l
+\px{\mu}\Xsc^a\Gamma^\mu\Gamma_a\pSUSY_r
+(\bXsp\Gamma^\mu\pSUSY_r)\px{\mu}\Xsp
\,,\qquad
\delta\Xsc^a =
\bXsp\Gamma^a\pSUSY_r
+\px{\mu}\Xsc^a\bXsp\Gamma^\mu\pSUSY_r
\,.
$$
The spinors are real and $\bXsp\equiv\Xsp^t\Gamma^0.$
{}From the point of view of the
2+1D worldvolume theory, the $\pSUSY_r$ parameters
generate worldvolume supersymmetry transformations,
while $\pSUSY_l$ generate the $\kappa$-symmetry \cite{Becker:1995kb}.
%Note that these transformations generate the 10+1D
%supersymmetry algebra,
%$$
%\lbrack\delta_1,\delta_2 \rbrack =
%i\bpSUSY_1 \Gamma^I\pSUSY_2 P_I
%$$
%which acts on the low-energy fields of the M$2$-brane as:
%$$
%\lbrack\delta_1,\delta_2 \rbrack\Xsc^a =
%\bpSUSY_1 \Gamma^a\pSUSY_2
%+\px{\mu}\Xsc^a\bpSUSY_1 \Gamma^\mu\pSUSY_2
%\,,\qquad
%\lbrack\delta_1,\delta_2 \rbrack\Xsp =
%\px{\mu}\Xsp(\bpSUSY_1 \Gamma^\mu\pSUSY_2)
%$$
%where we used the equations of motion
%$$
%\Gamma^\mu\px{\mu}\Xsp = 0
%\,.
%$$
%The term $\bpSUSY_1 \Gamma^a\pSUSY_2$ signifies
%the fact that $\Xsc^a$ corresponds to the $x^a$
%transverse coordinate of the M$2$-brane, and hence
%translations act on it affinely.
If we compactify this M$2$-brane
on $T^2$ by making directions $1,2$ periodic,
the spinors $\Xsp$ will have $16$ linearly independent
zero modes, which generate a multiplet of $256$ states.
These states are invariant under all supersymmetries with
$\pSUSY_l=0$, but not invariant under supersymmetries
with $\pSUSY_r$. As is customary, we refer to the
supersymmetry transformation with $\pSUSY_l=0$ as
the ``unbroken supersymmetries.''

Now consider an M$2$-brane stretched along directions $2,3$,
which upon reduction to type-IIA on direction $2$ will become an F$1$ in direction $3.$ At low-energy there are $8$ scalar fields in the vector representation $8_v$ of the group $SO(8)$ of rotations
in transverse directions $1,4,5,6,7,8,9,10$, as well as their
superpartners which are spinors in $(2,8_s)$ of $SO(2,1)\times SO(8).$ These spinors $\psi$ satisfy
$$
\psi = -\Gamma^{023}\psi = -\Gamma^{1456789\natural}\psi
\,.
$$
Upon reduction to type-IIA we write
$$
\psi = \psi_{L}+\psi_{R}
$$
where
$$
\psi_L = \frac{1}{2}(1+\Gamma^2)\psi\,,
\qquad
\psi_R = \frac{1}{2}(1-\Gamma^2)\psi\,,
$$
satisfy
$$
\Gamma^{03}\psi_L = \psi_L
\,,\quad
\Gamma^{03}\psi_R = -\psi_R
\,.
$$
which become left- and right-moving massless fields
along the string. Note that in type-IIA both $\psi_L$ and $\psi_R$
are in $8_s$ of $SO(8).$
Now, consider an M$2$-brane along directions $9,10$,
which becomes a D$2$-brane in type-IIA,
and compactify direction $10$ as well.
This M$2$-brane has low-energy fermions $\chi$ satisfying
$$
\chi = -\Gamma^{09\tenx}\chi = -\Gamma^{12345678}\chi
\,.
$$
Now compactify direction $x_{10}.$
At low-energy, below the $x_{10}$ compactification scale,
the wrapped D$2$-brane looks like a string which has
left- and right-moving massless fields $\chi_{L,R}$ along it.
The are defined by:
$$
\chi = \chi_{R}+\chi_{L}
\,,
$$
where
$$
\chi_L = \tfrac{1}{2}(1+\Gamma^\tenx)\chi\,,
\qquad
\chi_R = \tfrac{1}{2}(1-\Gamma^\tenx)\chi\,,
\qquad
\Gamma^{09}\chi_L = \chi_L
\,,\qquad
\Gamma^{09}\chi_R = -\chi_R
\,.
$$
Next, we consider a configuration where such a D$2$-brane
has two F$1$ strings (with the same orientation) emanating from a point on it: one string in the positive $x_3$ direction
and one string in the negative $x_3$ direction.
Note that the total charge at the point of origin is zero,
since the charge of the endpoint of one string cancels
the charge of the endpoint of the other string.
Denote the low-energy fields on the string in the positive
$x_3$-direction by $\psi^{(>)}$, and denote the low-energy fields on the string in the negative $x_3$-direction by $\psi^{(<)}$.
Similarly, the wrapped D$2$-brane, at energies below
the $x_{10}$ compactification scale, has
low-energy fields $\chi^{(<)}$ for the $x_9<0$ side
and $\chi^{(>)}$ for the $x_9>0$ side.
We are interested in the boundary conditions
that connect the values of the $4$ fields
$\psi^{(<,>)},\chi^{(<,>)}$ at the intersection point.

If we lift this system back to M-theory we get
two M$2$-branes that intersect at a point.
Perhaps the easiest way to derive the requisite boundary
conditions is to deform this system to a smooth M$2$-brane
that extends along a surface that, in appropriate complex
coordinates described below, is a holomorphic curve.
The low-energy reduction, below the $x_2,x_{10}$ compactification
scales, looks like a $(p,q)$-web as in figure
\figref{fig:D2F1junction}(b) (see \cite{Aharony:1997bh,Bergman:1998gs} for some examples).
The smooth geometry can be described by
techniques similar to those developed
in \cite{Witten:1997sc}. We define two complex coordinates
$$
u\equiv e^{\frac{i x_2 + x_3}{\xL_2}}
\,,\qquad
v\equiv e^{\frac{i x_{10} + x_9}{\xL_{10}}}
$$
where $\xL_2,\xL_{10}$ are the radii of directions $2$ and $10.$
The smooth holomorphic curve is given by
\be\label{eqn:uvCurve}
(u-1)(v-1)=C
\ee
where $C\neq 0$ is a constant.
Note that this is a deformation of the singular curve
$(u-1)(v-1)=0.$
An M$2$-brane that wraps this holomorphic curve
will have a low-energy fermionic field $\lambda$ on it.
We are looking for zero-modes of this field
which have a finite limit at the $4$ semi-infinite directions
$x_3\rightarrow\pm\infty$ and $x_9\rightarrow\pm\infty.$
Below, we explain how to make the connections:
\be\label{eqn:chipsibc}
\lambda(x_9=-\infty)\rightarrow
\chi^{(<)}
\,,\quad
\lambda(x_9=\infty)\rightarrow
\chi^{(>)}
\,,\quad
\lambda(x_3=-\infty)\rightarrow
\psi^{(<)}
\,,\quad
\lambda(x_3=\infty)\rightarrow
\psi^{(>)}
\,.
\ee
The linear algebraic relations among these four limit values
will constitute the requisite boundary conditions.

Consider a part of the M$2$-brane near $x_3\rightarrow\infty.$
It is approximately at constant $x_9,x_{10}$ and stretches
in directions $x_2,x_3.$ The spinor can be decomposed
according to the eigenvalue of $\Gamma^{23}$ as
\be\label{eqn:lambdaDecomp}
\lambda = \eta_{+}+\eta_{-}
\,,\qquad
\eta_\pm\equiv \tfrac{1}{2}(1\pm i\Gamma^{23})\lambda
\,,\qquad
\Gamma^{23}\eta_\pm =\mp i\eta_\pm
\,,\qquad\text{(near $x_3\rightarrow\infty$)}
\,.
\ee
We then calculate the zero-mode equation
$$	
0 = (\Gamma^2\partial_2+\Gamma^3\partial_3)\eta_\pm
= \Gamma^2(\partial_2\mp i\partial_3)\eta_\pm
$$
and so $\eta_{+}$ is holomorphic in $x_3+i x_2$, and hence in $u$,
while $\eta_{-}$ is anti-holomorphic.
When extending to the bulk of the holomorphic curve,
we have to keep track of the tangent and normal bundles
of the M$2$-brane surface given by \eqref{eqn:uvCurve}.
At an arbitrary point $p$ on this surface the
tangent plane $T_p$ can be thought of as a subspace of
$\R^4$ that is the constant tangent space in the
$x_2,x_3,x_9,x_{10}$ directions. As $p$ varies the
embedding $T_p\subset\R^4$ varies. Locally, we can pick
a rotation $\Omega_p\in U(2)\subset \Spin(4)$ that maps
the tangent  plane $T_p$ to a common plane, which we
choose to be the $x_2-x_3$ plane, and also varies
smoothly with $p.$ At any fixed point $p$ on the surface,
this rotation $\Omega_p$ is unique up to
$SO(2)\times SO(2)$ (rotations in the $x_2-x_3$ and
$x_9-x_{10}$ planes separately). Near $x_9\rightarrow\infty$,
for example, $T_p$ is the $x_9-x_{10}$ plane and we can take
the rotation in spinor representation to be
\be\label{eqn:Omega239ten}
\Omega = e^{\frac{\pi}{4}(\Gamma^{2\tenx}+\Gamma^{93})}
=e^{\frac{\pi}{4}(1+\Gamma^{239\tenx})\Gamma^{93}}
= \tfrac{1}{2} (1+\Gamma^{2\tenx}) (1+\Gamma^{93}) .
\ee
If we decompose the fermionic field near
$x_9\rightarrow\infty$ as
\be\label{eqn:chipm}
\chi^{(>)}=\chi^{(>)}_{R} + \chi^{(>)}_{L}
\,,\qquad
\chi^{(>)}_R=\tfrac{1}{2}(1 + \Gamma^{239\tenx})\chi^{(>)}
\,,\qquad
\chi^{(>)}_L=\tfrac{1}{2}(1 - \Gamma^{239\tenx})\chi^{(>)}
\ee
the components $\chi^{(>)}_R$ and $\chi^{(>)}_L$, after rotation of
the $x_9-x_{10}$ plane into the $x_2-x_3$ plane, are
\be
\Omega \chi^{(>)}_R=\chi^{(>)}_R
\,,\qquad
\Omega \chi^{(>)}_L=\Gamma^{93} \chi^{(>)}_L
=e^{\frac{\pi}{2}\Gamma^{93}} \chi^{(>)}_L
=\Gamma^{2\tenx} \chi^{(>)}_L
=e^{\frac{\pi}{2}\Gamma^{2\tenx}} \chi^{(>)}_L.
\ee
Thus, using $\Omega$ we can map chiral spinors at any point on the surface to a common space, and thus extend \eqref{eqn:lambdaDecomp}
by setting
\be\label{eqn:etapmOmega}
\eta_{\pm}=\frac{1}{2}(1\pm i\Gamma^{23})\Omega\lambda.
\ee
Let $\Klbd$ be the canonical bundle (i.e.,
the bundle whose sections are holomorphic $(1,0)$-forms
on the curve), and let $\Nlbd=\Klbd^{-1}$ be the normal bundle (where we embed the curve in $\C^2$ in directions $2,3,9,10$).
The modes $\eta_{+}$ transform as sections of $\Klbd^{1/2}\otimes(\Nlbd^{1/2}\oplus\Nlbd^{-1/2})=\Olbd\oplus\Klbd$, where $\Olbd$ is the trivial bundle. The relation \eqref{eqn:etapmOmega} thus maps a spinor $\lambda$ to a section of $\Olbd\oplus\Klbd$ (times a trivial spinor bundle in the transverse directions).
So, altogether, we can decompose zero modes into
$$
\lambda = \lambda_{R}+\lambda_{L}\,,\qquad
\lambda_R\equiv
\tfrac{1}{2}(1+\Gamma^{239\tenx})\lambda
\,,\qquad
\lambda_L\equiv
\tfrac{1}{2}(1-\Gamma^{239\tenx})\lambda
\,.
$$
Then, zero-modes $\lambda_L$ are sections
of $\Klbd^{1/2}\otimes\Nlbd^{1/2}=\Olbd$ which is the trivial bundle, while zero-modes $\lambda_R$ are sections of
$\Klbd^{1/2}\otimes\Nlbd^{-1/2}=\Klbd.$
Thus, $\lambda_L$ is simply a holomorphic function
of $u$, with finite limits at the $4$ ends, while
$\lambda_R$ is a holomorphic $1$-form with finite
limits at the $4$ ends.

In terms of the coordinate $u$, the
curve \eqref{eqn:uvCurve} is mapped to the complex $u$-plane
with $4$ singular points: $u=0,\infty$ correspond to the
two ends of the F$1$-string, while $u=1-C,1$ correspond to
$v=0,\infty$, which are the two ends of the D$2$-brane.
Equation \eqref{eqn:chipsibc} becomes
\be\label{eqn:chipsibcc}
\lambda(u=1-C)\rightarrow
\chi^{(<)}
\,,\qquad
\lambda(u=1)\rightarrow
\chi^{(>)}
\,,\qquad
\lambda(u=0)\rightarrow
\psi^{(<)}
\,,\qquad
\lambda(u=\infty)\rightarrow
\psi^{(>)}
\,.
\ee
We denote
$$
\chi^{(<,>)}_R\equiv\tfrac{1}{2}(1+\Gamma^{239\tenx})\chi^{(<,>)}
\,,\qquad
\chi^{(<,>)}_L\equiv\tfrac{1}{2}(1-\Gamma^{239\tenx})\chi^{(<,>)}
\,.
$$
and
$$
\psi^{(<,>)}_R\equiv\tfrac{1}{2}(1+\Gamma^{239\tenx})\psi^{(<,>)}
\,,\qquad
\psi^{(<,>)}_L\equiv\tfrac{1}{2}(1-\Gamma^{239\tenx})\psi^{(<,>)}
\,.
$$
As $\lambda_L$ modes are sections of the trivial line-bundle, and are therefore constant functions, their boundary conditions must be:
\be\label{eqn:chipsibcplus}
\chi^{(<)}_L =
\chi^{(>)}_L =
\psi^{(<)}_L =
\psi^{(>)}_L
\,.
\ee
On the other hand, the $\lambda_R$ modes are sections
of the canonical bundle. They correspond to holomorphic
$1$-forms which we denote by $\omega(u)du.$
Being constant near $x_3\rightarrow\infty$ means that $\omega du$
is proportional at $u=\infty$ to $d\log u = du/u$, and so has a first-order zero there. Similar analysis of the behavior
near the other three singular points $u=0,1-C,1,$
shows that the $1$-form needs to have at most a simple pole,
and since it vanishes at $u=\infty$, the $1$-form is of the form
$$
\omega(u) =
\left(\frac{\alpha}{u-(1-C)}
+\frac{\beta}{u-1}
+\frac{\gamma}{u}
\right)du
\,.
$$
Here $\chi^{(<)}_R$ is proportional to the constant $\alpha,$
$\chi^{(>)}_R$ is proportional to the constant $\beta,$ $\psi^{(<)}_L$ is proportional to the constant $\gamma,$ and $\psi^{(>)}_L$ is proportional to the constant $-(\alpha+\beta+\gamma).$
%We need to discuss how to convert $\alpha,\beta,\gamma$ to spinors.  When we converted spinors to holomorphic $1$-forms, we restricted to spinors that satisfy $\psi_R$ and $\chi_R$.  But when we discuss holomorphic sections we had near $u=\infty$ spinors that satisfy
%$$
%i\psi^{(>)}_R =
%-\Gamma^{23}\psi^{(>)}_R=
%\Gamma^{9\tenx}\psi^{(>)}_R
%\,.
%$$
%On the other hand, near $u=1$ we have
%$$
%i\chi^{(>)}_R =
%\Gamma^{23}\chi^{(>)}_R
%=-\Gamma^{9\tenx}\chi^{(>)}_R
%\,.
%$$
%This means that $\Gamma^{39}\chi^{(>)}_R$ and $\psi^{(>)}_R$
%are of the same chirality and can appear in the same equation.
%This factor of $\Gamma^{39}$ is related to
%$\Omega$ in \eqref{eqn:Omega239ten},
%which acts as $e^{\frac{\pi}{2}\Gamma^{39}}=\Gamma^{39}$
%on modes of negative $\Gamma^{239\tenx}$
%chirality and as the identity on modes of positive
%$\Gamma^{239\tenx}$ chirality.
Converting back to spinors using \eqref{eqn:etapmOmega}, we find:
\be\label{eqn:chipsibcminus}
0 =
\Gamma^{39}\chi^{(>)}_R -
\Gamma^{39}\chi^{(<)}_R+
\psi^{(>)}_R -
\psi^{(<)}_R
\,.
\ee
Equations \eqref{eqn:chipsibcplus}-\eqref{eqn:chipsibcminus}
are the requisite boundary conditions!

% ==============================================================

\section{Additional Combinatorics}
\label{app:combinatorics}

% --------------------------------------------------------------
\subsection{The number of decongested binding matrices}
\label{app:fnm}

In \eqref{eqn:fnm} we quoted the number $f(n,m)$ of non-equivalent decongested binding matrices that have at least one nonzero entry in each of the $n$ rows. We will now derive this expression. We can easily find a recursion formula for $f(n,m)$ by noting that we can uniquely relabel the strings so that the $m^{th}$ D$2$-brane is attached to the $n^{th}$ string. Suppose there are $0\le l\le m-n$ additional D$2$-branes attached to the $n^{th}$ string, then the remaining $(n-1)$ strings have $f(n-1,m-l-1)$ configurations, and therefore
\be\label{eqn:fnmRec}
f(n,m)=\sum_{l=0}^{m-n}
 \begin{pmatrix}m-1\\ l\\ \end{pmatrix}
 f(n-1,m-l-1)
=\sum_{j=n-1}^{m-1}
 \begin{pmatrix}m-1\\ j\\ \end{pmatrix}
 f(n-1,j).
\ee
Define the generating function
$$
f_n(u)\equiv \sum_{m=n}^\infty f(n,m)u^{-m}
\,.
$$
Then, \eqref{eqn:fnmRec} implies
% using
%$$
%\sum_{m=l+1}^\infty \begin{pmatrix}m-1\\ l\\ \end{pmatrix}u^{-m}
%=\frac{1}{(u-1)^{l+1}}\,,
%$$
\bear
f_n(u)&=&
%\sum_{m=n}^\infty\sum_{l=0}^{m-n}
% \begin{pmatrix}m-1\\ m-l-1\\ \end{pmatrix}
% f(n-1,m-l-1)u^{-m} =
\sum_{m=n}^\infty\sum_{j=n-1}^{m-1}
 \begin{pmatrix}m-1\\ j\\ \end{pmatrix}
 f(n-1,j)u^{-m}
=\sum_{j=n-1}^\infty\sum_{m=j+1}^\infty
 \begin{pmatrix}m-1\\ j\\ \end{pmatrix}
 f(n-1,j)u^{-m}
\nn\\
&=&
\sum_{j=n-1}^\infty(u-1)^{-(j+1)} f(n-1,j)
=\frac{1}{u-1}f_{n-1}(u-1)\,.
\nn
\eear
It follows that
$$
f_n(u)=\frac{1}{\prod_{j=1}^n (u-j)}
=\sum_{j=1}^n\frac{(-1)^{n-j}}{(j-1)!(n-j)!(u-j)}
\,,
$$
and therefore
$$
f(n,m)=\sum_{j=1}^n \frac{(-1)^{n-j}}{j!(n-j)!} j^m\,.
$$

% --------------------------------------------------------------

\subsection{A generating function for the number of sectors and Fibonacci numbers}
\label{app:GenFun}

In the following, we will count the number of sectors for all $m\geq1$. Remarkably, it turns out that the number of sectors for consecutive $m$'s follows a generalized Fibonacci sequence $=\{ 3, 7, 18, 47, 123 \ldots \}$.
To solve this combinatorial problem, we begin by introducing another set of notations to describe the binding matrices $\PMat$. We denote any continuous stretch of rows of $\PMat$ using letters $[j]$ defined as:
\be
\left( \begin{array}{c} 1\\0 \end{array} \right) \equiv [1],\,\,\,\,\,\,\,\left( \begin{array}{cc} 1&1\\1&0 \end{array}\right) \equiv [2],\,\ldots\,\,\,\,\,\, \underbrace{\left(\begin{array}{ccccc} 1&1&1&\ldots&1\\1&1&1&\ldots&0 \end{array} \right)}_{j\, \text{rows}} \equiv [j]\,\qquad
\ee
and similarly so when the zeroes are located in the first row.
Each letter represents two possible configurations: $[1]$ represents both $\bigl( \begin{array}{c} 1\\0 \end{array} \bigr)$ and $\bigl( \begin{array}{c} 0\\1 \end{array} \bigr)$;
$[2]$ represents both $\bigl( \begin{array}{cc} 1&1\\1&0 \end{array} \bigr)$ and $\bigl( \begin{array}{cc} 1&0\\1&1 \end{array} \bigr)$; and so on. Thus, for example,
\be\label{eqn:ExWord1}
\PMat = \left( \begin{array}{ccccccccccc}  1&1&1&0&1&1&1&0&1&1&1 \\   1&1&1&1&1&0&0&1&1&0&0                 \end{array} \right)\,
\text{is translated to the word $[4][2][1][1][2][1]$.}
\ee
In general we cannot recover $\PMat$ from the word, but note that different $\PMat$'s can be made equivalent after relabeling of the open strings. In particular, whenever there is a column with two $1$'s, we can relabel the open strings that are to the right of that column. This will change the matrix $\PMat$, and potentially also the accompanying permutation $\perm$, but will give an equivalent physical sector. For example, if we relabel all strings after the $5^{th}$ column of \eqref{eqn:ExWord1} we get [in the notation of \eqref{eqn:m=3sectors} of \secref{subsubsec:n=2;m=3}]:
$$
\PMat_\perm = \left( \begin{array}{ccccccccccc}
1&1&1&0&1&1&1&0&1&1&1 \\
1&1&1&1&1&0&0&1&1&0&0\end{array}\right)_{+1}
\sim
\PMat'_{\perm'} = \left( \begin{array}{ccccccccccc}
1&1&1&0&1&0&0&1&1&0&0 \\
1&1&1&1&1&1&1&0&1&1&1\end{array}\right)_{-1}
\,.
$$
This means that up to changing $\perm$ we can always assume that after a string of $\bigl( \begin{array}{c} 1\\1 \end{array} \bigr)$ columns there appears $\bigl( \begin{array}{c} 1\\0 \end{array} \bigr).$ The letters $[2], [3],\ldots$ thus translate back to a unique sequence of columns in $\PMat.$ But the letter $[1]$ can translate back to either $\bigl( \begin{array}{c} 1\\0 \end{array} \bigr)$ or $\bigl( \begin{array}{c} 0\\1 \end{array} \bigr).$

We denote the number of $[1]$ letters in a given word by $p$. Then, there are $2^p$ ways to translate the word back to $\PMat$. There are also two possibilities for $\perm$, which gives $2^{p+1}$ possibilities, but now each sector is counted twice because we can exchange the entire two rows of $\PMat$ to get equivalent sectors. Altogether, we find that a word with $p$ letters $[1]$ corresponds to $2^p$ sectors.
%Note that exchanging the two rows yields an equivalent sector and is naturally mapped to the same word. With regards to the S-R-twist, we observe that each $\PMat$ which contains one zero in every column gives rise to two distinct sectors by virtue of $\perm$ (see for example, \eqref{eqn:m=3sectors}). Other $\PMat$'s are either equivalent under $\perm$ or to a different permutation of another distinct $\PMat$.

We can now count the number of sectors for a generic $m>1$ as follows.  Let us consider words that have $p$ letters $[1]$ and $r\ge 0$ other letters, which makes $(r+p)$ letters in total. The total number of different ways to fill in the $[1]$'s is then
$\frac{(r+p)!}{r!p!}$. We need to compute also the total number of ways to write $(m-p)$ as a sum of $r$ numbers from $2,3,4,\dots$. This is calculated by subtracting $2$ from each letter and then computing the number of ways to write $(m-p-2r)$ in this form as a sum of $r$ nonnegative integers, which is simply equal to $\frac{(m-p-r-1)!}{(m-p-2r)!(r-1)!}$.
Putting these facts together, the total number of different configurations $d_m$ represented by this class of words is then
\be
\label{eqn:dm1}
d_m=\sum_{r,p}
\frac{(r+p)!(m-p-r-1)!}{(m-p-2r)!(r-1)!r! p!}2^{p}
\ee
Finally, we note that the set of $\PMat$'s that can be represented by a word lacks those which end with a
$\bigl( \begin{array}{c} 1\\1 \end{array} \bigr)$ column, i.e., $\PMat_{1m}=\PMat_{2m}=1.$ Including such configurations of which there are $d_{m-1}$, the total number of different sectors $D_m$ is then found to be
\be
\label{eqn:Dm}
D_m=d_m + d_{m-1}\,.
\ee
To find a closed form for $D_m$, we can sum  over $m,r,p$ to write down a rational generating function whose Taylor coefficents will yield $D_m$. It is convenient to consider first
\begin{eqnarray}
\label{eqn:F(t)}
G(t)&=&\sum_{m=0}^{\infty} d_m t^m
=\sum_{r,p,m} \frac{(r+p)!(m-p-r-1)!}{(m-p-2r)!(r-1)!r! p!}2^{p}t^m \cr
&=&\sum_{r,p,m}\frac{(r+p)!(m-p-r-1)!}{(m-p-2r)!(r-1)!r! p!}2^{p}t^{m-p-2r}t^{p+2r}
=\sum_{r,p}\frac{(r+p)!}{r! p!}2^{p}(1-t)^{-r}t^{p+2r}\cr
&=&\sum_p (2t)^p\bigl( 1-\frac{t^2}{1-t} \bigr)^{-p-1}
=\frac{1-t}{1-3t+t^2}\,\,.
\end{eqnarray}
Note that $d_0=1$. Now, \eqref{eqn:Dm} and \eqref{eqn:F(t)} allow us to construct the full generating function for $D_m$, defined for $m\geq 1$.
This gives us
\begin{eqnarray}
\label{eqn:generatingf}
F(t)&=&\sum_{m=1}^{\infty} D_m t^m
=\sum_{m=1}^{\infty}(d_m +d_{m-1}) t^m
=\sum_{m=1}^{\infty}(d_m t^m) + t \sum_{m=0}^{\infty} d_{m} t^m
\cr &=&
(1+t)G(t) -1
=\frac{3t-2t^2}{1-3t+t^2}
\,\,.
\end{eqnarray}
We can thus compute $D_m$ for all $m\geq 1$ easily from \eqref{eqn:generatingf}, and we get:
\be
\label{eqn:DmRes}
D_m={3, 7, 18, 47, 123, \ldots}\,,
\ee
which is a sequence that is constructed by taking the even-numbered terms of a Fibonacci sequence $L_n$ that starts with the first two seed values $L_0=2, L_1=1$. This sequence (Lucas numbers) is related to the standard Fibonacci sequence $F_n$, and we can write down an exact expression\footnote{One can also use a binomial-Fibonacci identity to write $D_m= \sum_{k=0}^m   {}^{2m-k}C_ {k} + \sum_{k=0}^{m-1}   {}^{2(m-1)-k}C_ {k} .$ Invoking \eqref{eqn:Dm}, this gives us a simple closed form for \eqref{eqn:dm1}. } for $D_m$:
\be
\label{eqn:Dm2}
D_m=L_{2m-1} + L_{2m-2} = F_{2m+1} + F_{2m-1}=\phi^{2m} + (1-\phi)^{2m}
\ee
where $\phi=(1+\sqrt{5})/2$ is the Golden ratio.

Equation \eqref{eqn:Dm2} can also be derived more directly. For this purpose, consider the family of sectors for $(m-1)$ D$2$-branes. To enumerate the sectors for $m$ D$2$-branes, we can add another column to the right\footnote{We can also add another column to the left, but to avoid over-counting, one can choose to add in only one direction.} of $\PMat$, i.e. any of $(1\,\,0)^\top$, $(0\,\,1)^\top$ and $(1\,\,1)^\top$. This gives a new set of $\PMat$'s which includes all the sectors for $m$ as a subset. Thus, we can write
\be
\label{eqn:Dm3}
D_m = 3D_{m-1} - \mathcal{O}_m
\ee
where $\mathcal{O}_m$ counts the sectors that have been over-counted. It turns out that $\mathcal{O}_m$ is exactly $D_{m-2}$.

To prove this, we observe that the family of sectors for $m-1$ D$2$-branes can always be divided into two classes: (i)those which are invariant under $\perm$ and (ii) those which are not. Class (i) matrices are bounded at both ends by at least one $(1\,\,1)^\top$, whereas for class (ii), no $(1\,\,1)^\top$ appears at either the left or right end. Now, for class (ii), when we add $(1\,\,1)^\top$ to the right end, the resulting $\PMat$ is now invariant under $\perm$, and thus $3 D_{m-1}$ over-counts by one for each distinct case. By removing the $1^{st}$ column, each such over-counted matrix can be mapped to a matrix of class (i) with $(m-2)$ D$2$-branes which end with either $(1\,\,0)^\top$ or $(0\,\,1)^\top$. Similarly, for class (i), consider each pair of terms generated by adding $(1\,\,0)^\top$ or $(0\,\,1)^\top$ at the right end. They can be easily seen to be equivalent, and thus each such matrix can be mapped to a matrix of class (i) with $(m-2)$ D$2$-branes and which end with $(1\,\,1)^\top$. Taking into account the over-counting for both classes, we see that the total number of over-counts can be mapped to precisely $D_{m-2}$.

We conclude therefore that $\mathcal{O}_m = D_{m-2}$. This means that we have, from \eqref{eqn:Dm3},
\be
\label{eqn:Dm3ii}
D_m = 3D_{m-1} - D_{m-2}
\ee
This is precisely the recurrence relation for the generalized Fibonacci sequence we have found in \eqref{eqn:Dm2}. Given $D_1$ and $D_2$, we can generate the rest of the sequence.

Finally, we note that in the case of $\lvk=2$, the number of states in each sector is always $4.$ (This is not the case for $\lvk=1,3$ as can be seen from our earlier computations.) For any $m$, we can verify this straightforwardly using the methods discussed in this section. Below, we present a short inductive derivation for $\perm=1$ sectors.

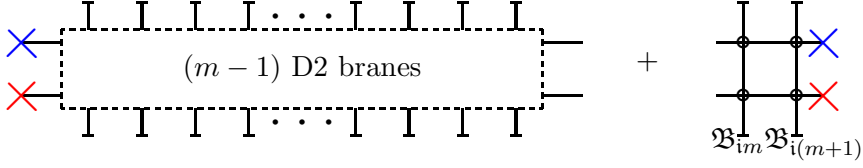
\begin{figure}[h!]
\begin{picture}(100,60)
\put(0,0){\begin{picture}(300,60)
\thicklines
%F1
\color{black}
\put(0,15){\line(1,0){15}}
\put(0,35){\line(1,0){15}}
\put(195,15){\line(1,0){15}}
\put(195,35){\line(1,0){15}}

% D2
\color{black}
\multiput(25,0)(20,0){4}{\line(0,1){10}}
\multiput(23,0)(20,0){4}{\line(1,0){4}}
\multiput(25,40)(20,0){4}{\line(0,1){10}}
\multiput(23,50)(20,0){4}{\line(1,0){4}}

\multiput(185,0)(-20,0){4}{\line(0,1){10}}
\multiput(185,40)(-20,0){4}{\line(0,1){10}}
\multiput(183,0)(-20,0){4}{\line(1,0){4}}
\multiput(183,50)(-20,0){4}{\line(1,0){4}}

%endpoints
\put(95,5){\circle*{2}}
\put(105,5){\circle*{2}}
\put(115,5){\circle*{2}}
\put(95,45){\circle*{2}}
\put(105,45){\circle*{2}}
\put(115,45){\circle*{2}}
%dashbox
\thicklines
\put(15,10){\dashbox{2}(180,30){$(m-1)$ D$2$ branes}}
\put(230,25){$+$}
\put(258,-5){$\PMat_{\iSt m}$}
\put(278,-5){$\PMat_{\iSt (m+1)}$}
%F1
\color{black}
\put(260,15){\line(1,0){40}}
\put(260,35){\line(1,0){40}}
\put(270,15){\circle{4}}
\put(290,15){\circle{4}}
\put(270,35){\circle{4}}
\put(290,35){\circle{4}}

% D2
\color{black}
\put(270,0){\line(0,1){50}}
\put(290,0){\line(0,1){50}}
\put(268,0){\line(1,0){4}}
\put(288,0){\line(1,0){4}}
\put(268,50){\line(1,0){4}}
\put(288,50){\line(1,0){4}}

%SR
\color{red}
\put(-4.6,19.6){\line(1,-1){10}}
\put(-5.0,10){\line(1,1){10}}
\color{blue}
\put(-4.6,39.6){\line(1,-1){10}}
\put(-5.0,30){\line(1,1){10}}

\color{red}
\put(295.4,19.6){\line(1,-1){10}}
\put(295.0,10){\line(1,1){10}}
\color{blue}
\put(295.4,39.6){\line(1,-1){10}}
\put(295.0,30){\line(1,1){10}}
\end{picture}}
\end{picture}
\caption{The above decomposes the $\perm=1$ class of sectors (with $(m+1)$ D$2$-branes) in a particular way. The circles on the last two D$2$-branes indicate various possibilities for the last
two columns of $\PMat$, giving possibly different sectors.}
\label{fig:Counting4states}
\end{figure}

To be definite, let us consider an arbitrary $\lvk=2$ sector in $m+1$ which begins with $\PMat_{i1}=(1\,1)^\top$. Such a sector can be thought of as an $(m-1)$ sector augmented by two more columns of $\PMat$ as represented in \figref{fig:Counting4states}.

% --------------------------------------------------------------

\noindent
% --------------------------------------------------------------
There are five different possibilities for the last two columns of $\PMat$ that we need to consider, the rest being related by symmetries. Below, we assume that $\PMat_{\iSt 1}=1$ for $\iSt=1,2$. Also, we denote the action that corresponds to the configuration before adding the last column by $I_{m}$, and the resulting action to be $I_{m+1}$. After some algebra, we simplify the various actions to be, in each case,
\begin{enumerate}

\item{$\PMat_{\iSt m}=(1\,1)^\top,
\PMat_{\iSt(m+1)}=(1\,1)^\top.$}

$I_{m+1} = I_{m} +\int \left( \pvar^{m+1} - \pvar^m \right) \left( 2d\yvar^{1m} - d\yvar^{1(m+1)} - dq^{2m}    \right)$.

\item{$\PMat_{\iSt m}=(1\,1)^\top,
\PMat_{\iSt(m+1)}=(1\,0)^\top.$}

$I_{m+1} = I_{m} + \int \left( \pvar^{m+1} - \pvar^m \right) \left( d\yvar^{1m} - d\yvar^{1(m+1)} \right)$.

\item{$\PMat_{\iSt m}=(1\,0)^\top,
\PMat_{\iSt(m+1)}=(1\,0)^\top.$}

$I_{m+1} = I_{m} + \int \left( \pvar^{m+1} - \pvar^m \right) \left( d\yvar^{1m} - d\yvar^{1(m+1)}    \right)$.

\item{$\PMat_{\iSt m}=(1\,0)^\top,
\PMat_{\iSt(m+1)}=(0\,1)^\top.$}

$I_{m+1} = I_{m} + \int \left( \pvar^{m+1} - \pvar^l \right) \left( d\yvar^{1m} - d\yvar^{1(m+1)}    \right)$,
where $\PMat_{2l}$ is the second rightmost column of $\PMat_{2\iDb}$ that is unity.

\item{$\PMat_{\iSt m}=(1\,0)^\top,
\PMat_{\iSt(m+1)}=(1\,1)^\top.$}

$I_{m+1} = I_{m} + \int \left( \pvar^{l} - \pvar^{m+1} \right) \left( d\yvar^{2m} - d\yvar^{1m}  \right)+\left( \pvar^{m} - \pvar^{m+1} \right) \left( d\yvar^{1(m+1)} - d\yvar^{1m}  \right)$.

\end{enumerate}
Thus, we see that in all of these cases, the new action (for $m+1$ branes) differs from the previous one (for $m$ branes) by a term
dependent on new conjugate pairs of $(\pvar,\qvar)$. This implies that the determinant, and hence the number of states remains the same.
Assuming other choices of $\PMat_{\iSt 1}$ and $\perm=-1$ leads to a similar conclusion too, but we will leave details to the interested reader.
Essentially, only in the $\lvk=2$ case, the number of states $= 4$ for all sectors with $m=2$ branes. The calculation above, together with similar ones for other choices of $\PMat_{\iSt 1}$ and $\perm=-1$, then implies that the number of states $= 4 \forall\,m$ by induction.

\section{An alternative set of boundary conditions using tilted D$3$-branes}
\label{app:tiltedD3}

At the end of \secref{subsec:WittenIndex} we mentioned an alternative possibly useful set of boundary conditions for the fields $\scX_\iStIIB^\mu,\fpsi_\iStIIB$ ($\iStIIB=1,\dots,2m$) at the $x_9=\DXiv$ or $x_9=-\DXiv$ end of the $2m$ open strings, which we will now describe.
The boundary conditions that we used in the main text are formally realized by D$5$-branes. Here we will instead realize the boundary conditions by D$3$-branes. By tilting the D$3$-branes,
these boundary conditions can be made to preserve one real supercharge. They also have the advantage that they can be realized more comfortably in string theory, avoiding the complications mentioned below \eqref{eqn:D5bcfpsi}.
However, they suffer from additional fermionic zero-modes which render the Witten Index identically zero. We discuss this construction below.

% --------------------------------------------------------------
\begin{figure}[t]

\begin{picture}(400,220)

\put(10,10){\begin{picture}(400,200)

\color{black}
\put(0,180){\framebox{type-IIB}}
\thicklines
\color{black}
\put(0,100){\line(1,0){350}}
\put(10,102){$n$ D$3$-branes}

\multiput(100,100)(150,0){2}{\begin{picture}(0,0)
\thicklines
\color{black}
\put(-20,100){\line(1,-1){40}}
\put(-10,92){D$3$}

\thinlines
\color{blue}
\multiput(0,2)(0,20){4}{
  \qbezier(0,0)(5,5)(0,10)
  \qbezier(0,10)(-5,15)(0,20)}
\put(0,80){\circle*{4}}

\color{red}
\put(0,0){\circle*{4}}
\put(-5,-8){$q$}

\end{picture}}

\multiput(150,100)(150,0){2}{\begin{picture}(0,0)
\thicklines
\color{black}
\put(20,-100){\line(-1,1){40}}
\put(12,-92){D$3$}

\thinlines
\color{blue}
\multiput(0,-2)(0,-20){4}{
  \qbezier(0,0)(5,-5)(0,-10)
  \qbezier(0,-10)(-5,-15)(0,-20)}
\put(0,-80){\circle*{4}}

\color{red}
\put(0,0){\circle*{4}}
\put(-2,6){$\overline{q}$}

\end{picture}}

\end{picture}}
\end{picture}
\caption{
External quark and anti-quark sources
are realized as endpoints of fundamental strings.
$2m$ D$3$-branes ($m=2$ in the picture)
control the $(x_1,x_2)$ coordinates
of the sources. $(x_1,x_2)$ are along the direction
of the $n$ D$3$-branes.
}
\label{fig:puppeteer}
\end{figure}
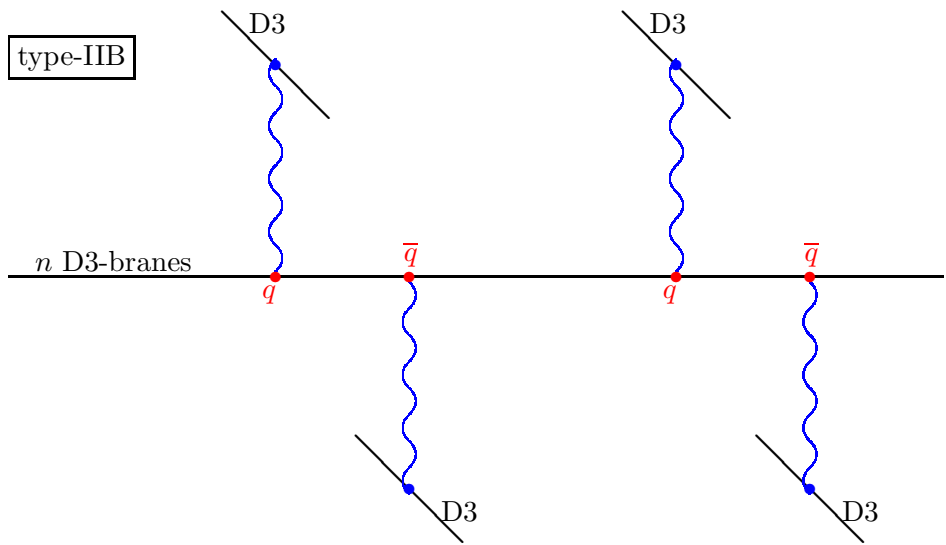
% --------------------------------------------------------------
% --------------------------------------------------------------
% --------------------------------------------------------------
% --------------------------------------------------------------
In this alternative set-up, we realize the $2m$ sources
as endpoints of $2m$ strings that end on the $n$ D$3$-branes.
The coordinates of one end of the $j^{th}$ string are
thus given by
$$
(x_0=t,
x_1=a_1^{(j)},
x_2=a_2^{(j)},
x_3=x_4=x_5=x_6=x_7=x_8=x_9=0).
$$
We control the coordinates $(x_1,x_2)$ of the endpoint
of the $j^{th}$ string by letting its
other endpoint lie on another D$3$-brane
whose $(x_1,x_2)$ position is fixed.
Thus, for each $j=1,\dots,2m$ we introduce
a D$3$-brane which controls the $j^{th}$ source
(see \figref{fig:puppeteer}).\footnote{This is reminiscent of the way Wilson loops were calculated in the topological string realization of Chern--Simons theory by Ooguri and Vafa \cite{Ooguri:1999bv}. We are grateful to Kevin Schaeffer for pointing out to us the connection with that work.}.
We take the $2m$ D$3$-branes
to be parallel to each other,
and let the $\iStIIB^{th}$ one occupy the locus
\be\label{eqn:manipulator}
x_1-x_4=a_1^{(j)}\,,\quad
x_2-x_7=a_2^{(j)}\,,\quad
x_3 = 0\,,\quad
x_9 = \DXiv_j\,,\quad
x_5=x_6=0\,.
\ee
We will assume that
$$
\DXiv_{j+m}=-\DXiv_j<0\,,\qquad
j=1,\dots,m,
$$
so that a D$3$-brane
that controls a quark ($j\le m$) is at a positive $x_9$
and a D$3$-brane that controls an anti-quark
($j>m$) is at a negative $x_9$,
and all strings have a nonzero mass.
We will also set $\DXiv_1=\cdots=\DXiv_m=\DXiv$ for simplicity.
As we argued below \eqref{eqn:2pdq}, $\DXiv$ will
not affect the low-energy description, and in fact
the mass of the string is an irrelevant operator in the IR.

Equation \eqref{eqn:manipulator} describes a D$3$-brane
that extends along the $x_8$ direction and along the diagonals
of the $x_1-x_4$ and $x_2-x_7$ planes.
The directions of the D-branes are summarized in \tabref{tab:DD}.
They are designed so that the combined system of original
$n$ D$3$-branes and the additional $2m$ D$3$-branes preserves
some amount of supersymmetry. More precisely,
we find $4$ unbroken supersymmetries
that are preserved by this combined system
\cite{Berkooz:1996km,Kachru:1999vj,Mihailescu:2000dn,Witten:2000mf}.
Including the fundamental string
and the S-R-twist we find that there is only one unbroken
real supercharge.
Another way of saying this is that out of the original
$12$ supercharges that are preserved by the $n$ D$3$-branes
and the twist, $11$ are broken by the $2m$ branes
and the fundamental strings.

\begin{table}
\begin{tabular}{|l|l|l|c|c|c|c|c|c|}
\hline\hline
Brane & type &(number) &
$1,4$ & $2,7$ & $3$ & $5,6$ & $8$ & $9$
\\ \hline\hline
Original & D3   & ($n$) &
$\longrightarrow$&$\longrightarrow$&\tx&&&
\\ \hline
String & F1   &  ($2m$) &
&&&&&\ox
\\ \hline
additional & D3   &  ($2m$) &
$\nearrow$&$\nearrow$&&&\wx&
\\ \hline\hline
\end{tabular}
\caption{
Open strings end on the original $n$ D$3$ branes
and additional $2m$ D$3$-branes.
$\nearrow$ denotes a brane that extends along the diagonal
of the corresponding plane (such as $x_1-x_4=\text{const.}$)
and $\longrightarrow$
denotes a brane that extends along the first direction
($x_4=\text{const.}$).
}
\label{tab:DD}
\end{table}
Recall that the original $n$ D$3$-branes
extend in directions $x_1,x_2,x_3$ and occupy the locus
\be\label{eqn:original}
x_4=x_5=x_6=x_7=x_8=x_9=0\,.
\ee
Therefore, an open string with one endpoint on the original
D$3$-branes and the other endpoint on one of the $2m$ D$3$-branes
will have minimal length (of $\DXiv$) if and only if
all its coordinates except $x_9$ are constant:
$$
x_1 = a_1^{(\iStIIB)}\,,\quad
x_2 = a_2^{(\iStIIB)}\,,\quad
x_3=x_4=x_5=x_6=x_7=x_8=0\,.
$$
The positions of the $2m$ D$3$-branes therefore control the positions of the $2m$ quarks and anti-quarks.

Now we transform the system to type-IIA by applying
the U-duality transformation described in \tabref{tab:Dualities}.
After the series of dualities of \tabref{tab:Dualities}
the $2m$ D$3$-branes turn into type-IIA NS$5$-branes
that wrap directions $x_1,x_4,x_7,x_8,x_{10}.$
The parameters $(a_1^{(\iStIIB)},a_2^{(\iStIIB)})$ that enter into
the conditions \eqref{eqn:manipulator} are encoded in
the compact scalar $\Phi$ and $2$-form $B$
that are part of the low-energy tensor multiplet
of the NS$5$-brane.
We have
\be\label{eqn:BandPhi}
B= (x_4 + a_1^{(j)})dx_1\wedge dx_{10}
-x_7 dx_0\wedge dx_8
\,,
\qquad
\Phi = x_7 + a_2^{(j)}.
\ee
The type-IIA system is described in \tabref{tab:DNS2}.

\begin{table}[t]
\begin{tabular}{|l|c|c|c|c|c|c|c|c|c|}
\hline\hline
Brane &
$1$ & $3$ & $4$ & $5$ & $6$ & $7$ & $8$ & $9$ & $10$
\\ \hline\hline
 F1   &
    & \tx &     &     &     &     &     &     &
\\ \hline
 D2   &
    &     &     &     &     &     &     & \ox & \wx
\\ \hline
 NS5  &
\wx &     & \wx &     &     & \wx & \wx &     & \wx
\\ \hline\hline
\end{tabular}
\caption{In the type IIA dual of \tabref{tab:DD},
the D$2$-branes end on NS$5$-branes.
Appropriate low-energy
bachground fields on the
NS$5$-brane world volumes control the position
of the Wilson loop in the type-IIB picture.
}
\label{tab:DNS2}
\end{table}

Similarly to the set-up in the main text, we have to connect each D$2$-brane that corresponds to a quark with a D$2$-brane that corresponds to an anti-quark
and glue them into a smooth D$2$-brane that ends on one
NS$5$-brane at $x_9=\DXiv$ ($j=1,\dots,m$) and another NS$5$-brane at $x_9=-\DXiv$.

For the specific purpose of computing the Witten Index however, this configuration is not so useful because, in an analogous computation as was done in \secref{subsec:fzeromodesn=1}, we found that there are fermionic zero modes that will make the contribution to the Witten Index vanish. Nonetheless, we also found that this configuration preserves one real supercharge, and thus it may turn out to be useful in understanding other aspects of the problem.

\end{appendix}

% ==============================================================
% ==============================================================
% ==============================================================
% ==============================================================
% ==============================================================

\bibliographystyle{my-h-elsevier}

\end{document}